\documentclass[longauth]{aa}
\usepackage[varg]{txfonts}
\usepackage{natbib}
\usepackage{graphicx,multirow}
\usepackage{color}

\bibpunct{(}{)}{;}{a}{}{,}

\begin{document}

\titlerunning{The CALIFA survey II. First public data release }
\authorrunning{Husemann et al.}
\title{CALIFA, the Calar Alto Legacy Integral Field Area survey}
\subtitle{II. First public data release\thanks{Based on observations collected at the Centro Astron\'omico
Hispano Alem\'an (CAHA) at Calar Alto, operated jointly by the Max-
Planck-Institut f\"ur Astronomie (MPIA) and the Instituto de Astrof\'isica de
Andaluc\'ia (CSIC)}}

\author{B.~Husemann\inst{\ref{inst1}}
   \and K.~Jahnke\inst{\ref{inst3}}
   \and S.~F.~S\'anchez\inst{\ref{inst6},\ref{inst2}}
   \and D.~Barrado\inst{\ref{inst2}}
   \and S.~Bekerait{\.e}\inst{\ref{inst1}}
   \and D.~J.~Bomans\inst{\ref{inst16},\ref{inst17}}
   \and A.~Castillo-Morales\inst{\ref{inst13}}
   \and C.~Catal{\'a}n-Torrecilla\inst{\ref{inst13}}
   \and R.~Cid~Fernandes\inst{\ref{inst7}}
   \and J.~Falc\'on-Barroso\inst{\ref{inst4},\ref{inst5}}
   \and R.~Garc{\'\i}a-Benito\inst{\ref{inst6}}
   \and R.~M.~Gonz\'alez Delgado \inst{\ref{inst6}}
   \and J.~Iglesias-P\'aramo\inst{\ref{inst6},\ref{inst2}}
   \and B.~D.~Johnson\inst{\ref{inst14}}
   \and D.~Kupko\inst{\ref{inst1}}
   \and R.~L\'opez-Fernandez\inst{\ref{inst6}}
   \and M.~Lyubenova\inst{\ref{inst3}}
   \and R.~A.~Marino\inst{\ref{inst10}}
   \and D.~Mast\inst{\ref{inst2}}
   \and A.~Miskolczi\inst{\ref{inst16}}
   \and A.~Monreal-Ibero\inst{\ref{inst6}}
   \and A.~Gil~de~Paz\inst{\ref{inst13}} 
   \and E.~P\'erez\inst{\ref{inst6}}
   \and I.~P\'erez\inst{\ref{inst24},\ref{inst25}}
   \and F.~F.~Rosales-Ortega\inst{\ref{inst8},\ref{inst125}}
   \and T.~Ruiz-Lara\inst{\ref{inst24}}
   \and U.~Schilling\inst{\ref{inst16}}
   \and G.~van~de~Ven\inst{\ref{inst3}}
   \and J.~Walcher\inst{\ref{inst1}}
   \and J.~Alves\inst{\ref{inst18}}
   \and A.~L.~de~Amorim\inst{\ref{inst7}}
   \and N.~Backsmann\inst{\ref{inst1}}
   \and J.~K.~Barrera-Ballesteros\inst{\ref{inst4}}
   \and J.~Bland-Hawthorn\inst{\ref{inst40}}
    \and C.~Coritjo\inst{\ref{inst6}}
   \and R.-J.~Dettmar\inst{\ref{inst16},\ref{inst17}}
   \and M.~Demleitner\inst{\ref{inst9}}
   \and A.~I.~D\'iaz\inst{\ref{inst8}}
   \and H.~Enke\inst{\ref{inst1}} 
   \and E.~Florido\inst{\ref{inst24},\ref{inst25}}
   \and H. Flores\inst{\ref{inst23}} 
   \and L.~Galbany\inst{\ref{inst11}}
   \and A.~Gallazzi\inst{\ref{inst20}}
   \and B.~Garc\'ia-Lorenzo\inst{\ref{inst4},\ref{inst5}}
   \and J.~M.~Gomes\inst{\ref{inst12}}
   \and N.~Gruel\inst{\ref{inst28}}
   \and T.~Haines\inst{\ref{inst15}}
   \and L.~Holmes\inst{\ref{inst29}}  
   \and B.~Jungwiert\inst{\ref{inst43}}
   \and V.~Kalinova\inst{\ref{inst3}}
   \and C.~Kehrig\inst{\ref{inst6}}
   \and R.~C.~Kennicutt Jr\inst{\ref{inst21}}
   \and J.~Klar\inst{\ref{inst1}}
   \and M.~D.~Lehnert\inst{\ref{inst23}} 
   \and {\'A}.~R.~L\'{o}pez-S\'{a}nchez\inst{\ref{inst41},\ref{inst42}}
   \and A.~de~Lorenzo-C{\'a}ceres\inst{\ref{inst5}}
   \and E.~M\'armol-Queralt\'o\inst{\ref{inst4},\ref{inst5}}
   \and I.~M\'arquez\inst{\ref{inst6}}
   \and J.~Mendez-Abreu\inst{\ref{inst4},\ref{inst5}}
   \and M.~Moll{\'a}\inst{\ref{inst10}} 
   \and A.~del~Olmo\inst{\ref{inst6}}
   \and S.~E.~Meidt\inst{\ref{inst3}}
   \and P.~Papaderos\inst{\ref{inst12}}
   \and J.~Puschnig\inst{\ref{inst18}}
   \and A.~Quirrenbach\inst{\ref{inst26}}
   \and M.~M.~Roth\inst{\ref{inst1}}
   \and P.~S\'anchez-Bl\'azquez\inst{\ref{inst8}}
   \and K.~Spekkens\inst{\ref{inst29}}
   \and R.~Singh\inst{\ref{inst3}}
   \and V.~Stanishev\inst{\ref{inst11}}
   \and S.~C.~Trager\inst{\ref{inst22}}
   \and J.~M.~Vilchez\inst{\ref{inst6}}
   \and V.~Wild\inst{\ref{inst27}}
   \and L.~Wisotzki\inst{\ref{inst1}}
   \and S.~Zibetti\inst{\ref{inst19}}
   \and B.~Ziegler\inst{\ref{inst18}}
}

\institute{Leibniz-Institut f\"ur Astrophysik Potsdam (AIP), An der Sternwarte 16, D-14482 Potsdam, Germany,\label{inst1}
\email{bhusemann@aip.de}
\and 
Max-Planck-Institut f\"ur Astronomie, K\"onigstuhl 17, D-69117 Heidelberg, Germany\label{inst3}
\and 
Instituto de Astrof\'isica de Andaluc\'ia (IAA/CSIC), Glorieta de la Astronom\'{\i}a s/n Aptdo. 3004, E-18080 Granada, Spain\label{inst6}
\and
Centro Astron\'omico Hispano Alem\'an de Calar Alto (CSIC-MPG), C/ Jes\'us Durb\'an Rem\'on 2-2, E-4004 Almer\'ia, Spain\label{inst2}
\and
Astronomisches Institut, Ruhr-Universit{\"a}t Bochum, Universit{\"a}tsstr. 150, D-44801 Bochum, Germany\label{inst16}
\and
RUB Research Department Plasmas with Complex Interactions\label{inst17}
\and
Departamento de Astrof{\'i}sica y CC. de la Atm{\'o}sfera, Universidad Complutense de Madrid, E-28040, Madrid, Spain\label{inst13}
\and
Departamento de F\'{\i}sica, Universidade Federal de Santa Catarina, P.O. Box 476, 88040-900, Florian\'opolis, SC, Brazil\label{inst7}
\and 
Instituto de Astrof\'isica de Canarias, V\'ia L\'actea s/n, La Laguna, Tenerife, Spain\label{inst4}
\and
Departamento de Astrof\'isica, Universidad de La Laguna, E-38205 La Laguna, Tenerife, Spain\label{inst5}
\and 
Institute d'astrophysique de Paris, CNRS, UPMC 98bis Bd Arago Paris 75014, France\label{inst14}
\and 
CEI Campus Moncloa, UCM-UPM, Departamento de Astrof\'{i}sica y CC$.$ de la Atm\'{o}sfera, Facultad de CC$.$ F\'{i}sicas, Universidad Complutense de Madrid, Avda.\,Complutense s/n, E-28040 Madrid, Spain\label{inst10}
\and
Departamento de F\'{\i}sica Te\'orica y del Cosmos, University of Granada, Facultad de Ciencias (Edificio Mecenas), E-18071 Granada, Spain\label{inst24}
\and
Instituto Carlos I de F\'{\i}sica Te\'orica y Computaci\'on\label{inst25}
\and
Departamento de F\'{\i}sica Te\'orica, Universidad Aut\'onoma de Madrid, E-28049, Madrid, Spain\label{inst8}
\and
Instituto Nacional de Astrof{\'i}sica, {\'O}ptica y Electr{\'o}nica, Luis E. Erro 1, 72840 Tonantzintla, Puebla, Mexico\label{inst125}
\and
University of Vienna, Department of Astrophysics, T\"urkenschanzstr. 17, 1180 Vienna, Austria\label{inst18}
\and 
Sydney Institute for Astronomy, School of Physics A28, University of Sydney, NSW 2006, Australia\label{inst40}
\and
Astronomisches Rechen-Institut, Zentrum für Astronomie der Universit\"at Heidelberg, Mönchhofstraße 12-14, D-69120 Heidelberg\label{inst9}
\and
GEPI, Observatoire de Paris, UMR 8111, CNRS, Universit{\'e} Paris Diderot, 5 place Jules Janssen, 92190, Meudon\label{inst23}
\and
CENTRA - Centro Multidisciplinar de Astrof\'isica, Instituto Superior T\'ecnico, Av. Rovisco Pais 1, 1049-001 Lisbon, Portugal\label{inst11}
\and
Dark Cosmology Centre, Niels Bohr Institute- University of Copenhagen, Juliane Mariesvej 30, DK-2100 Copenhagen, Denmark\label{inst20}
\and
Centro de Astrof{\'i}sica and Faculdade de Ci\^{e}ncias, Universidade do Porto, Rua das Estrelas, 4150-762 Porto, Portugal\label{inst12}
\and
University of Sheffield, Department of Physics and Astronomy, Hounsfield Road Sheffield, S3 7RH, UK\label{inst28}
\and
Department of Physics and Astronomy University of Missouri - Kansas City, Kansas City, MO 64110, USA\label{inst15}
\and 
Department of Physics, Royal Military College of Canada, PO Box 17000, Station Forces, Kingston, Ontario, K7K 7B4, Canada\label{inst29}
\and
Astronomical Institute, Academy of Sciences of the Czech Republic, Bo\v{c}n\'{i} II 1401/1a, CZ-141 00 Prague 4, Czech Republic\label{inst43}
\and
Institute of Astronomy, University of Cambridge, Madingley Road, Cambridge CB3 0HA, UK\label{inst21}
\and 
Australian Astronomical Observatory, PO Box 915, North Ryde, NSW 1670, Australia\label{inst41}
\and 
Department of Physics and Astronomy, Macquarie University, NSW 2109, Australia\label{inst42}
\and
Landessternwarte, Zentrum für Astronomie der Universit\"at Heidelberg, K\"onigstuhl 12, D-69117 Heidelberg, Germany\label{inst26}
\and
Kapteyn Astronomical Institute, University of Groningen,Postbus 800, NL-9700 AV Groningen, Netherlands\label{inst22}
\and
School of Physics and Astronomy, University of St Andrews, North Haugh, St Andrews, KY16 9SS, UK (SUPA)\label{inst27}
\and
INAF-Osservatorio Astrofisico di Arcetri - Largo Enrico Fermi, 5 - I-50125 Firenze, Italy\label{inst19}
}

\abstract{
We present the first public data release (DR1) of the Calar Alto Legacy Integral Field Area (CALIFA) survey. It consists of science-grade optical datacubes for the first 100 of eventually 600 nearby ($0.005<z<0.03$) galaxies, obtained with the integral-field spectrograph PMAS/PPak mounted on the 3.5m telescope at the Calar Alto observatory. The galaxies in DR1 already cover a wide range of properties in color-magnitude space, morphological type, stellar mass, and gas ionization conditions. This offers the potential to tackle a variety of open questions in galaxy evolution using spatially resolved spectroscopy. Two different spectral setups are available for each galaxy, (i) a low-resolution V500 setup covering the nominal wavelength range 3745--7500\AA\ with a spectral resolution of 6.0\AA\ (FWHM), and (ii) a medium-resolution V1200 setup covering the nominal wavelength range 3650--4840\AA\ with a spectral resolution of 2.3\AA\ (FWHM). We present the characteristics and data structure of the CALIFA datasets that should be taken into account for scientific exploitation of the data, in particular the effects of vignetting, bad pixels and spatially correlated noise. The data quality test for all 100 galaxies showed that we reach a median limiting continuum sensitivity of $1.0\times10^{-18}\,\mathrm{erg}\,\mathrm{s}^{-1}\,\mathrm{cm}^{-2}\,\mathrm{\AA}^{-1}\,\mathrm{arcsec}^{-2}$ at 5635\AA\ and $2.2\times10^{-18}\,\mathrm{erg}\,\mathrm{s}^{-1}\,\mathrm{cm}^{-2}\,\mathrm{\AA}^{-1}\,\mathrm{arcsec}^{-2}$  at 4500\AA\ for the V500 and V1200 setup respectively, which corresponds to limiting $r$ and $g$ band surface brightnesses of $23.6\,\mathrm{mag}\,\mathrm{arcsec}^{-2}$ and $23.4\,\mathrm{mag}\,\mathrm{arcsec}^{-2}$, or an unresolved emission-line flux detection limit of roughly $1\times10^{-17}\,\mathrm{erg}\,\mathrm{s}^{-1}\,\mathrm{cm}^{-2}\,\mathrm{arcsec}^{-2}$ and $0.6\times10^{-17}\,\mathrm{erg}\,\mathrm{s}^{-1}\,\mathrm{cm}^{-2}\,\mathrm{arcsec}^{-2}$, respectively. The median spatial resolution is 3\farcs7, and the absolute spectrophotometric calibration is better than 15\% ($1\sigma$). We also describe the available interfaces and tools that allow easy access to this first public CALIFA data at \url{http://califa.caha.es/DR1}.
}


\maketitle

\section{Introduction}
The Calar Alto Legacy Integral Field Area (CALIFA) survey \citep[][hereafter S12]{Sanchez:2012a} is an ongoing large project of the Centro Astron\'omico Hispano-Alem\'an at the Calar Alto observatory to obtain spatially resolved spectra for 600 local (0.005\,$<$\,$z$\,$<$\,0.03) galaxies by means of integral field spectroscopy (IFS). CALIFA observations started in June 2010 with the Potsdam Multi Aperture Spectrograph \citep[PMAS,][]{Roth:2005}, mounted to the 3.5\,m telescope, utilizing the large ($74\arcsec\times64\arcsec$) hexagonal field-of-view (FoV) offered by the PPak fiber bundle \citep{Verheijen:2004,Kelz:2006}. A diameter-selected sample of 939 galaxies were drawn from the 7th data release of the Sloan Digital Sky Survey \citep[SDSS,][]{Abazajian:2009} which will be described in Walcher et al. (in prep., hereafter W12. From this mother sample the 600 target galaxies are randomly selected. 

Combining the techniques of imaging and spectroscopy through optical IFS provides a more comprehensive view of individual galaxy properties than any traditional survey. CALIFA-like observations were collected during the feasibility studies \citep{Marmol-Queralto:2011,Viironen:2012} and the PPak IFS Nearby Galaxy Survey \citep[PINGS,][]{Rosales-Ortega:2010}, a predecessor of this survey. First results based on those datasets already explored their information content \citep[e.g.][]{Sanchez:2011,Rosales-Ortega:2011b,Alonso-Herrero:2012, Sanchez:2012b, Rosales-Ortega:2012}. CALIFA can therefore be expected to make a substantial contribution to our understanding of galaxy evolution in various aspects including, (i) the relative importance and consequences of merging and secular processes, (ii) the evolution of galaxies across the color-magnitude diagram, (iii) the effects of the environment on galaxies, (iv) the AGN-host galaxy connection,  (v) the internal dynamical processes in galaxies, and (vi) the global and spatially resolved star formation history of various galaxy types. 

In this article, we introduce the first data release (DR1) of CALIFA which grants public access to high-quality data for a set of 100 galaxies. The properties of the galaxies in the DR1 sample are summarized in Sect.~\ref{sect:DR1_sample}. We describe the data characteristics (Sect.~\ref{sect:data}), data structure (Sect.~\ref{sect:data_format}), and data quality (Sect.~\ref{sect:QC}) of the distributed CALIFA data as essential information for any scientific analysis. Several interfaces are available to access the CALIFA DR1 data, which are introduced in Sect.~\ref{sect:DR1_access}. 

\section{The CALIFA DR1 sample}\label{sect:DR1_sample}
\begin{figure}
 \resizebox{\hsize}{!}{\includegraphics{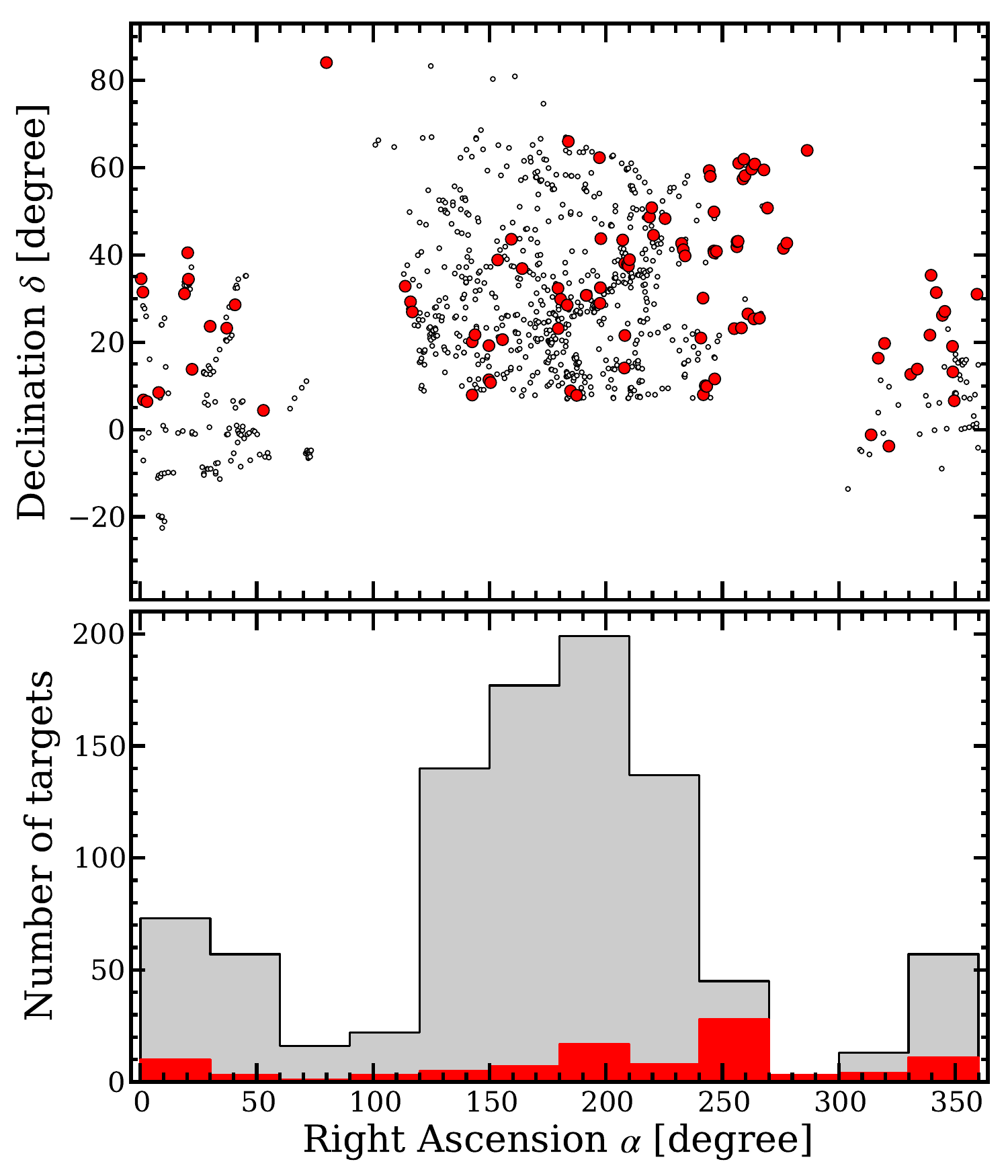}}
  \caption{Distribution of galaxies in the CALIFA mother sample on the sky (upper panel) and as a function of right ascension (lower panel). Galaxies in the CALIFA DR1 sample are highlighted by filled symbols with red color. The number distribution in bins of 30\degr\ along the right ascension is shown in the lower panel for the mother sample (gray area) and the DR1 sample (red shaded area).}
  \label{fig:DR1_dist}
\end{figure}
A sample of 939 potential CALIFA galaxies, also known as the ``CALIFA mother sample'', was drawn from the photometric catalog SDSS DR7 as outlined in S12 and W12. The primary CALIFA selection criterion was the angular isophotal diameter ($45\arcsec < D_{25}<80\arcsec$) of the galaxies, which was complemented by limiting redshifts, $0.005<z<0.03$. These redshift limits were imposed to ensure that all interesting spectral features can be observed with a fixed spectral setup of the instrument and that the sample would not be dominated by dwarf galaxies. Redshift information was taken from SIMBAD for all galaxies where SDSS DR7 spectra were unavailable. The reader is referred to W12 for a detailed characterization of the CALIFA mother sample and a thorough evaluation of the selection effects implied by the chosen selection criteria. From the CALIFA mother sample, 600 galaxies are randomly selected for observation purely based on visibility, and we refer to these galaxies as the virtual final CALIFA sample hereafter. 

\begin{figure}
 \resizebox{\hsize}{!}{\includegraphics{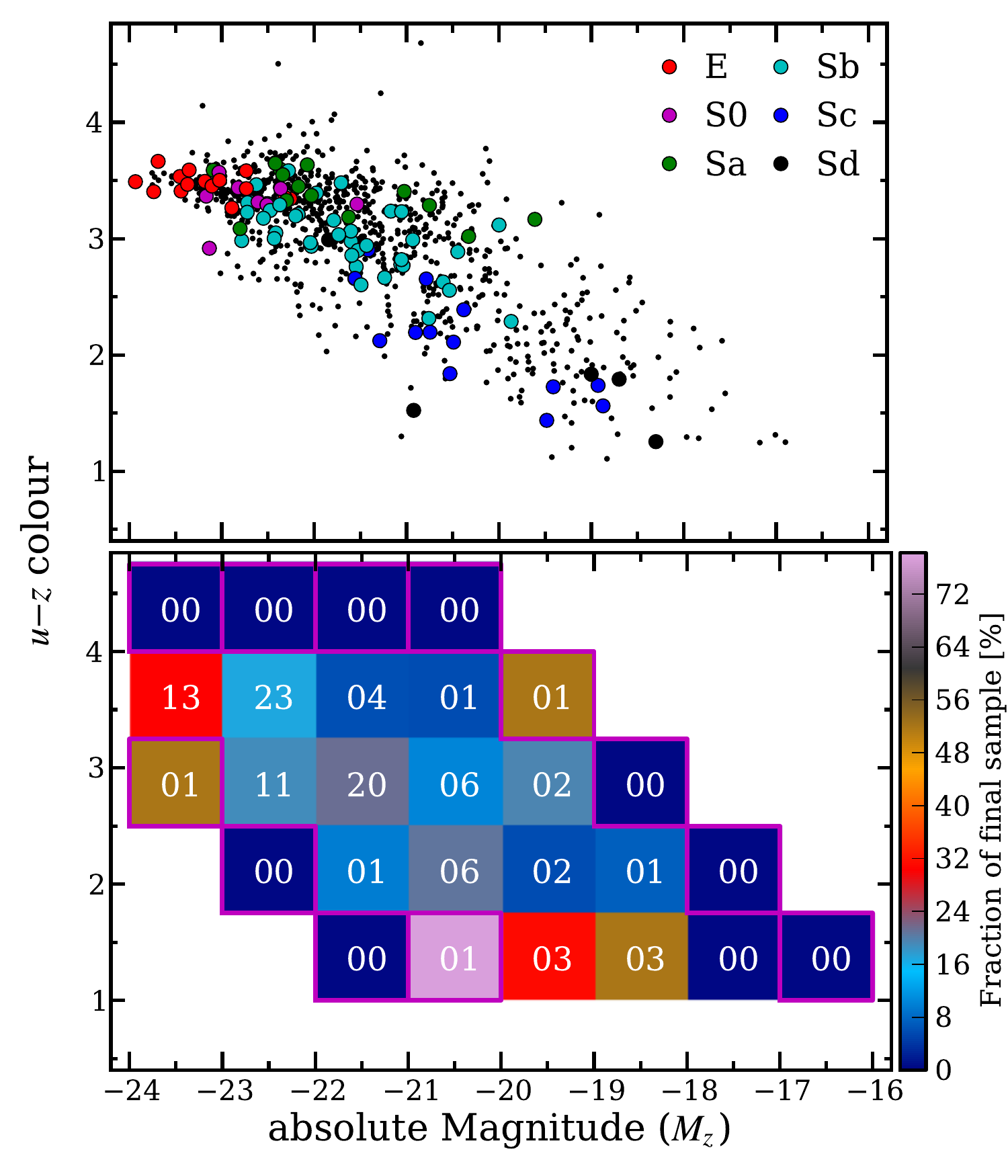}}
  \caption{\textit{Upper panel:} Distribution of CALIFA galaxies in the $u-z$ vs.
$M_z$ color-magnitude diagram. Black dots indicate galaxies in the CALIFA mother sample (S12, W12) and colored symbols denote CALIFA DR1 galaxies. Different colors represent the morphological classification
of galaxies which range from ellipticals (E) to late-type spirals (Sd). \textit{Lower panel:} The fraction of galaxies in the DR1 sample with respect to the expected final CALIFA sample distribution in bins of 1\,mag in $M_z$ and 0.75\,mag in $u-z$. The total number of galaxies per bin in the DR1 sample is written in each bin. The bins for which the number of galaxies in the mother sample is less than 5 are prone to low-number statistics and enclosed by a magenta square for clarity.}
  \label{fig:DR1_CM_diag}
\end{figure}

The first 100 public DR1 galaxies were observed in both spectral setups from the start of observations in June 2010 until June 2012. We list these galaxies in Table~\ref{tab:DR1_sample} together with their primary characteristics. The distribution of galaxies in the sky follows the underlying SDSS footprint (Fig.~\ref{fig:DR1_dist}). An exception is the additional cut in $\delta >7\degr$ that was applied to the region of the Northern Galactic Cap, given the sufficiently large number of objects with better visibility from Calar Alto. The number of galaxies in DR1 is not homogeneous as a function of right ascension, $\alpha$(J2000), and has a clear peak around $\alpha\sim255\degr$. This can be explained by the unexpected downtime of the 3.5\,m telescope from August 2010 until April 2011 due to operational reasons at the observatory, which delayed the survey roughly by 8 months. CALIFA observations, therefore, so far span three summer seasons at Calar Alto, but only a single winter season. Nevertheless, the distribution of physical properties is nearly random, as expected, and covers galaxies with a wide range of properties as discussed below.

\begin{longtab}
\begin{longtable}{lcccccccccc}
\caption{CALIFA DR1 galaxies and their characteristics. }\\\hline\hline
\label{tab:DR1_sample}
\small{Name} &  \small{ID\tablefootmark{a}} & \small{$\alpha$(J2000)\tablefootmark{b}} & \small{$\delta$(J2000)\tablefootmark{b}} & \small{$z$\tablefootmark{c}} & \small{$m_g$\tablefootmark{d}} & \small{$m_z$\tablefootmark{d}} & \small{$m_u-m_z$\tablefootmark{d}} & \small{type\tablefootmark{e}} & \small{bar\tablefootmark{f}}  & \small{$b/a$\tablefootmark{g}}\\\hline
\endfirsthead
\caption{continued.}\\\hline\hline
\small{Name} &  \small{ID\tablefootmark{a}} & \small{$\alpha$(J2000)\tablefootmark{b}} & \small{$\delta$(J2000)\tablefootmark{b}} & \small{$z$\tablefootmark{c}} & \small{$m_g$\tablefootmark{d}} & \small{$m_z$\tablefootmark{d}} & \small{$m_u-m_z$\tablefootmark{d}} & \small{type\tablefootmark{e}} & \small{bar\tablefootmark{f}}  & \small{$b/a$\tablefootmark{g}}\\\hline
\endhead
\hline
\endfoot
\small{\object{IC5376}} & \small{001} & \small{00:01:19.778} & \small{+34:31:32.40} & \small{0.0168} & \small{14.21} & \small{12.79} &\small{3.25} & \small{Sb } & \small{A} & \small{0.27} \\
\small{\object{NGC7819}} & \small{003} & \small{00:04:24.505} & \small{+31:28:19.22} & \small{0.0167} & \small{14.07} & \small{13.61} &\small{2.17} & \small{Sc } & \small{A} & \small{0.53} \\
\small{\object{UGC00036}} & \small{007} & \small{00:05:13.881} & \small{+06:46:19.30} & \small{0.0210} & \small{14.13} & \small{12.61} &\small{3.41} & \small{Sab } & \small{AB} & \small{0.60} \\
\small{\object{NGC0036}} & \small{010} & \small{00:11:22.297} & \small{+06:23:21.66} & \small{0.0203} & \small{13.46} & \small{12.28} &\small{2.83} & \small{Sb } & \small{B} & \small{0.65} \\
\small{\object{UGC00312}} & \small{014} & \small{00:31:23.921} & \small{+08:28:00.23} & \small{0.0145} & \small{13.86} & \small{13.61} &\small{1.58} & \small{Sd } & \small{B} & \small{0.35} \\
\small{\object{NGC0444}} & \small{039} & \small{01:15:49.562} & \small{+31:04:50.24} & \small{0.0161} & \small{14.55} & \small{13.90} &\small{2.07} & \small{Scd } & \small{A} & \small{0.24} \\
\small{\object{NGC0477}} & \small{042} & \small{01:21:20.483} & \small{+40:29:17.33} & \small{0.0196} & \small{14.37} & \small{13.20} &\small{2.86} & \small{Sbc } & \small{AB} & \small{0.66} \\
\small{\object{IC1683}} & \small{043} & \small{01:22:38.929} & \small{+34:26:13.65} & \small{0.0162} & \small{14.04} & \small{12.75} &\small{3.05} & \small{Sb } & \small{AB} & \small{0.59} \\
\small{\object{UGC01057}} & \small{053} & \small{01:28:53.253} & \small{+13:47:37.67} & \small{0.0212} & \small{14.44} & \small{13.49} &\small{2.42} & \small{Sc } & \small{AB} & \small{0.30} \\
\small{\object{NGC0776}} & \small{073} & \small{01:59:54.525} & \small{+23:38:39.39} & \small{0.0164} & \small{13.38} & \small{12.31} &\small{2.94} & \small{Sb } & \small{B} & \small{0.69} \\
\small{\object{UGC01938}} & \small{088} & \small{02:28:22.136} & \small{+23:12:52.65} & \small{0.0213} & \small{14.50} & \small{13.40} &\small{2.76} & \small{Sbc } & \small{AB} & \small{0.25} \\
\small{\object{NGC1056}} & \small{100} & \small{02:42:48.312} & \small{+28:34:26.96} & \small{0.0052} & \small{12.95} & \small{11.50} &\small{3.13} & \small{Sa } & \small{A} & \small{0.57} \\
\small{\object{NGC1349}} & \small{127} & \small{03:31:27.511} & \small{+04:22:51.24} & \small{0.0220} & \small{13.39} & \small{12.05} &\small{3.32} & \small{E6 } & \small{A} & \small{0.89} \\
\small{\object{UGC03253}} & \small{146} & \small{05:19:41.885} & \small{+84:03:09.43} & \small{0.0138} & \small{13.56} & \small{12.57} &\small{2.77} & \small{Sb } & \small{B} & \small{0.62} \\
\small{\object{NGC2410}} & \small{151} & \small{07:35:02.261} & \small{+32:49:19.56} & \small{0.0156} & \small{13.29} & \small{12.00} &\small{3.05} & \small{Sb } & \small{AB} & \small{0.32} \\
\small{\object{UGC03995}} & \small{155} & \small{07:44:09.127} & \small{+29:14:50.75} & \small{0.0159} & \small{13.13} & \small{11.98} &\small{3.85} & \small{Sb } & \small{B} & \small{0.46} \\
\small{\object{NGC2449}} & \small{156} & \small{07:47:20.299} & \small{+26:55:48.70} & \small{0.0163} & \small{13.48} & \small{12.12} &\small{3.46} & \small{Sab } & \small{AB} & \small{0.50} \\
\small{\object{IC2487}} & \small{273} & \small{09:30:09.166} & \small{+20:05:27.04} & \small{0.0145} & \small{13.87} & \small{13.02} &\small{3.21} & \small{Sc } & \small{AB} & \small{0.22} \\
\small{\object{IC0540}} & \small{274} & \small{09:30:10.338} & \small{+07:54:09.90} & \small{0.0069} & \small{14.28} & \small{12.91} &\small{3.16} & \small{Sab } & \small{AB} & \small{0.30} \\
\small{\object{NGC2916}} & \small{277} & \small{09:34:57.601} & \small{+21:42:18.94} & \small{0.0124} & \small{13.10} & \small{11.94} &\small{2.80} & \small{Sbc } & \small{A} & \small{0.59} \\
\small{\object{UGC05358}} & \small{306} & \small{09:58:47.135} & \small{+11:23:19.31} & \small{0.0097} & \small{15.21} & \small{14.99} &\small{1.46} & \small{Sd } & \small{B} & \small{0.33} \\
\small{\object{UGC05359}} & \small{307} & \small{09:58:51.646} & \small{+19:12:53.91} & \small{0.0283} & \small{14.66} & \small{13.58} &\small{2.70} & \small{Sb } & \small{B} & \small{0.37} \\
\small{\object{UGC05396}} & \small{309} & \small{10:01:40.484} & \small{+10:45:23.13} & \small{0.0181} & \small{14.38} & \small{13.66} &\small{2.71} & \small{Sbc } & \small{AB} & \small{0.27} \\
\small{\object{NGC3160}} & \small{319} & \small{10:13:55.114} & \small{+38:50:34.53} & \small{0.0229} & \small{14.54} & \small{12.98} &\small{3.45} & \small{Sab } & \small{AB} & \small{0.29} \\
\small{\object{UGC05598}} & \small{326} & \small{10:22:14.003} & \small{+20:35:21.87} & \small{0.0188} & \small{14.72} & \small{13.68} &\small{2.61} & \small{Sb } & \small{A} & \small{0.30} \\
\small{\object{UGC05771}} & \small{341} & \small{10:37:19.339} & \small{+43:35:15.32} & \small{0.0248} & \small{14.09} & \small{12.56} &\small{3.40} & \small{E6 } & \small{A} & \small{0.71} \\
\small{\object{UGC06036}} & \small{364} & \small{10:55:55.261} & \small{+36:51:41.46} & \small{0.0218} & \small{14.07} & \small{12.44} &\small{3.50} & \small{Sa } & \small{A} & \small{0.29} \\
\small{\object{NGC3991}} & \small{475} & \small{11:57:30.959} & \small{+32:20:13.28} & \small{0.0108} & \small{13.84} & \small{13.37} &\small{1.38} & \small{Sm } & \small{A} & \small{0.22} \\
\small{\object{NGC4003}} & \small{479} & \small{11:57:59.033} & \small{+23:07:29.63} & \small{0.0219} & \small{14.00} & \small{12.59} &\small{3.27} & \small{S0a } & \small{B} & \small{0.42} \\
\small{\object{UGC07012}} & \small{486} & \small{12:02:03.145} & \small{+29:50:52.73} & \small{0.0102} & \small{14.35} & \small{14.01} &\small{1.41} & \small{Scd } & \small{AB} & \small{0.54} \\
\small{\object{NGC4185}} & \small{515} & \small{12:13:22.192} & \small{+28:30:39.46} & \small{0.0130} & \small{13.18} & \small{12.30} &\small{3.30} & \small{Sbc } & \small{AB} & \small{0.64} \\
\small{\object{NGC4210}} & \small{518} & \small{12:15:15.842} & \small{+65:59:07.15} & \small{0.0091} & \small{13.19} & \small{12.25} &\small{2.84} & \small{Sb } & \small{B} & \small{0.73} \\
\small{\object{IC0776}} & \small{528} & \small{12:19:03.120} & \small{+08:51:22.15} & \small{0.0081} & \small{14.77} & \small{15.71} &\small{0.47} & \small{Sdm } & \small{A} & \small{0.56} \\
\small{\object{NGC4470}} & \small{548} & \small{12:29:37.778} & \small{+07:49:27.12} & \small{0.0079} & \small{12.81} & \small{12.18} &\small{1.72} & \small{Sc } & \small{A} & \small{0.66} \\
\small{\object{NGC4676A}} & \small{577} & \small{12:46:10.107} & \small{+30:43:54.89} & \small{0.0222} & \small{14.71} & \small{13.36} &\small{2.89} & \small{Sdm(x)} & \small{AB} & \small{0.28} \\
\small{\object{UGC08234}} & \small{607} & \small{13:08:46.505} & \small{+62:16:18.09} & \small{0.0270} & \small{13.54} & \small{12.37} &\small{2.86} & \small{S0 } & \small{A} & \small{0.63} \\
\small{\object{NGC5000}} & \small{608} & \small{13:09:47.486} & \small{+28:54:24.99} & \small{0.0187} & \small{13.84} & \small{12.90} &\small{2.70} & \small{Sbc } & \small{B} & \small{0.60} \\
\small{\object{UGC08250}} & \small{609} & \small{13:10:20.137} & \small{+32:28:59.47} & \small{0.0176} & \small{15.15} & \small{14.28} &\small{2.26} & \small{Sc } & \small{A} & \small{0.19} \\
\small{\object{UGC08267}} & \small{610} & \small{13:11:11.334} & \small{+43:43:34.78} & \small{0.0242} & \small{14.82} & \small{13.28} &\small{3.54} & \small{Sb } & \small{AB} & \small{0.20} \\
\small{\object{UGC08733}} & \small{657} & \small{13:48:38.994} & \small{+43:24:44.82} & \small{0.0078} & \small{14.54} & \small{14.03} &\small{1.56} & \small{Sdm } & \small{B} & \small{0.49} \\
\small{\object{IC0944}} & \small{663} & \small{13:51:30.867} & \small{+14:05:31.95} & \small{0.0234} & \small{13.84} & \small{12.31} &\small{3.73} & \small{Sab } & \small{A} & \small{0.30} \\
\small{\object{UGC08778}} & \small{664} & \small{13:52:06.668} & \small{+38:04:01.27} & \small{0.0108} & \small{14.15} & \small{12.95} &\small{2.89} & \small{Sb } & \small{A} & \small{0.21} \\
\small{\object{UGC08781}} & \small{665} & \small{13:52:22.745} & \small{+21:32:21.66} & \small{0.0253} & \small{13.82} & \small{12.64} &\small{3.17} & \small{Sb } & \small{B} & \small{0.52} \\
\small{\object{NGC5378}} & \small{676} & \small{13:56:51.013} & \small{+37:47:50.05} & \small{0.0100} & \small{13.41} & \small{12.36} &\small{3.14} & \small{Sb } & \small{B} & \small{0.63} \\
\small{\object{NGC5394}} & \small{680} & \small{13:58:33.652} & \small{+37:27:12.54} & \small{0.0114} & \small{14.13} & \small{13.20} &\small{3.12} & \small{Sbc(x)} & \small{B} & \small{0.74} \\
\small{\object{NGC5406}} & \small{684} & \small{14:00:20.119} & \small{+38:54:55.52} & \small{0.0180} & \small{13.08} & \small{11.91} &\small{3.24} & \small{Sb } & \small{B} & \small{0.88} \\
\small{\object{NGC5682}} & \small{758} & \small{14:34:44.978} & \small{+48:40:12.83} & \small{0.0076} & \small{14.38} & \small{13.97} &\small{1.67} & \small{Scd } & \small{B} & \small{0.31} \\
\small{\object{NGC5720}} & \small{764} & \small{14:38:33.281} & \small{+50:48:54.87} & \small{0.0260} & \small{13.97} & \small{12.85} &\small{2.88} & \small{Sbc } & \small{B} & \small{0.65} \\
\small{\object{UGC09476}} & \small{769} & \small{14:41:32.028} & \small{+44:30:45.97} & \small{0.0109} & \small{13.56} & \small{12.85} &\small{2.26} & \small{Sbc } & \small{A} & \small{0.63} \\
\small{\object{UGC09665}} & \small{783} & \small{15:01:32.464} & \small{+48:19:10.92} & \small{0.0085} & \small{14.15} & \small{12.89} &\small{2.99} & \small{Sb } & \small{A} & \small{0.23} \\
\small{\object{UGC09873}} & \small{797} & \small{15:29:50.650} & \small{+42:37:44.10} & \small{0.0188} & \small{15.09} & \small{14.32} &\small{2.60} & \small{Sb } & \small{A} & \small{0.21} \\
\small{\object{UGC09892}} & \small{798} & \small{15:32:51.947} & \small{+41:11:29.28} & \small{0.0189} & \small{14.73} & \small{13.75} &\small{2.59} & \small{Sbc } & \small{A} & \small{0.29} \\
\small{\object{NGC5966}} & \small{806} & \small{15:35:52.108} & \small{+39:46:08.04} & \small{0.0151} & \small{13.36} & \small{11.95} &\small{3.40} & \small{E4 } & \small{A} & \small{0.60} \\
\small{\object{NGC6032}} & \small{820} & \small{16:03:01.124} & \small{+20:57:21.32} & \small{0.0145} & \small{13.92} & \small{12.85} &\small{3.33} & \small{Sbc } & \small{B} & \small{0.30} \\
\small{\object{UGC10205}} & \small{822} & \small{16:06:40.180} & \small{+30:05:56.65} & \small{0.0219} & \small{13.73} & \small{12.36} &\small{3.27} & \small{S0a } & \small{A} & \small{0.58} \\
\small{\object{NGC6063}} & \small{823} & \small{16:07:12.993} & \small{+07:58:44.36} & \small{0.0095} & \small{13.57} & \small{12.85} &\small{2.54} & \small{Sbc } & \small{A} & \small{0.60} \\
\small{\object{IC1199}} & \small{824} & \small{16:10:34.346} & \small{+10:02:25.32} & \small{0.0158} & \small{13.80} & \small{12.60} &\small{2.79} & \small{Sb } & \small{AB} & \small{0.41} \\
\small{\object{NGC6081}} & \small{826} & \small{16:12:56.857} & \small{+09:52:01.57} & \small{0.0171} & \small{13.64} & \small{12.10} &\small{3.55} & \small{S0a } & \small{A} & \small{0.46} \\
\small{\object{UGC10331}} & \small{828} & \small{16:17:21.123} & \small{+59:19:12.46} & \small{0.0152} & \small{14.38} & \small{13.85} &\small{1.84} & \small{Sc(x)} & \small{AB} & \small{0.26} \\
\small{\object{NGC6125}} & \small{829} & \small{16:19:11.535} & \small{+57:59:02.89} & \small{0.0154} & \small{12.94} & \small{11.50} &\small{3.38} & \small{E1 } & \small{A} & \small{0.91} \\
\small{\object{NGC6146}} & \small{832} & \small{16:25:10.330} & \small{+40:53:34.32} & \small{0.0292} & \small{13.32} & \small{11.87} &\small{3.38} & \small{E5 } & \small{A} & \small{0.77} \\
\small{\object{NGC6154}} & \small{833} & \small{16:25:30.483} & \small{+49:50:24.93} & \small{0.0199} & \small{13.73} & \small{12.77} &\small{2.98} & \small{Sab } & \small{B} & \small{0.65} \\
\small{\object{NGC6150}} & \small{835} & \small{16:25:49.965} & \small{+40:29:19.41} & \small{0.0292} & \small{14.00} & \small{12.50} &\small{3.52} & \small{E7 } & \small{A} & \small{0.48} \\
\small{\object{UGC10384}} & \small{837} & \small{16:26:46.684} & \small{+11:34:48.96} & \small{0.0165} & \small{14.54} & \small{13.31} &\small{2.42} & \small{Sb } & \small{A} & \small{0.22} \\
\small{\object{NGC6173}} & \small{840} & \small{16:29:44.875} & \small{+40:48:41.96} & \small{0.0294} & \small{13.20} & \small{11.78} &\small{3.48} & \small{E6 } & \small{A} & \small{0.65} \\
\small{\object{UGC10650}} & \small{843} & \small{17:00:14.582} & \small{+23:06:22.83} & \small{0.0099} & \small{15.22} & \small{22.71} &\small{-5.68} & \small{Scd } & \small{A} & \small{0.20} \\
\small{\object{UGC10693}} & \small{845} & \small{17:04:53.020} & \small{+41:51:55.76} & \small{0.0280} & \small{13.50} & \small{12.15} &\small{3.47} & \small{E7 } & \small{AB} & \small{0.68} \\
\small{\object{UGC10695}} & \small{846} & \small{17:05:05.573} & \small{+43:02:35.36} & \small{0.0280} & \small{14.06} & \small{12.67} &\small{3.29} & \small{E5 } & \small{A} & \small{0.67} \\
\small{\object{UGC10710}} & \small{847} & \small{17:06:52.521} & \small{+43:07:19.96} & \small{0.0280} & \small{14.38} & \small{13.02} &\small{3.30} & \small{Sb } & \small{A} & \small{0.25} \\
\small{\object{NGC6310}} & \small{848} & \small{17:07:57.480} & \small{+60:59:24.56} & \small{0.0114} & \small{13.52} & \small{12.23} &\small{3.29} & \small{Sb } & \small{A} & \small{0.22} \\
\small{\object{NGC6314}} & \small{850} & \small{17:12:38.715} & \small{+23:16:12.29} & \small{0.0221} & \small{13.55} & \small{12.23} &\small{3.09} & \small{Sab } & \small{A} & \small{0.51} \\
\small{\object{NGC6338}} & \small{851} & \small{17:15:22.976} & \small{+57:24:40.28} & \small{0.0274} & \small{13.42} & \small{11.90} &\small{3.82} & \small{E5 } & \small{A} & \small{0.66} \\
\small{\object{UGC10796}} & \small{852} & \small{17:16:47.724} & \small{+61:55:12.43} & \small{0.0102} & \small{14.57} & \small{14.53} &\small{1.43} & \small{Scd } & \small{AB} & \small{0.49} \\
\small{\object{UGC10811}} & \small{854} & \small{17:18:43.725} & \small{+58:08:06.43} & \small{0.0291} & \small{14.49} & \small{13.20} &\small{2.98} & \small{Sb } & \small{B} & \small{0.42} \\
\small{\object{IC1256}} & \small{856} & \small{17:23:47.285} & \small{+26:29:11.48} & \small{0.0159} & \small{13.87} & \small{12.89} &\small{2.57} & \small{Sb } & \small{AB} & \small{0.59} \\
\small{\object{NGC6394}} & \small{857} & \small{17:30:21.423} & \small{+59:38:23.61} & \small{0.0284} & \small{14.48} & \small{13.19} &\small{3.08} & \small{Sbc } & \small{B} & \small{0.29} \\
\small{\object{UGC10905}} & \small{858} & \small{17:34:06.437} & \small{+25:20:38.28} & \small{0.0265} & \small{13.71} & \small{12.23} &\small{3.42} & \small{S0a } & \small{A} & \small{0.53} \\
\small{\object{NGC6411}} & \small{859} & \small{17:35:32.849} & \small{+60:48:48.25} & \small{0.0123} & \small{12.78} & \small{11.54} &\small{3.53} & \small{E4 } & \small{A} & \small{0.68} \\
\small{\object{NGC6427}} & \small{860} & \small{17:43:38.598} & \small{+25:29:38.17} & \small{0.0108} & \small{13.34} & \small{11.90} &\small{3.18} & \small{S0 } & \small{AB} & \small{0.61} \\
\small{\object{NGC6497}} & \small{863} & \small{17:51:17.966} & \small{+59:28:15.14} & \small{0.0105} & \small{13.59} & \small{12.34} &\small{3.57} & \small{Sab } & \small{B} & \small{0.65} \\
\small{\object{NGC6515}} & \small{864} & \small{17:57:25.195} & \small{+50:43:41.24} & \small{0.0228} & \small{13.54} & \small{12.22} &\small{3.29} & \small{E3 } & \small{A} & \small{0.78} \\
\small{\object{UGC11228}} & \small{865} & \small{18:24:46.259} & \small{+41:29:33.85} & \small{0.0194} & \small{13.81} & \small{12.36} &\small{3.40} & \small{S0 } & \small{B} & \small{0.57} \\
\small{\object{UGC11262}} & \small{866} & \small{18:30:35.698} & \small{+42:41:33.70} & \small{0.0186} & \small{15.01} & \small{14.59} &\small{2.19} & \small{Sc } & \small{A} & \small{0.39} \\
\small{\object{NGC6762}} & \small{867} & \small{19:05:37.089} & \small{+63:56:02.79} & \small{0.0098} & \small{13.87} & \small{12.49} &\small{3.13} & \small{Sab } & \small{A} & \small{0.49} \\
\small{\object{UGC11649}} & \small{872} & \small{20:55:27.619} & \small{-01:13:30.87} & \small{0.0127} & \small{13.57} & \small{12.44} &\small{3.26} & \small{Sab } & \small{B} & \small{0.88} \\
\small{\object{NGC7025}} & \small{874} & \small{21:07:47.336} & \small{+16:20:09.22} & \small{0.0166} & \small{12.87} & \small{11.38} &\small{3.67} & \small{S0a } & \small{A} & \small{0.72} \\
\small{\object{UGC11717}} & \small{877} & \small{21:18:35.412} & \small{+19:43:07.39} & \small{0.0212} & \small{14.36} & \small{12.81} &\small{3.72} & \small{Sab } & \small{A} & \small{0.47} \\
\small{\object{MCG-01-54-016}} & \small{878} & \small{21:25:59.970} & \small{-03:48:32.26} & \small{0.0098} & \small{14.92} & \small{14.67} &\small{1.52} & \small{Scd } & \small{A} & \small{0.13} \\
\small{\object{NGC7194}} & \small{881} & \small{22:03:30.937} & \small{+12:38:12.41} & \small{0.0272} & \small{13.66} & \small{12.15} &\small{3.45} & \small{E3 } & \small{A} & \small{0.79} \\
\small{\object{UGC11958}} & \small{883} & \small{22:14:46.882} & \small{+13:50:27.13} & \small{0.0262} & \small{14.15} & \small{12.84} &\small{4.59} & \small{S0(x)} & \small{A} & \small{0.74} \\
\small{\object{NGC7321}} & \small{887} & \small{22:36:28.022} & \small{+21:37:18.35} & \small{0.0238} & \small{13.45} & \small{12.35} &\small{2.94} & \small{Sbc } & \small{B} & \small{0.69} \\
\small{\object{UGC12127}} & \small{888} & \small{22:38:29.421} & \small{+35:19:46.89} & \small{0.0275} & \small{13.59} & \small{12.19} &\small{3.86} & \small{E1 } & \small{A} & \small{0.85} \\
\small{\object{UGC12185}} & \small{890} & \small{22:47:25.063} & \small{+31:22:24.67} & \small{0.0222} & \small{14.12} & \small{12.99} &\small{3.07} & \small{Sb } & \small{B} & \small{0.47} \\
\small{\object{NGC7436B}} & \small{893} & \small{22:57:57.546} & \small{+26:09:00.01} & \small{0.0246} & \small{13.40} & \small{11.84} &\small{3.92} & \small{E2(x)} & \small{A} & \small{0.90} \\
\small{\object{NGC7466}} & \small{896} & \small{23:02:03.464} & \small{+27:03:09.34} & \small{0.0251} & \small{14.09} & \small{12.92} &\small{2.93} & \small{Sbc } & \small{A} & \small{0.53} \\
\small{\object{NGC7549}} & \small{901} & \small{23:15:17.270} & \small{+19:02:30.43} & \small{0.0157} & \small{13.59} & \small{12.53} &\small{2.88} & \small{Sbc } & \small{B} & \small{0.75} \\
\small{\object{NGC7563}} & \small{902} & \small{23:15:55.927} & \small{+13:11:46.03} & \small{0.0143} & \small{13.30} & \small{11.83} &\small{3.62} & \small{Sa } & \small{B} & \small{0.68} \\
\small{\object{NGC7591}} & \small{904} & \small{23:18:16.259} & \small{+06:35:08.86} & \small{0.0165} & \small{13.44} & \small{12.18} &\small{2.90} & \small{Sbc } & \small{B} & \small{0.59} \\
\small{\object{UGC12864}} & \small{935} & \small{23:57:23.920} & \small{+30:59:31.45} & \small{0.0156} & \small{14.35} & \small{13.95} &\small{2.51} & \small{Sc } & \small{B} & \small{0.32} \\
\small{\object{NGC4676B}\tablefootmark{h}} & \small{939} & \small{12:46:11.235} & \small{+30:43:21.87} & \small{0.0218} & \small{15.03} & \small{13.68} &\small{3.94} & \small{...(x)} & \small{...} & \small{0.82} \\
\end{longtable}
\tablefoot{
\tablefoottext{a}{CALIFA unique ID number for the galaxy.}
\tablefoottext{b}{Equatorial coordinates of the galaxies as provided by NED.}
\tablefoottext{c}{Redshift of the galaxies based on SDSS DR7 spectra or complemented with SIMBAD information if SDSS spectra are not available.}
\tablefoottext{d}{Petrosian magnitudes as given by SDSS DR7 database corrected for Galactic extinction.}
\tablefoottext{e}{Morphological type from our own visual classification (see W12 for details).}
\tablefoottext{f}{Bar strength of the galaxy as an additional outcome of our visual classification.}
\tablefoottext{g}{Ratio between the semi-minor and semi-major axis based on a detailed re-analysis of the SDSS images (see W12 for details).}
\tablefoottext{h}{A visual morphological classification of this particular galaxy NGC 4676B is missing.}
}
\end{longtab}

\begin{figure}
\resizebox{\hsize}{!}{\includegraphics{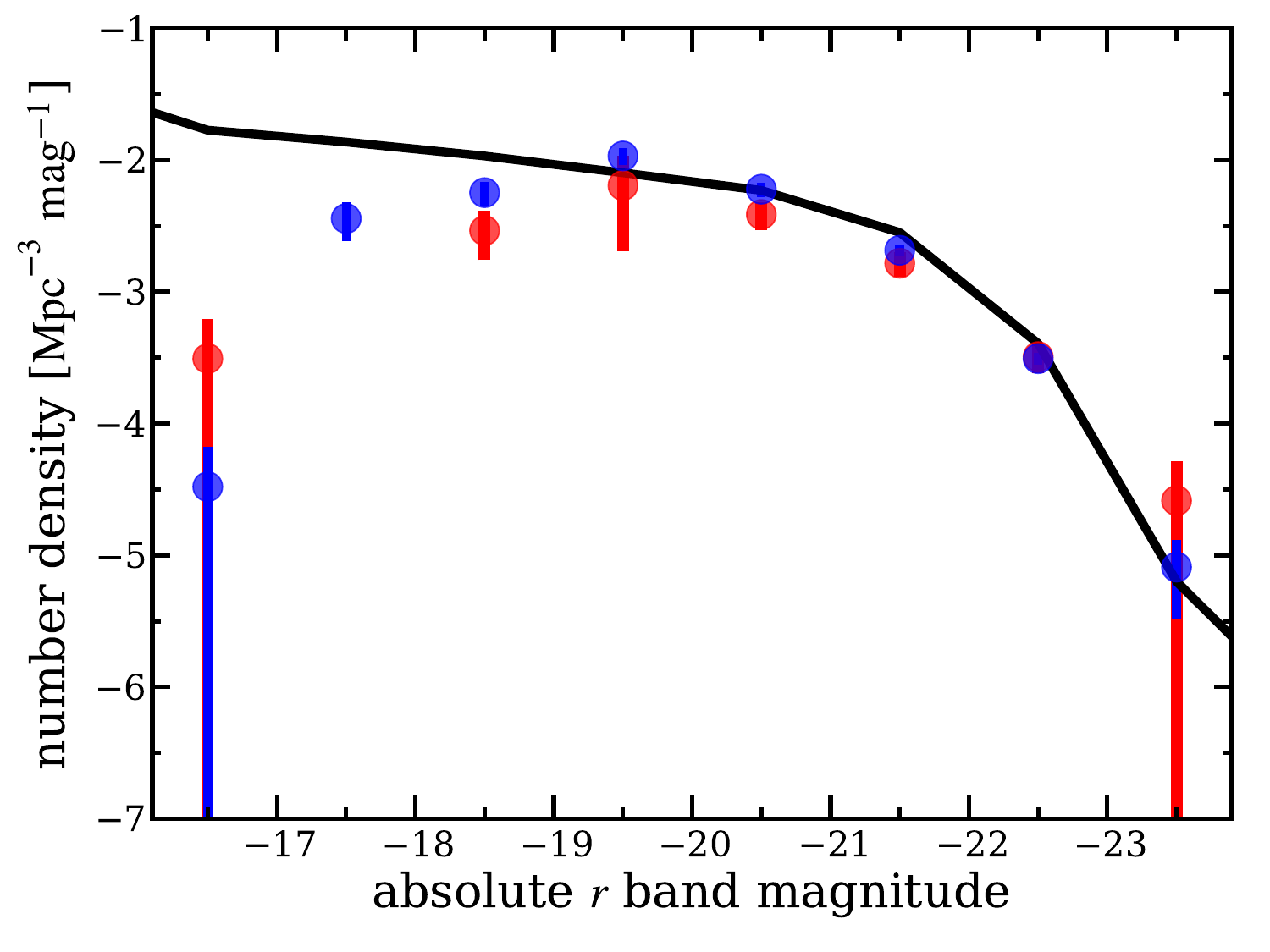}}
 \caption{Luminosity functions in the $r$  band of the DR1 sample (red points), the CALIFA mother sample (blue points), and the low-redshift sample of the NYU value added catalog \citep[black line,][]{Blanton:2005}. Despite the small size of the DR1 sample the luminosity function is reproduced well.}
 \label{fig:DR1_LF}
\end{figure}

The distribution of galaxies in the color-magnitude diagram (Fig.~\ref{fig:DR1_CM_diag}) shows that the DR1 sample covers almost homogeneously  the full range of the CALIFA mother sample. On average, the DR1 targets comprises  $\sim$20\% per color-magnitude bin of the total expected number when CALIFA is completed.  However, there is currently still a deficit of low luminosity galaxies with intermediate colors. In other color-magnitude bins, especially within those where the mother sample contains few galaxies, fluctuations can be explained by the effect of low number statistics. Figure~\ref{fig:DR1_CM_diag} highlights the need to further increase the numbers to the full CALIFA sample to obtain enough galaxies in each bin for a meaningful multi-dimensional statistical analysis. In Fig.~\ref{fig:DR1_LF}, we show the $r$ band luminosity function (LF) of the DR1 sample as compared to the mother sample and the reference SDSS sample of \citet{Blanton:2005}. All technical details on how we obtained the LFs are described in W12. We simply note here that despite the small sample size of DR1, we already reproduce the LF reasonably well. The turnover of the LF at $M_r > -19.5$ is entirely expected and understood.

\begin{figure}
 \resizebox{\hsize}{!}{\includegraphics{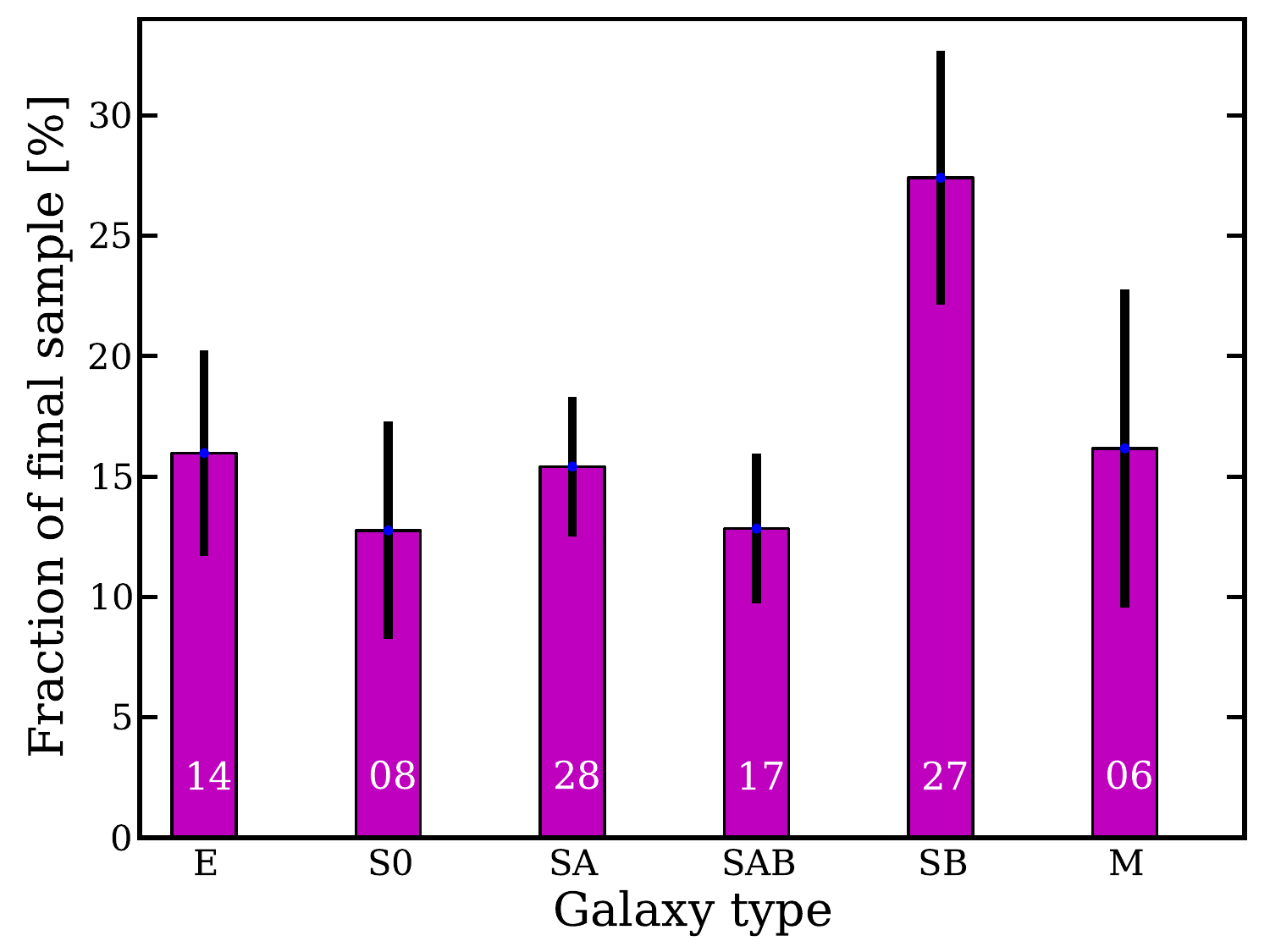}}
  \caption{The fraction of galaxies in the DR1 sample with respect to the expected final CALIFA sample distribution, split by visually classified morphology. We divide the galaxies into ellipticals (E), lenticulars (S0), non-barred spirals (SA), weakly barred spirals (SAB), strongly barred spirals (SB), and ongoing mergers (M) of any type. The morphological distribution of the DR1 sample lies close to that of the mother sample. The total number of galaxies in the DR1 for each morphology type is written on the bar. Error bars are computed from the Poisson errors of the associated DR1 number counts.}
  \label{fig:DR1_morph}
\end{figure}

\begin{figure}
 \resizebox{\hsize}{!}{\includegraphics{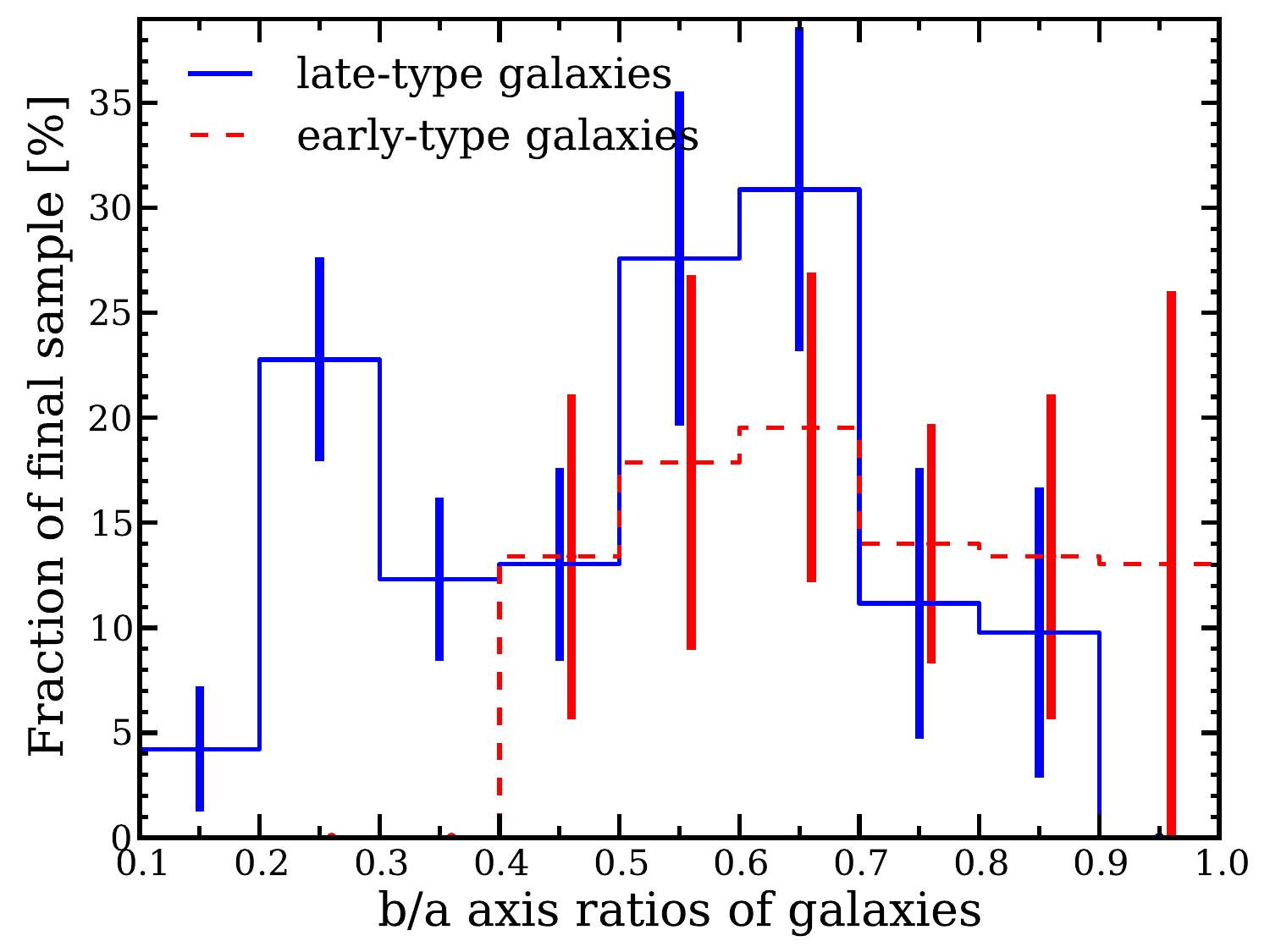}}
  \caption{The fraction of galaxies in the DR1 sample with respect to the expected final CALIFA sample distribution, as a function of axis ratio ($b/a$). Galaxies were separated into early-type galaxies (E+S0) and late-type galaxies (Sa and later). The CALIFA mother sample does not include any elliptical galaxies with $b/a<0.3$ or any spiral galaxies with $b/a>0.9$. Error bars are computed from the Poisson errors of the associated DR1 number counts.}
  \label{fig:DR1_ba_ratios}
\end{figure}

Galaxy morphologies were inferred by combining the independent visual classifications of several collaboration members as described in W12. In Fig.~\ref{fig:DR1_morph} we show the fraction of DR1 galaxies with respect to the expected final sample distribution for different morphological types grouped into elliptical, lenticular, spiral galaxies (separated by different bar strength) as well as ongoing galaxy mergers. A more detailed classification of spirals into early- and late-types is available, but we do not distinguish between them here because of the still modest number of galaxies within DR1.  The DR1 coverage seems to be consistent with a random selection because the fraction of DR1 galaxies with respect to the expected final sample is almost constant for all types. Axis ratios ($b/a$) were measured from the SDSS $r$ band image by calculating light moments after proper sky subtraction and masking of foreground stars (see W12 for details). The fractions of axis ratios, which can be used as an indication of the inclination of spiral galaxies, covered by the DR1 sample is homogeneous with respect to the final sample (Fig.~\ref{fig:DR1_ba_ratios}).  Based on a Kolmogorov-Smirnov test, we further quantified that the morphology and the axis-ratio distribution of DR1 is consistent with being randomly drawn from the CALIFA mother sample with $>$95\% confidence. 

\begin{figure}
 \resizebox{\hsize}{!}{\includegraphics{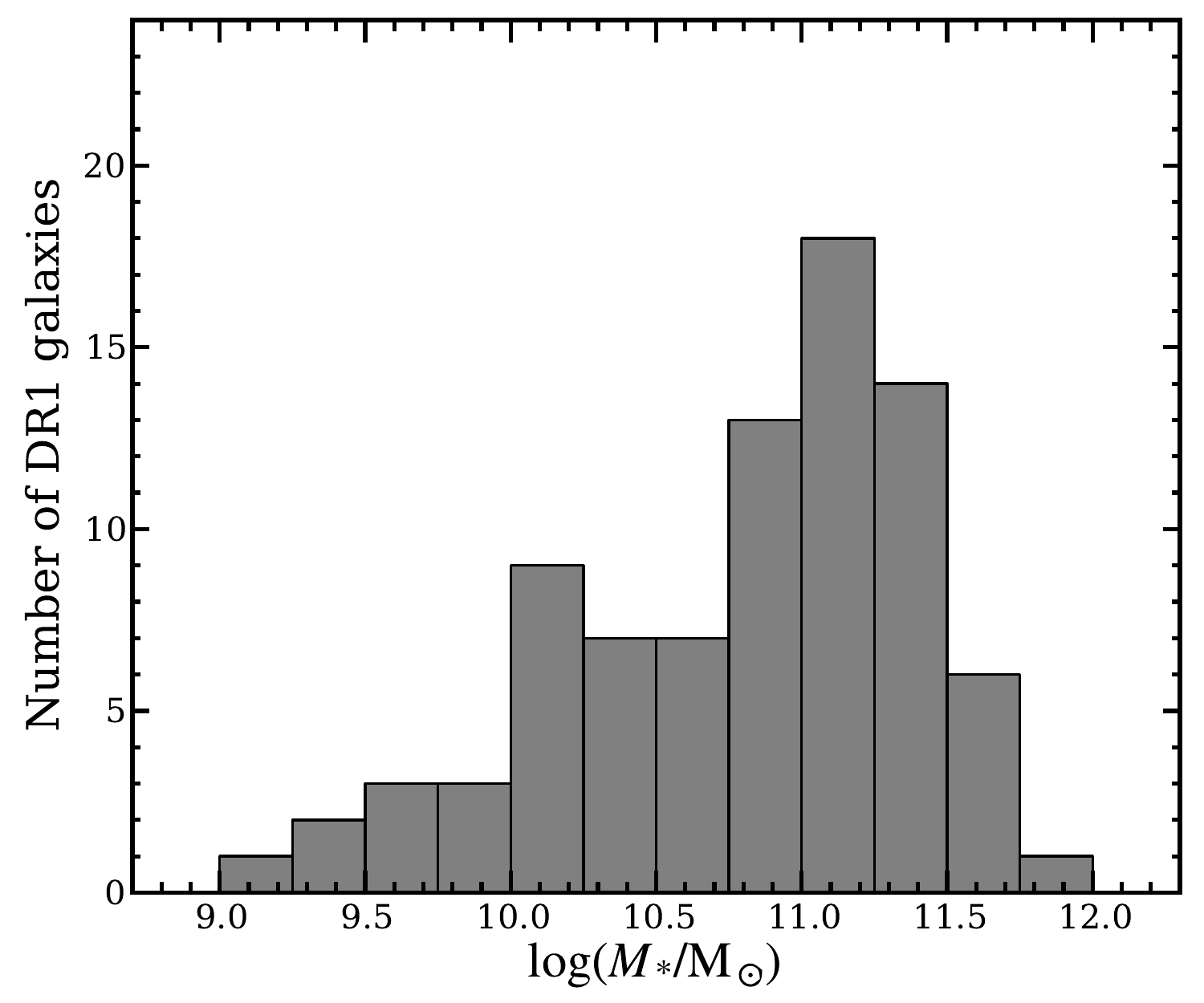}}
  \caption{Distribution of stellar masses for galaxies in the DR1 sample. The stellar masses have been determined from the CALIFA data (see text for details).}
  \label{fig:DR1_mass_dist}
\end{figure}

In Fig.~\ref{fig:DR1_mass_dist}, we present the distribution of stellar masses for the DR1 galaxies. They were directly inferred from the CALIFA data by applying spatially resolved spectral synthesis using the {\sc Starlight} code \citep{CidFernandes:2005} with a combination of single stellar populations (SSPs) from the libraries of \citet{Gonzalez-Delgado:2005} and \citet{Vazdekis:2010}, which both adopt a Salpeter initial mass function. Details of these measurements will be presented in Gonz\'alez Delgado et al. (in prep.). The DR1 galaxies cover intermediate to high-mass galaxies, including more than 5 galaxies per 0.25\,dex bin between $10^{10}$ and $10^{12}M_{\sun}$ and a median value close to $10^{11}\mathrm{M}_{\sun}$. The asymmetric distribution is expected from the distribution in absolute magnitudes (see Fig.~\ref{fig:DR1_CM_diag}) and is inherited from the CALIFA mother sample due to its selection criteria (see W12 for details). 

\begin{figure}
 \resizebox{\hsize}{!}{\includegraphics{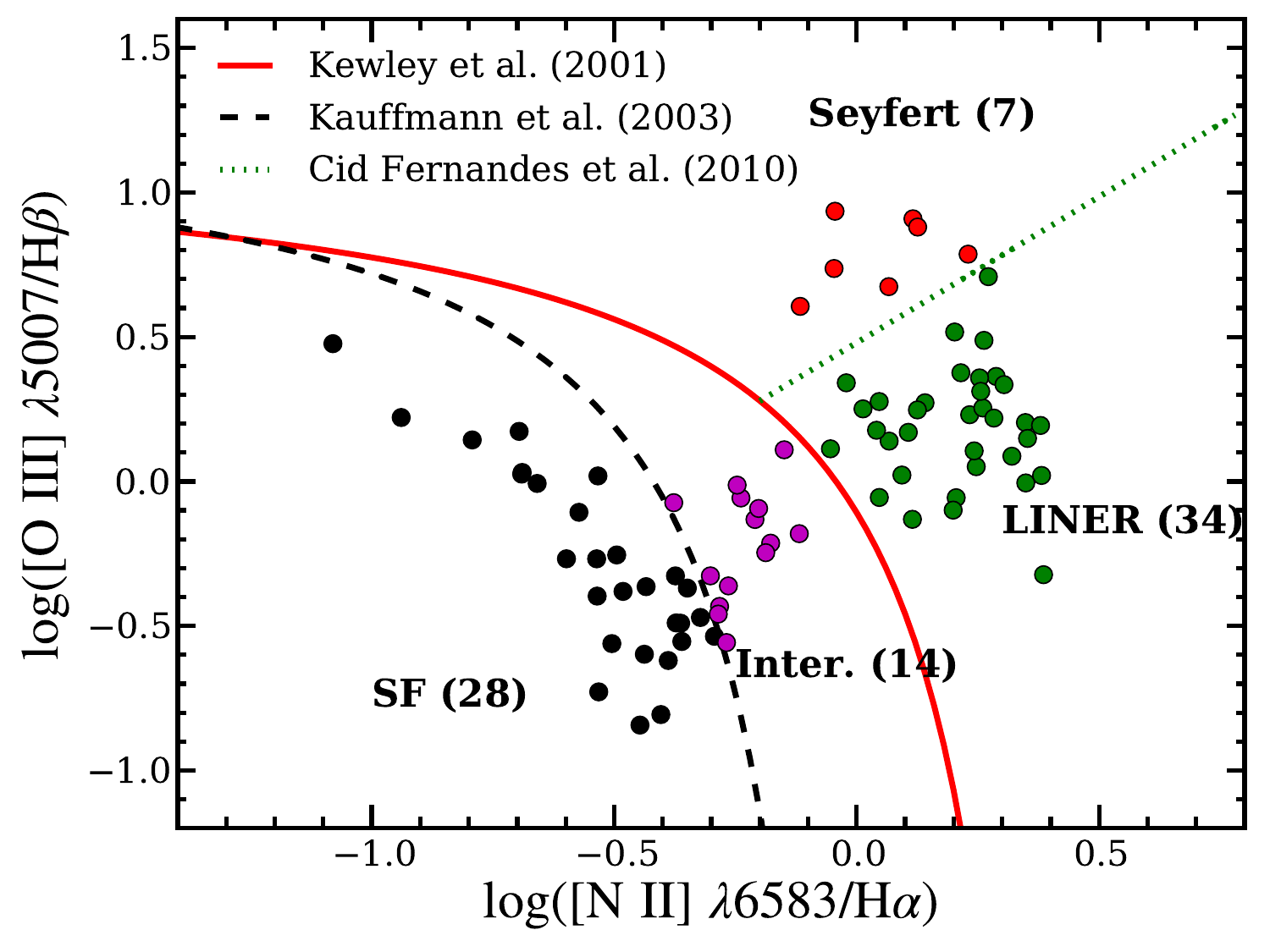}}
  \caption{Emission-line diagnostic diagram for the most central spaxel of each DR1 datacube. Only objects for which all required emission lines have $\mathrm{S/N}>3$ are shown. The demarcation lines of \citet{Kewley:2001}, \citet{Kauffmann:2003}, and \citet{CidFernandes:2010} are used to classify the galaxies into star forming (SF), intermediate, Seyfert, and LINER-type galaxies, which are denoted with black, magenta, red, and green symbols, respectively.}
  \label{fig:DR1_BPT_diagram}
\end{figure}

Many different kinds of galaxies are already covered in the DR1 sample in sufficient numbers to perform spatially resolved comparison studies. In addition to global galaxy parameters presented previously, we show the $\mathrm{[\ion{O}{iii}]}/\mathrm{H}\beta$ vs. $\mathrm{[\ion{N}{ii}]}/\mathrm{H}\alpha$ emission-line diagnostic diagram \citep{Baldwin:1981, Veilleux:1987} for the nucleus in Fig.~\ref{fig:DR1_BPT_diagram}. It is constructed from the most central spaxel at the coordinates of each CALIFA galaxy center (Table~\ref{tab:DR1_sample} excluding objects with a wrong astrometry in the data as discussed later in Sect.~\ref{sect:spatial_res}). After removing the stellar continuum using a spectral synthesis approach, we measured the emission line fluxes by fitting Gaussians to their profiles in the residual spectra. A variety of different ionization mechanisms can be found in the DR1 galaxy sample, ranging from pure star formation to Seyfert-type active galactic nuclei (AGN), with a significant number of galaxies in between.  Also, Low-Ionization Nuclear Emission-line Regions (LINERs) \citep{Heckman:1980}, which are predominantly hosted by bulge-dominated or elliptical galaxies, appear to be frequent. Robust classification of individual galaxies according to such a scheme can be complicated, i.e., in border-line cases or due to systematic uncertainties in the emission-line measurements, as well as adopting other types of diagnostic diagrams \citep[e.g.][]{CidFernandes:2010}, or a different classification scheme. This has been considered in the \citet{Kehrig:2012} study of the ionization source of the interstellar medium in two LINER-like early-type CALIFA galaxies. Rather than to suggest or impose a certain classification of individual galaxy nuclei for the DR1 sample, we present one particular representation here only to demonstrate the diversity of the DR1 galaxy sample.

\section{Data processing and error propagation}\label{sect:data}
The instrument characteristics and observing strategy of the CALIFA survey define the requirements for the data reduction scheme. These requirements are thoroughly described in S12 and only briefly summarized here for completeness. The PPak fiber bundle of the PMAS instrument comprises 382 fibers, each with a diameter of $2.7\arcsec$ projected on the sky. The primary FoV has a hexagonal shape of $74\arcsec\times 62\arcsec$ in size, sampled by 331 fibers with a filling factor of $\sim$60\%. A set of 36 fibers are dedicated to sample the sky background and are distributed in 6 smaller bundles along a circle of $\sim 72\arcsec$ radius around the FoV center. The remaining 15 fibers are coupled to the calibration unit of the instrument. 

Each CALIFA galaxy is targeted twice with different spectral settings, a low-resolution (V500) setup covering the nominal wavelength range 3745--7500\AA\ at spectral resolution of $R\sim850$, and a mid-resolution (V1200)
grism covering the nominal wavelength range 3400--4840\AA\ at a spectral resolution of $R\sim1650$. The useful wavelength range is, however, reduced by internal vignetting within the spectrograph for several fibers  to 4240--7140\AA\ and 3650--4620\AA\ in the worst case for the V500 and V1200, respectively. Three dithered pointings are taken for each object to reach a filling factor of 100\% across the entire FoV. The exposure time per pointing is fixed to 900\,s for the V500 and 1800\,s for the V1200. The latter is further split into 2 or 3 individual exposures. 

We are continuously upgrading the CALIFA pipeline (see S12 for details). The main improvements to the data reduction pipeline used to produce the DR1 data are briefly mentioned in the next section followed by a detailed characterization of the propagated noise.

\subsection{Improvements on the CALIFA data reduction scheme}
At the first stages of the project,  we employed the {\tt R3D} data reduction package \citep{Sanchez:2006a} as the basis to develop a fully automatic IFS data reduction pipeline dedicated to the CALIFA data. A description of that pipeline and the individual reduction steps was presented by S12. With the aim of improving the flexibility, portability, maintenance and capabilities of the pipeline,  most of the data reduction tasks have been transformed to a {\tt Python}-based architecture. 

The main driver for the pipeline improvements was to implement the propagation of the Poisson plus read-out noise as well as bad pixels caused by cosmic ray hits, bad CCD columns, or the effect of vignetting, from the raw data to the final CALIFA data product in a consistent way. This was not possible with the reduction scheme adopted by the previous pipeline version (V1.2), because the numerous interpolation and resampling steps did not allow a reliable error propagation.  Details of the most important improvements for the current  pipeline version (V1.3c) are provided in Appendix~\ref{apx:data_reduction}. 

The final output of the V1.3c pipeline are cubes of intensity, errors and bad pixels. Details on the data format are described in Sect.~\ref{sect:data_format}. Below we evaluate specifically the accuracy of the pipeline-produced noise estimates and characterize effects of correlated noise that should be properly taken into account in any scientific analysis.

\subsection{Accuracy of the propagated noise}\label{sect:noise}
\begin{figure}
 \resizebox{\hsize}{!}{\includegraphics{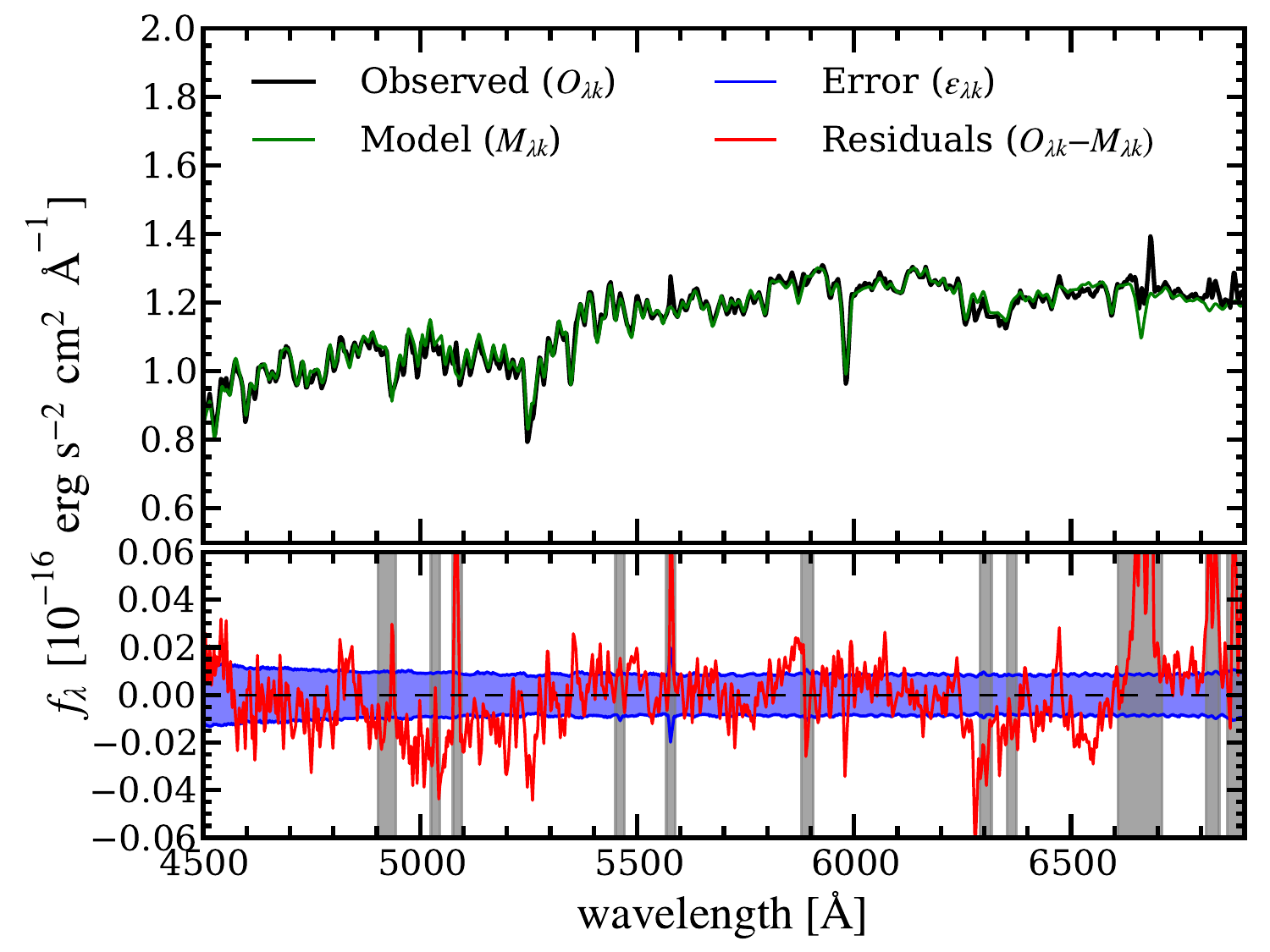}}
  \caption{Example of a stellar population fit with {\sc Starlight} to the central V500 spectrum of NGC 5966, restricted to the unvignetted part. The observed, modeled and residual spectrum are shown together with the pipeline produced $\pm1\sigma$ error. Sky and galaxy emission line regions masked during the fitting are indicated by the gray shaded areas.}
  \label{fig:spec_fit}
\end{figure}
The derivation of error spectra in IFS data is a complex task, for which there are no well established recipes. It is therefore relevant to verify the reliability of the pipeline errors ($\epsilon_\lambda$). An approximate assessment of the error spectra was carried out with the aid of full spectral continuum fitting described by Cid Fernandes et al. (in prep.), and Gonzalez Delgado et al. (in prep.). The CALIFA spectra are decomposed in terms of combinations of SSP spectra from \citet{Gonzalez-Delgado:2005} and \citet{Vazdekis:2010} including a dust extinction term following the \citet{Cardelli:1989} law. An example of such a fit is shown for a central  V500 spectrum in Fig.~\ref{fig:spec_fit} for illustrative purposes. For a certain spectrum $k$ it presents the observation ($O_{\lambda k}$), the best-fit model ($M_{\lambda k}$), the residuals ($R_{\lambda k}= O_{\lambda k}-M_{\lambda k}$), and the provided pipeline error spectrum ($\epsilon_{\lambda k}$).  

\begin{figure}
 \resizebox{\hsize}{!}{\includegraphics{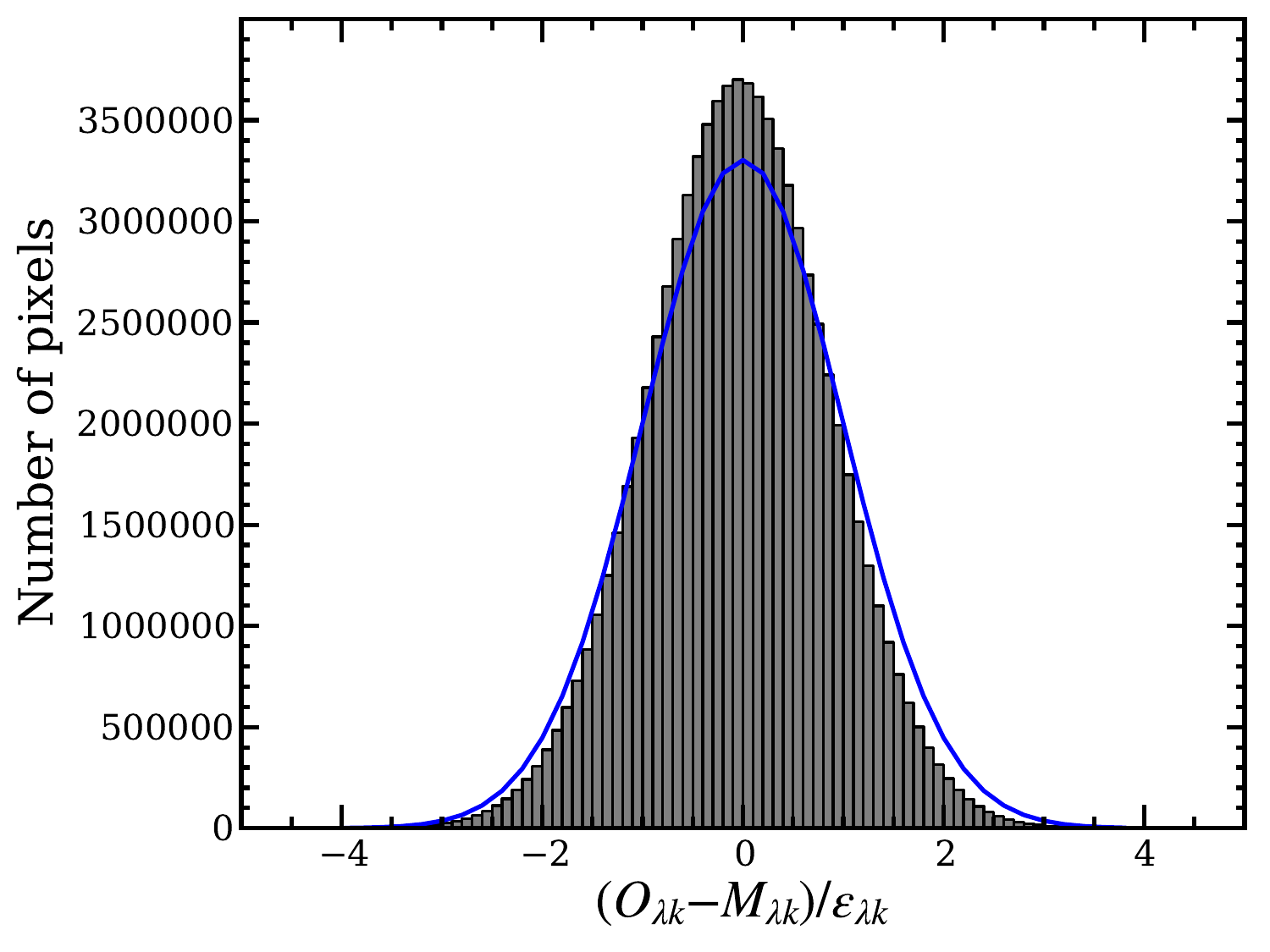}}
 \caption{Histogram of the observed ($O_{\lambda k}$) minus model ($M_{\lambda k}$) residuals at each wavelength normalized by the formal flux error ($\epsilon_{\lambda k}$). The model spectra are obtained from {\sc Starlight} fits for the stellar continuum, excluding the main emission lines, as well as bad pixels. The histogram describes almost a Gaussian centered at 0 with a dispersion of $\sim 0.8$. The blue line indicates a Gaussian distribution centered at 0 with a dispersion of 1, normalized to the total number of pixels for comparison. On the assumption that the spectral residuals are due to noise (i.e., neglecting model imperfections), this figure demonstrates that the pipeline error estimates are reliable.}
\label{fig:U_hist}
\end{figure}

Neglecting systematic deviations of the model spectra with respect to the real data in some wavelength regions, one would expect the residuals $R_{\lambda k}$ to be typically of the order of $\epsilon_{\lambda k}$. Indeed, we find that this is the case. The histogram of $U_{\lambda k} = R_{\lambda k} / \epsilon_{\lambda k}$ (Fig.~\ref{fig:U_hist}) was obtained from over $10^8$ data points of nearly $10^5$ spectra for the V500 spectra distributed within this DR1. Emission lines and faulty pixels were excluded from the statistics (shaded areas in Fig.~\ref{fig:spec_fit}), as the spectral fits are only meant to reproduce the stellar component.  The histogram follows nearly a Gaussian shape, centered at $-0.04$ with a standard deviation of 0.8. The expected standard deviation would be 1.0 if the error spectra would statistically agree with the noise in the residual spectra and assuming that the stellar population model is a perfect fit. Selecting different wavelength ranges for this comparison to avoid regions with systematic template mismatches mainly changes the centroid of the distribution whereas the standard deviation is nearly constant. Thus, the propagated errors are overestimated by $\sim$20\% in the mean. This was also verified for the V1200 setup using the same method.  The overall conclusion of this test is that the errors provided in DR1 are robust, with a slight systematic overestimation of $\sim$20\% that can be taken into account during any analysis if needed.

\subsection{Characterization of spatially correlated noise}\label{sect:correlation}
It is often necessary to spatially co-add spaxels in the datacubes to achieve a minimum S/N in the spectra for a specific application. For CALIFA we adopt an inverse-distance weighted image reconstruction scheme so far, which averages the flux among all fibers within 5\arcsec\ for a given spaxel in the final datacube by assuming a 2D Gaussian profile with a dispersion of 1\arcsec\ for the individual weighting factors (see S12 for details). Like many other image resampling schemes it introduces significant correlation between the spaxels in the final datacubes. This can be understood from simple arguments. Each CALIFA dataset contains 993 physically independent spectra from the fibers of the three dithered pointings, but the final datacube consists of more than 4000 spectra at a sampling of $1\arcsec\times1\arcsec$ per spaxel. This  automatically implies that the final spaxels \emph{cannot be completely independent from each other} within the CALIFA datacubes through the complex correlation of signal and noise between neighboring spaxels. 
\begin{figure}
  \resizebox{\hsize}{!}{\includegraphics{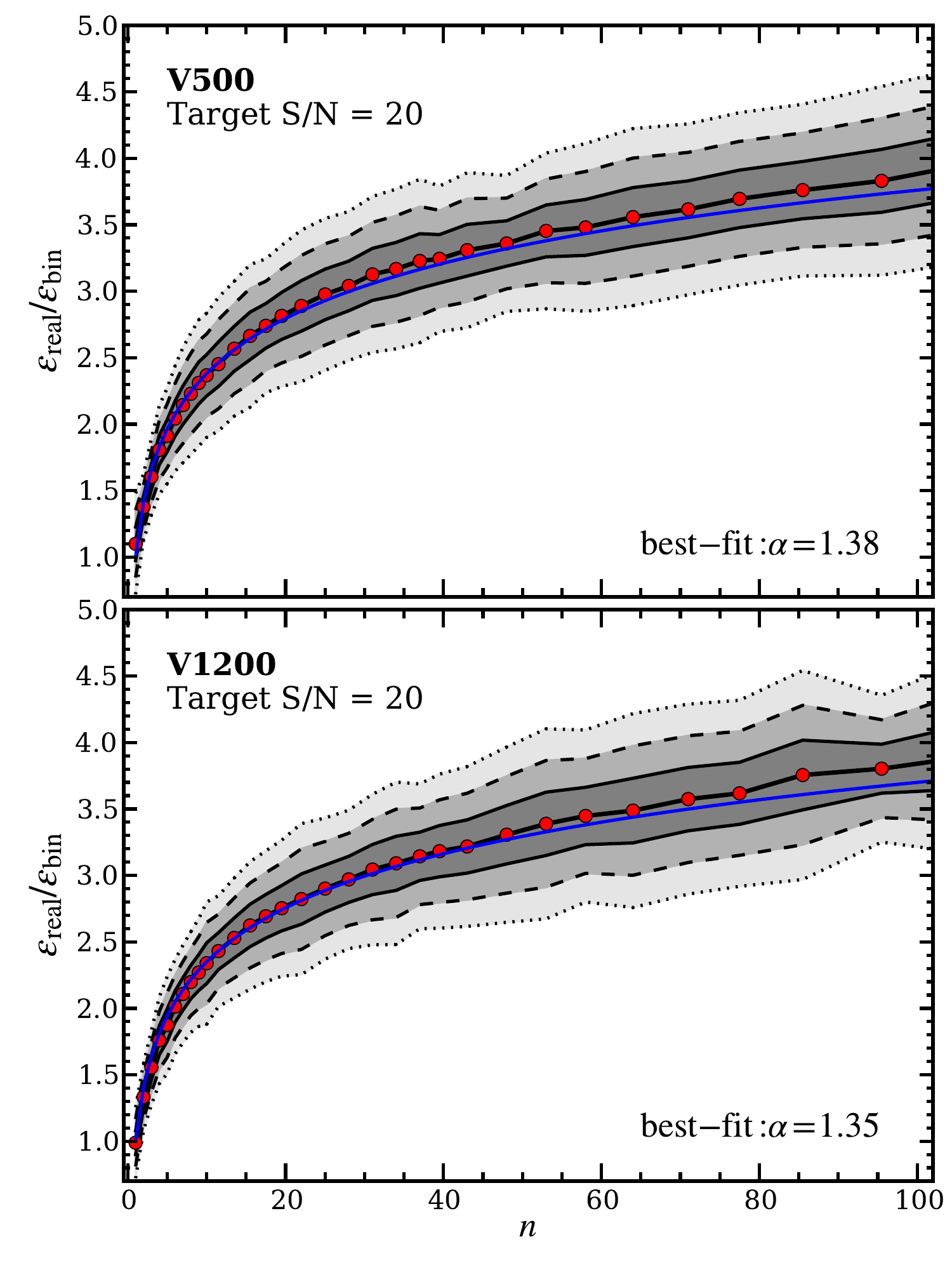}}
  \caption{Ratio of the real error ($\epsilon_\mathrm{real}$) to the analytically propagated error ($\epsilon_\mathrm{bin}$) as a 
function of number of spaxels per bin for all the V500 (upper panel) and V1200 (lower panel) data of DR1 at a 
target S/N of 20. Grey shades mark the 1$\sigma$, 2$\sigma$ and 3$\sigma$ levels. 
The blue lines represent the best fit logarithmic function with $\alpha=1.38$ and $\alpha=1.35$, respectively.}
  \label{fig:DR1_noise_correlation}
\end{figure}

In the limit case of co-adding the spectra of the entire cube, the pipeline analytically calculates an error weighting factor for each pixel such that the formal error of the co-added spectrum is identical to the one obtained by co-adding the individual fiber spectra. Of course, this is an unrealistic case, because only spectra within a small zone are typically co-added to preserve some spatial resolution. A popular method for adaptive binning is the Voronoi-binning scheme implemented for optical IFS data by \citet{Cappellari:2003}. It assumes that the spectra are completely independent of each other to compute the required bin size for a given target S/N. Blindly adopting such a binning scheme to CALIFA datacubes will lead to incorrect results, because the assumption that the spectra are independent is not valid. Either the bin sizes will be smaller than required to achieve the target S/N, or alternatively expressed, the error of the co-added spectra will be higher than formally expected given the error of individual spaxels.

Here, we characterize the effect of the correlated noise by determining the ratio of the ``real'' error ($\epsilon_\mathrm{real}$), directly estimated from the residuals $R_{\lambda k}$, to the analytically propagated error ($\epsilon_\mathrm{bin}$) of binned spectra as a function of bin for a certain target S/N. The results obtained for all DR1 datasets are shown in Fig.~\ref{fig:DR1_noise_correlation} for the two instrumental setups with a target S/N level of 20. The observed trends can be sufficiently described by a simple logarithmic function\footnote{An independent estimate of this empirical relation, proposed by Cid Fernandes et al. (in prep.), is completely consistent within the estimated uncertainties.} 
\begin{equation}
 \epsilon_\mathrm{real}=\epsilon_\mathrm{bin}\left[1+\alpha\log n\right]\quad ,\label{eq:correlation}
\end{equation}
with $n$ the number of spaxels per bin.

The values for the slope $\alpha$ range from 1.35 to 1.45 with a mean of $\alpha=1.4$, for target S/N values between 10 and 60. The fits to the data result in less accurate exponents  for target S/N ratios of 10 and 60 given the poor sampling towards large- and small-size bins, but the overall shape is preserved.  This means that $\alpha$ mainly depends on the number of spaxels per bin and not on the target S/N.

\begin{figure*}
    \includegraphics[width=\textwidth]{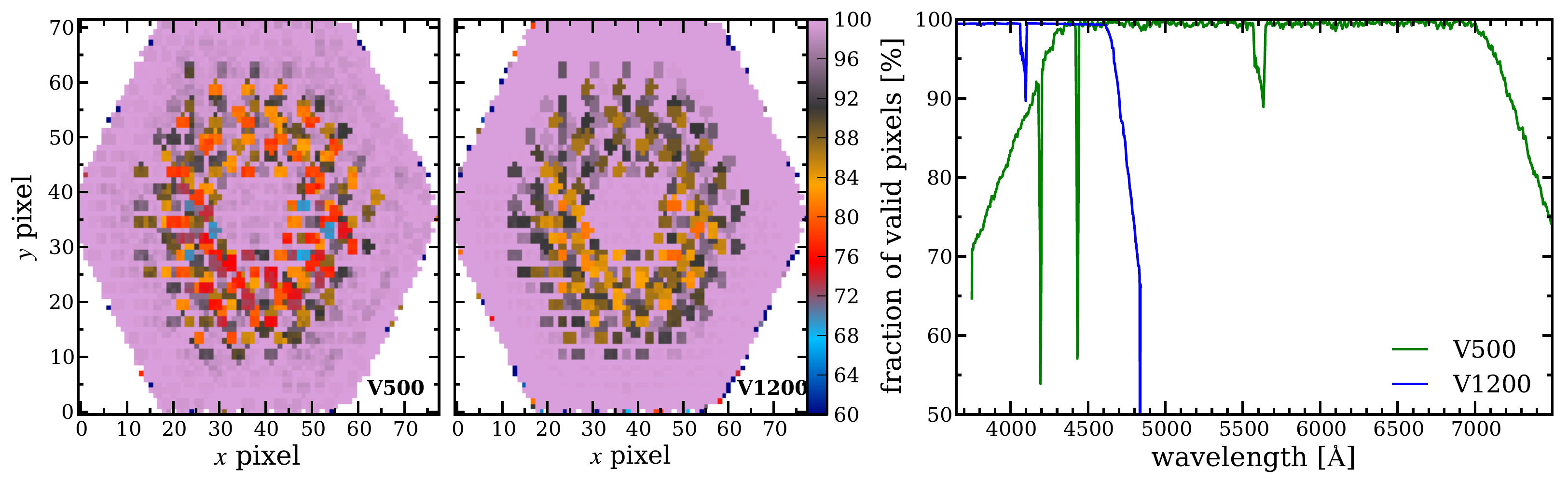}
    \caption{Fraction of valid pixels in each spectrum across the CALIFA FoV (left panels) and as a function of wavelength (right panel) for the V500 and V1200 setup. The spaxels most severely affected by the vignetting in the blue and red part of the spectra lead to the ring like structure around the FoV center. The blue part of the V1200 data does not show a vignetting effect here because we cut off the wavelength range in the final data already at 3650\AA\ due to the very low sensitivity in the blue. Four bad CCD columns are visible that lead to a significantly reduced fraction of valid pixels at narrow wavelength regions.}
  \label{fig:badpix_frac}
\end{figure*}

\section{CALIFA data format and characteristics}\label{sect:data_format}
The CALIFA data are stored and delivered as datacubes (three-dimensional data) in the standard binary FITS format and consist of four FITS Header/Data units (HDU). These datacubes represent (1) the measured flux densities, corrected for Galactic extinction as described in S12, in units of $10^{-16}\,\mathrm{erg}\,\mathrm{s}^{-1}\,\mathrm{cm}^{-2}\,\mathrm{\AA}^{-1}$ (primary datacube), (2) associated errors, (3) error weighting factors, and (4) bad pixels flags (Table~\ref{tab:HDUs}). They allow users to properly take into account the characteristics of each dataset concerning bad pixels and noise for their specific analysis. The first two axes of the cubes correspond to the spatial dimension along right ascension and declination with a $1\arcsec\times1\arcsec$ sampling. The third dimension represents the wavelength along a linear grid. The dimensions of each datacube ($N_\alpha$, $N_\delta$, and $N_\lambda$), as well as the spectral sampling ($d_\lambda$) and constant resolution ($\delta_\lambda$) along the entire wavelength range, are summarized in Table~\ref{tab:cube_dimension}. Note that the spatial resolution is worse than the actual seeing during the observation because the large aperture of each fiber (2\farcs7) strongly undersamples the point spread function. The final CALIFA spatial resolution is mainly set by the dither pattern and the image reconstruction scheme rather than the seeing. We evaluate this as part of our quality control tests discussed below in Sect.~\ref{sect:spatial_res}.

\subsection{Error datacubes}
The error datacubes in the first FITS extension correspond to the associated $1\sigma$ noise level of each pixel as formally propagated by the pipeline. As validation of our efforts to optimize the pipeline, we verified that the measured noise is systematically lower by $\sim$20\% than the formal error vector for individual spaxels (see Sect.~\ref{sect:noise}). In the case of bad pixels, we assigned an error value that is roughly ten orders of magnitude higher than the typical value for the considered dataset. Any analysis based on the $\chi^2$ statistics will implicitly take into account bad pixels if the error vector is considered. The correlation of noise becomes important only in cases where the CALIFA data need to be spatially binned as discussed in Sect.~\ref{sect:correlation}. The second FITS extension reports the error scaling factor for each pixel in the limiting case that all valid spaxels of the cube would be co-added. Suggestions about how to deal with the correlated noise and proper usage of the error scaling factors were described above in detail. 

\begin{table*}
\caption{CALIFA FITS file structure}
\label{tab:HDUs}
\begin{tabular}{clll}\hline\hline
HDU & Extension name & Format & Content\\\hline
0 & Primary & 32-bit float & flux density in units of $10^{-16}\,\mathrm{erg}\,\mathrm{s}^{-1}\,\mathrm{cm}^{-2}\,\mathrm{\AA}^{-1}$\\
1 & ERROR & 32-bit float & $1\sigma$ error on the flux density\\
2 & ERRWEIGHT & 32-bit float & error weighting factor\\
3 & BADPIX &   8-bit integer & bad pixel flags (1=bad, 0=good)\\ \hline
\end{tabular}
\end{table*}

\begin{table}
\caption{Dimension and sampling of CALIFA datacubes}
\label{tab:cube_dimension}
\begin{tabular}{lccccccc}\hline\hline
Setup & $N_\alpha$\tablefootmark{a} & $N_\delta$\tablefootmark{a} & $N_\lambda$\tablefootmark{a} & $\lambda_\mathrm{start}$\tablefootmark{b} & $\lambda_\mathrm{end}$\tablefootmark{c} & $d_\lambda$\tablefootmark{d} & $\delta_\lambda$\tablefootmark{e} \\\hline
V500  &  78        & 73         &  1877       & 3749\AA                  & 7501\AA                & 2.0\AA      & 6.0\AA           \\
V1200 &  78        & 73         &  1701       & 3650\AA                  & 4840\AA                & 0.7\AA      & 2.3\AA           \\\hline
\end{tabular}
\tablefoot{\tablefoottext{a}{Number of pixels in each dimension.}
\tablefoottext{b}{Wavelength of the first pixel on the wavelength direction.}
\tablefoottext{c}{Wavelength of the last pixel on the wavelength direction.}
\tablefoottext{d}{Wavelength sampling per pixel.}
\tablefoottext{e}{Homogenized spectral resolution (FWHM) over the entire wavelength range.}
}
\end{table}

\subsection{Bad pixel datacubes}
Bad pixel datacubes are produced by the pipeline and stored in the third FITS extension. They report pixels for which no sufficient information is available in the raw data because of cosmic rays, bad CCD columns, or the effect of vignetting. The data at bad pixels have been interpolated, and \emph{it is considered not usable} even if the spectrum looks ``good'' at those locations. The V1200 data is less affected by cosmic rays compared to the V500 data, because several frames are observed per pointing. The data at a bad pixel are therefore restored from the other available frame(s) although this results in a lower S/N. 

The distribution of bad pixels is not homogeneous within the datacube because of the vignetting effect as noted in S12.  In Fig.~\ref{fig:badpix_frac}, we present the typical fraction of valid pixels along the spatial and spectral axis for the V500 and V1200 setup, respectively. Note that four bad CCD columns have been identified, which lead to four wavelength regions where the fraction of valid pixels is significantly reduced. The vignetting effect of the instrument results in a wavelength coverage reduced by up to $\sim25$\% for some fibers on the blue and/or red side of the spectral range. 

The hexagonal PPak FoV is resampled to a rectangular grid, so that the uncovered corners are filled with zeros. These are also flagged as bad pixels for consistency, whereas the residuals of bright night-sky emission lines are not flagged as bad pixels. The strength of their residuals is different for each dataset and might be handled differently depending on the specific data analysis. 

\subsection{FITS header information}
The FITS header contains the standard keywords that register the spatial axes to the standard World Coordinate System \citep[WCS,][]{Greisen:2002} and the wavelength to the spectral axis in a linear grid. Each CALIFA datacube  contains the full FITS header information of all raw frames from which it was created, where each header entry is expanded with a unique prefix for a given pointing/frame. The prefix consists of the string ``PPAK'' followed by the designation of the pointing \textit{PREF}, which can be ``P1'', ``P2'', or ``P3'' for the three pointings of the V500 setup, or ``P1F1'', ``P1F2'', ``P2F1'', etc..., for the individual frames taken for each pointing of the V1200 setup. 

The reduction pipeline also collects information regarding, e.g., sky brightness, flexure offsets, Galactic extinction, approximate limiting magnitude, etc., and adds it to the FITS header. Header keywords that may be of general interest for the analysis and/or evaluation of the data are summarized in Table~\ref{tab:header_keys} for convenience. Note that the systemic velocity of the galaxies (MED\_VEL) automatically estimated by the pipeline is not robustly measured from the V1200 data, because of the smaller wavelength coverage and lower S/N compared to the V500 data. 

\begin{table*}
 \caption{Main FITS header keywords and their meaning}
 \label{tab:header_keys}
 \centering
 \begin{tabular}{lll}\hline\hline
  Keyword & data type & Meaning\\\hline
  OBJECT & string & Name of the target galaxy \\
  CALIFAID & integer & ID of the galaxy \\
  NAXIS1   & integer & Number of pixels along right ascension axis  ($N_\alpha$)\\
  NAXIS2   & integer & Number of pixels along declination axis  ($N_\delta$)\\
  NAXIS3   & integer & Number of pixels along wavelength axis  ($N_\lambda$)\\
  CRPIX1   & float   & Reference pixel of the galaxy center in right ascension\\
  CRVAL1   & float   & Right ascension $\alpha$ (J2000) of the galaxy center in degrees\\
  CDELT1   & float   & Sampling along right ascension axis in arcsec\\
  CRPIX2   & float   & Reference pixel of the galaxy center in declination\\
  CRVAL2   & float   & Declination $\delta$ (J2000) of the galaxy center in degrees\\
  CDELT2   & float   & Sampling along declination axis in arcsec\\
  CRVAL3   & float   & Wavelength of the first pixel along the wavelength axis in \AA \\
  CDELT3   & float   & Sampling of the wavelength axis in \AA\\
  hierarch PIPE VERS & string & Version of the reduction pipeline\\
  hierarch PIPE UNITS & string & Units of the flux density\\
  hierarch PIPE GALEXT AV & float & Galactic $V$ band extinction along line-of-sight\\
  hierarch PIPE REDUDATE & string & Date/time the data were reduced\\
  hierarch PIPE VMAG 3SIG & float & Estimated $3\sigma$ surface brightness depth in $\mathrm{mag}\,\mathrm{arcsec}^{-2}$\\
  MED\_VEL & float & Estimated systemic velocity in km/s\\
  hierarch PPAK \textit{PREF} DATE-OBS & string & Date/time of the observation\\
  hierarch PPAK \textit{PREF} UT\_START & float & Start of the observation in seconds UT time\\
  hierarch PPAK \textit{PREF} UT\_END & float & End of the observation in seconds UT time\\
  hierarch PPAK \textit{PREF} MJD-OBS & float & Modified Julian date of the observation\\
  hierarch PPAK \textit{PREF} AIRMASS & float & Airmass during the observation\\
  hierarch PPAK \textit{PREF} EXPTIM & float & Exposure time of the observation\\  
  hierarch PPAK \textit{PREF} HVCOR & float & Line-of-sight difference to helocentric velocity\\
  hierarch PPAK \textit{PREF} EXT\_V & float & $V$ band atmospheric extinction in magnitudes\\                
  hierarch PPAK \textit{PREF} PIPE FLEX XOFF & float & Measured flexure offset in $x$-direction in CCD pixels\\
  hierarch PPAK \textit{PREF} PIPE FLEX YOFF & float & Measured flexure offset in $y$-direction in CCD pixels\\
  hierarch PPAK \textit{PREF} PIPE SPEC RES & float & Homogenized spectral resolution (FWHM) in \AA \\    
  hierarch PPAK \textit{PREF} PIPE SKY MEAN & float & Mean surface brightness ($\mu_\mathrm{sky}$) during observation \\\hline
 \end{tabular}
\end{table*}

\begin{longtab}
\onecolumn
\begin{longtable}{lccccccccccc}
\caption{CALIFA DR1 quality control parameters for the V500 data }\\\hline\hline
\label{tab:QC_par_V500}
\small{ID} & \small{airmass\tablefootmark{a}} & \small{$\mu_{V,\mathrm{sky}}$\tablefootmark{b}} & \small{$A_V$\tablefootmark{c}} & \small{seeing\tablefootmark{d}} & \small{flags(R)\tablefootmark{e}} & \small{$\delta_\lambda$\tablefootmark{f}} & \small{$\left(\frac{g_\mathrm{CALIFA}}{g_\mathrm{SDSS}}\right)$\tablefootmark{g}} & \small{$\left(\frac{r_\mathrm{CALIFA}}{r_\mathrm{SDSS}}\right)$\tablefootmark{h}} & \small{S/N($R_{50}$)\tablefootmark{i}} & \small{$I_{3\sigma}$\tablefootmark{j}} & \small{flags(QC)\tablefootmark{k}}\\\hline
\endfirsthead
\caption{continued.}\\\hline\hline
\small{ID} & \small{airmass\tablefootmark{a}} & \small{$\mu_{V,\mathrm{sky}}$\tablefootmark{b}} & \small{$A_V$\tablefootmark{c}} & \small{seeing\tablefootmark{d}} & \small{flags(R)\tablefootmark{e}} & \small{$\delta_\lambda$\tablefootmark{f}} & \small{$\left(\frac{g_\mathrm{CALIFA}}{g_\mathrm{SDSS}}\right)$\tablefootmark{g}} & \small{$\left(\frac{r_\mathrm{CALIFA}}{r_\mathrm{SDSS}}\right)$\tablefootmark{h}} & \small{S/N($R_{50}$)\tablefootmark{i}} & \small{$I_{3\sigma}$\tablefootmark{j}} & \small{flags(QC)\tablefootmark{k}}\\\hline
\endhead
\hline
\endfoot
\small{001} & \small{$1.02\pm0.01$} & \small{$21.4$} &  \small{$0.18$} & \small{...} & \small{0000} & \small{$5.95\pm0.33$} & \small{1.01} & \small{0.94} & \small{35$\pm$6} & \small{0.8$\pm$0.2} & \small{0000}\\
\small{003} & \small{$1.01\pm0.01$} & \small{$20.8$} &  \small{$0.23$} & \small{$0.8\pm0.1$} & \small{0000} & \small{$6.46\pm0.98$} & \small{1.01} & \small{1.05} & \small{18$\pm$5} & \small{1.0$\pm$0.3} & \small{0000}\\
\small{007} & \small{$1.34\pm0.05$} & \small{$20.5$} &  \small{$0.17$} & \small{...} & \small{0000} & \small{$6.11\pm0.41$} & \small{0.92} & \small{0.95} & \small{32$\pm$5} & \small{1.1$\pm$0.3} & \small{0000}\\
\small{010} & \small{$1.31\pm0.05$} & \small{$21.0$} &  \small{$0.22$} & \small{...} & \small{0000} & \small{$5.70\pm0.19$} & \small{1.05} & \small{0.98} & \small{27$\pm$7} & \small{1.0$\pm$0.2} & \small{0000}\\
\small{014} & \small{$1.26\pm0.04$} & \small{$21.0$} &  \small{$0.21$} & \small{...} & \small{0000} & \small{$5.72\pm0.33$} & \small{1.43} & \small{1.31} & \small{28$\pm$7} & \small{0.6$\pm$0.2} & \small{0010}\\
\small{039} & \small{$1.03\pm0.01$} & \small{$20.8$} &  \small{$0.23$} & \small{$0.9\pm0.1$} & \small{0010} & \small{$6.02\pm0.18$} & \small{1.04} & \small{1.05} & \small{19$\pm$4} & \small{0.9$\pm$0.3} & \small{0000}\\
\small{042} & \small{$1.10\pm0.03$} & \small{$21.0$} &  \small{$0.13$} & \small{...} & \small{0000} & \small{$6.77\pm0.57$} & \small{1.06} & \small{1.00} & \small{17$\pm$3} & \small{1.1$\pm$0.3} & \small{0200}\\
\small{043} & \small{$1.01\pm0.01$} & \small{$21.3$} &  \small{$0.19$} & \small{...} & \small{0010} & \small{$6.12\pm0.22$} & \small{0.99} & \small{0.91} & \small{37$\pm$6} & \small{0.8$\pm$0.2} & \small{0000}\\
\small{053} & \small{$1.19\pm0.04$} & \small{$20.7$} &  \small{$0.17$} & \small{$0.8\pm0.1$} & \small{0000} & \small{$6.35\pm0.29$} & \small{1.04} & \small{1.05} & \small{32$\pm$7} & \small{1.0$\pm$0.2} & \small{0000}\\
\small{073} & \small{$1.04\pm0.01$} & \small{$21.1$} &  \small{$0.20$} & \small{...} & \small{0000} & \small{$6.47\pm0.45$} & \small{1.03} & \small{0.95} & \small{24$\pm$4} & \small{1.5$\pm$0.3} & \small{0000}\\
\small{088} & \small{$1.13\pm0.03$} & \small{$20.8$} &  \small{$0.17$} & \small{$1.0\pm0.4$} & \small{0000} & \small{$5.65\pm0.87$} & \small{1.04} & \small{0.99} & \small{27$\pm$5} & \small{1.2$\pm$0.3} & \small{0000}\\
\small{100} & \small{$1.37\pm0.26$} & \small{$20.7$} &  \small{$0.14$} & \small{...} & \small{0000} & \small{$6.39\pm0.37$} & \small{1.00} & \small{0.99} & \small{36$\pm$7} & \small{1.9$\pm$0.5} & \small{0000}\\
\small{127} & \small{$1.22\pm0.02$} & \small{$20.6$} &  \small{$0.28$} & \small{$0.9\pm0.1$} & \small{0000} & \small{$5.90\pm0.32$} & \small{1.00} & \small{1.00} & \small{19$\pm$3} & \small{2.0$\pm$0.5} & \small{0000}\\
\small{146} & \small{$1.47\pm0.00$} & \small{$20.7$} &  \small{$0.12$} & \small{...} & \small{0000} & \small{$4.86\pm0.79$} & \small{0.97} & \small{0.94} & \small{21$\pm$4} & \small{1.7$\pm$0.4} & \small{0000}\\
\small{151} & \small{$1.07\pm0.02$} & \small{$21.0$} &  \small{$0.15$} & \small{...} & \small{0000} & \small{$7.01\pm0.16$} & \small{1.07} & \small{1.03} & \small{26$\pm$7} & \small{1.6$\pm$0.4} & \small{0000}\\
\small{155} & \small{$1.03\pm0.01$} & \small{$21.1$} &  \small{$0.16$} & \small{...} & \small{0100} & \small{$6.34\pm0.28$} & \small{1.05} & \small{1.02} & \small{18$\pm$5} & \small{1.6$\pm$0.5} & \small{0000}\\
\small{156} & \small{$1.12\pm0.04$} & \small{$21.0$} &  \small{$0.16$} & \small{...} & \small{0100} & \small{$6.33\pm0.36$} & \small{0.99} & \small{0.94} & \small{22$\pm$14} & \small{1.4$\pm$0.4} & \small{0000}\\
\small{273} & \small{$1.06\pm0.01$} & \small{$21.1$} &  \small{$0.14$} & \small{...} & \small{0000} & \small{$6.67\pm0.19$} & \small{1.05} & \small{1.00} & \small{27$\pm$5} & \small{1.0$\pm$0.3} & \small{0000}\\
\small{274} & \small{$1.18\pm0.02$} & \small{$20.9$} &  \small{$0.17$} & \small{...} & \small{0000} & \small{$5.64\pm0.29$} & \small{1.03} & \small{0.95} & \small{34$\pm$7} & \small{0.9$\pm$0.2} & \small{0000}\\
\small{277} & \small{$1.60\pm0.12$} & \small{$20.3$} &  \small{$0.33$} & \small{...} & \small{1000} & \small{$6.69\pm0.47$} & \small{1.07} & \small{1.12} & \small{21$\pm$4} & \small{1.8$\pm$0.6} & \small{1000}\\
\small{306} & \small{$1.12\pm0.01$} & \small{$20.9$} &  \small{$0.16$} & \small{$1.0\pm0.1$} & \small{0000} & \small{$5.54\pm0.06$} & \small{1.05} & \small{0.99} & \small{13$\pm$5} & \small{0.8$\pm$0.3} & \small{0000}\\
\small{307} & \small{$1.06\pm0.01$} & \small{$21.2$} &  \small{$0.16$} & \small{...} & \small{0100} & \small{$3.84\pm0.71$} & \small{1.07} & \small{1.04} & \small{18$\pm$4} & \small{1.2$\pm$0.3} & \small{0000}\\
\small{309} & \small{$1.27\pm0.05$} & \small{$21.0$} &  \small{$0.15$} & \small{...} & \small{0000} & \small{$7.44\pm0.30$} & \small{1.11} & \small{1.04} & \small{15$\pm$4} & \small{1.6$\pm$0.4} & \small{0000}\\
\small{319} & \small{$1.11\pm0.03$} & \small{$20.5$} &  \small{$0.56$} & \small{...} & \small{0000} & \small{$5.82\pm0.39$} & \small{1.01} & \small{1.07} & \small{25$\pm$6} & \small{1.4$\pm$0.4} & \small{0000}\\
\small{326} & \small{$1.05\pm0.01$} & \small{$20.8$} &  \small{$0.17$} & \small{...} & \small{0000} & \small{$5.72\pm0.09$} & \small{1.01} & \small{0.96} & \small{31$\pm$7} & \small{0.8$\pm$0.2} & \small{0000}\\
\small{341} & \small{$0.78\pm0.68$} & \small{$14.0$} &  \small{$0.11$} & \small{$1.5\pm0.2$} & \small{1000} & \small{$6.24\pm0.41$} & \small{0.99} & \small{0.97} & \small{30$\pm$6} & \small{1.0$\pm$0.3} & \small{0000}\\
\small{364} & \small{$1.33\pm0.07$} & \small{$20.8$} &  \small{$0.12$} & \small{...} & \small{0000} & \small{$6.30\pm0.26$} & \small{1.00} & \small{0.95} & \small{39$\pm$9} & \small{1.0$\pm$0.2} & \small{0000}\\
\small{475} & \small{$1.35\pm0.07$} & \small{$20.8$} &  \small{$0.12$} & \small{...} & \small{0000} & \small{$6.42\pm0.56$} & \small{1.82} & \small{1.66} & \small{29$\pm$17} & \small{1.0$\pm$0.2} & \small{0210}\\
\small{479} & \small{$1.14\pm0.03$} & \small{$20.8$} &  \small{$0.17$} & \small{...} & \small{0000} & \small{$5.62\pm0.12$} & \small{1.01} & \small{0.96} & \small{32$\pm$6} & \small{0.8$\pm$0.2} & \small{0000}\\
\small{486} & \small{$0.70\pm0.61$} & \small{$14.0$} &  \small{$0.10$} & \small{$1.2\pm0.7$} & \small{0000} & \small{$5.88\pm0.48$} & \small{1.03} & \small{0.98} & \small{27$\pm$7} & \small{0.7$\pm$0.2} & \small{0000}\\
\small{515} & \small{$1.01\pm0.00$} & \small{$20.9$} &  \small{$0.14$} & \small{...} & \small{0000} & \small{$6.22\pm0.24$} & \small{1.12} & \small{1.06} & \small{21$\pm$4} & \small{1.7$\pm$0.3} & \small{0000}\\
\small{518} & \small{$1.16\pm0.01$} & \small{$20.9$} &  \small{$0.15$} & \small{...} & \small{0000} & \small{$6.52\pm0.30$} & \small{1.03} & \small{1.00} & \small{23$\pm$4} & \small{1.6$\pm$0.4} & \small{0000}\\
\small{528} & \small{$1.14\pm0.01$} & \small{$20.8$} &  \small{$0.17$} & \small{...} & \small{0001} & \small{$6.30\pm0.38$} & \small{1.17} & \small{1.16} & \small{9$\pm$3} & \small{0.9$\pm$0.3} & \small{0000}\\
\small{548} & \small{$1.16\pm0.01$} & \small{$20.8$} &  \small{$0.19$} & \small{$1.2\pm0.1$} & \small{0000} & \small{$6.04\pm0.50$} & \small{1.02} & \small{0.98} & \small{53$\pm$10} & \small{0.8$\pm$0.4} & \small{0000}\\
\small{577} & \small{$1.02\pm0.01$} & \small{$21.0$} &  \small{$0.15$} & \small{...} & \small{0000} & \small{$6.12\pm0.41$} & \small{1.07} & \small{1.00} & \small{17$\pm$11} & \small{0.8$\pm$0.2} & \small{0100}\\
\small{607} & \small{$1.22\pm0.03$} & \small{$21.3$} &  \small{$0.16$} & \small{...} & \small{0100} & \small{$6.26\pm0.17$} & \small{1.00} & \small{0.94} & \small{50$\pm$11} & \small{1.1$\pm$0.3} & \small{0000}\\
\small{608} & \small{$1.31\pm0.06$} & \small{$21.1$} &  \small{$0.17$} & \small{...} & \small{0000} & \small{$7.20\pm0.24$} & \small{1.12} & \small{1.08} & \small{17$\pm$4} & \small{1.3$\pm$0.3} & \small{0000}\\
\small{609} & \small{$1.07\pm0.02$} & \small{$21.3$} &  \small{$0.14$} & \small{...} & \small{0100} & \small{$4.56\pm1.09$} & \small{1.07} & \small{1.00} & \small{21$\pm$6} & \small{1.0$\pm$0.2} & \small{0000}\\
\small{610} & \small{$1.11\pm0.03$} & \small{$21.2$} &  \small{$0.17$} & \small{...} & \small{0000} & \small{$6.99\pm0.23$} & \small{1.12} & \small{1.06} & \small{26$\pm$7} & \small{1.2$\pm$0.3} & \small{0000}\\
\small{657} & \small{$1.02\pm0.01$} & \small{$21.2$} &  \small{$0.19$} & \small{$1.1\pm0.1$} & \small{0000} & \small{$6.44\pm0.84$} & \small{1.08} & \small{1.01} & \small{15$\pm$4} & \small{0.8$\pm$0.3} & \small{0000}\\
\small{663} & \small{$1.20\pm0.04$} & \small{$20.9$} &  \small{$0.17$} & \small{...} & \small{0000} & \small{$5.61\pm0.42$} & \small{1.06} & \small{0.99} & \small{29$\pm$7} & \small{1.0$\pm$0.3} & \small{0000}\\
\small{664} & \small{$1.04\pm0.02$} & \small{$21.2$} &  \small{$0.17$} & \small{$1.1\pm0.1$} & \small{0000} & \small{$6.46\pm0.96$} & \small{1.02} & \small{0.97} & \small{40$\pm$6} & \small{0.8$\pm$0.2} & \small{0000}\\
\small{665} & \small{$1.06\pm0.01$} & \small{$21.1$} &  \small{$0.17$} & \small{...} & \small{0000} & \small{$6.29\pm0.44$} & \small{1.03} & \small{0.96} & \small{22$\pm$3} & \small{1.1$\pm$0.4} & \small{0000}\\
\small{676} & \small{$1.03\pm0.03$} & \small{$21.0$} &  \small{$0.15$} & \small{...} & \small{1100} & \small{$6.62\pm1.04$} & \small{1.02} & \small{0.97} & \small{17$\pm$4} & \small{1.5$\pm$0.3} & \small{0000}\\
\small{680} & \small{$1.07\pm0.02$} & \small{$21.0$} &  \small{$0.17$} & \small{...} & \small{0000} & \small{$5.57\pm0.13$} & \small{1.00} & \small{0.94} & \small{19$\pm$9} & \small{0.7$\pm$0.2} & \small{0000}\\
\small{684} & \small{$1.05\pm0.02$} & \small{$21.0$} &  \small{$0.17$} & \small{...} & \small{0000} & \small{$5.48\pm0.11$} & \small{1.03} & \small{0.98} & \small{33$\pm$6} & \small{1.5$\pm$0.3} & \small{0000}\\
\small{758} & \small{$1.02\pm0.00$} & \small{$21.1$} &  \small{$0.24$} & \small{$1.1\pm0.1$} & \small{0000} & \small{$7.11\pm1.14$} & \small{1.13} & \small{1.08} & \small{23$\pm$6} & \small{0.7$\pm$0.3} & \small{0000}\\
\small{764} & \small{$1.29\pm0.05$} & \small{$21.1$} &  \small{$0.18$} & \small{$1.0\pm0.1$} & \small{0000} & \small{$5.43\pm0.31$} & \small{1.04} & \small{0.94} & \small{24$\pm$4} & \small{0.9$\pm$0.2} & \small{0000}\\
\small{769} & \small{$1.02\pm0.01$} & \small{$21.2$} &  \small{$0.17$} & \small{...} & \small{0000} & \small{$7.20\pm0.62$} & \small{1.05} & \small{1.00} & \small{22$\pm$4} & \small{1.2$\pm$0.3} & \small{0000}\\
\small{783} & \small{$1.08\pm0.02$} & \small{$21.3$} &  \small{$0.16$} & \small{...} & \small{0000} & \small{$5.76\pm0.91$} & \small{1.08} & \small{1.07} & \small{35$\pm$8} & \small{0.9$\pm$0.2} & \small{0000}\\
\small{797} & \small{$1.05\pm0.07$} & \small{$21.0$} &  \small{$0.15$} & \small{...} & \small{0000} & \small{$6.79\pm0.14$} & \small{1.01} & \small{0.98} & \small{23$\pm$5} & \small{0.7$\pm$0.2} & \small{0000}\\
\small{798} & \small{$1.05\pm0.02$} & \small{$21.3$} &  \small{$0.21$} & \small{...} & \small{0000} & \small{$5.68\pm0.46$} & \small{1.03} & \small{1.00} & \small{30$\pm$6} & \small{0.6$\pm$0.2} & \small{0000}\\
\small{806} & \small{$1.00\pm0.00$} & \small{$21.1$} &  \small{$0.25$} & \small{$1.1\pm0.1$} & \small{0000} & \small{$6.33\pm0.82$} & \small{1.15} & \small{1.08} & \small{22$\pm$5} & \small{1.0$\pm$0.2} & \small{0000}\\
\small{820} & \small{$1.07\pm0.02$} & \small{$21.2$} &  \small{$0.17$} & \small{...} & \small{0100} & \small{$4.49\pm1.45$} & \small{1.02} & \small{1.00} & \small{11$\pm$3} & \small{2.1$\pm$0.5} & \small{0100}\\
\small{822} & \small{$1.03\pm0.01$} & \small{$21.2$} &  \small{$0.17$} & \small{...} & \small{0000} & \small{$7.53\pm0.17$} & \small{1.32} & \small{1.22} & \small{19$\pm$5} & \small{1.4$\pm$0.3} & \small{0000}\\
\small{823} & \small{$1.15\pm0.00$} & \small{$20.9$} &  \small{$0.19$} & \small{...} & \small{0000} & \small{$5.40\pm0.08$} & \small{1.09} & \small{0.99} & \small{26$\pm$5} & \small{0.9$\pm$0.3} & \small{0000}\\
\small{824} & \small{$1.14\pm0.01$} & \small{$20.9$} &  \small{$0.17$} & \small{...} & \small{0010} & \small{$5.60\pm0.12$} & \small{1.42} & \small{1.13} & \small{38$\pm$7} & \small{0.8$\pm$0.3} & \small{0010}\\
\small{826} & \small{$1.18\pm0.03$} & \small{$21.1$} &  \small{$0.17$} & \small{...} & \small{0000} & \small{$6.54\pm0.38$} & \small{1.01} & \small{0.94} & \small{33$\pm$7} & \small{1.1$\pm$0.3} & \small{0000}\\
\small{828} & \small{$1.21\pm0.03$} & \small{$21.3$} &  \small{$0.26$} & \small{$0.9\pm0.1$} & \small{0000} & \small{$5.70\pm0.07$} & \small{1.43} & \small{1.28} & \small{33$\pm$9} & \small{0.6$\pm$0.2} & \small{0010}\\
\small{829} & \small{$1.09\pm0.01$} & \small{$21.2$} &  \small{$0.18$} & \small{...} & \small{0000} & \small{$5.85\pm0.21$} & \small{1.12} & \small{1.05} & \small{41$\pm$7} & \small{1.3$\pm$0.2} & \small{0000}\\
\small{832} & \small{$1.01\pm0.01$} & \small{$21.3$} &  \small{$0.17$} & \small{...} & \small{0000} & \small{$7.29\pm0.68$} & \small{1.01} & \small{0.99} & \small{44$\pm$10} & \small{1.2$\pm$0.4} & \small{0000}\\
\small{833} & \small{$1.12\pm0.03$} & \small{$21.2$} &  \small{$0.18$} & \small{$0.9\pm0.1$} & \small{0000} & \small{$6.16\pm0.15$} & \small{1.05} & \small{0.94} & \small{22$\pm$4} & \small{1.1$\pm$0.2} & \small{0000}\\
\small{835} & \small{$1.00\pm0.00$} & \small{$21.1$} &  \small{$0.17$} & \small{$1.1\pm0.2$} & \small{0000} & \small{$6.09\pm0.50$} & \small{1.01} & \small{0.97} & \small{43$\pm$8} & \small{0.9$\pm$0.2} & \small{0000}\\
\small{837} & \small{$1.11\pm0.00$} & \small{$21.1$} &  \small{$0.17$} & \small{$0.9\pm0.1$} & \small{0000} & \small{$5.82\pm0.12$} & \small{1.13} & \small{1.08} & \small{49$\pm$7} & \small{0.7$\pm$0.2} & \small{0000}\\
\small{840} & \small{$1.18\pm0.04$} & \small{$21.0$} &  \small{$0.28$} & \small{$0.9\pm0.1$} & \small{0000} & \small{$6.26\pm0.44$} & \small{1.08} & \small{0.99} & \small{32$\pm$6} & \small{1.3$\pm$0.3} & \small{0000}\\
\small{843} & \small{$1.45\pm0.09$} & \small{$21.0$} &  \small{$0.23$} & \small{$0.8\pm0.1$} & \small{0000} & \small{$6.04\pm0.92$} & \small{1.12} & \small{0.97} & \small{18$\pm$7} & \small{1.0$\pm$0.3} & \small{0100}\\
\small{845} & \small{$1.03\pm0.02$} & \small{$21.4$} &  \small{$0.19$} & \small{...} & \small{0000} & \small{$5.64\pm0.13$} & \small{1.08} & \small{1.02} & \small{40$\pm$9} & \small{0.8$\pm$0.3} & \small{0000}\\
\small{846} & \small{$1.06\pm0.02$} & \small{$21.0$} &  \small{$0.39$} & \small{$0.8\pm0.1$} & \small{0000} & \small{$5.72\pm0.11$} & \small{1.07} & \small{1.00} & \small{22$\pm$4} & \small{1.1$\pm$0.3} & \small{0000}\\
\small{847} & \small{$1.31\pm0.09$} & \small{$21.3$} &  \small{$0.15$} & \small{$0.7\pm0.1$} & \small{0100} & \small{$5.41\pm0.56$} & \small{0.96} & \small{0.81} & \small{28$\pm$11} & \small{0.9$\pm$0.3} & \small{0000}\\
\small{848} & \small{$1.10\pm0.00$} & \small{$21.0$} &  \small{$0.38$} & \small{$0.9\pm0.1$} & \small{0000} & \small{$5.89\pm0.21$} & \small{1.06} & \small{1.00} & \small{36$\pm$5} & \small{0.9$\pm$0.3} & \small{0000}\\
\small{850} & \small{$1.05\pm0.01$} & \small{$21.0$} &  \small{$0.29$} & \small{...} & \small{0000} & \small{$6.35\pm0.62$} & \small{1.02} & \small{0.99} & \small{37$\pm$8} & \small{1.0$\pm$0.4} & \small{0000}\\
\small{851} & \small{$1.32\pm0.04$} & \small{$20.5$} &  \small{$0.41$} & \small{$0.7\pm0.1$} & \small{0000} & \small{$5.80\pm0.08$} & \small{1.08} & \small{1.05} & \small{17$\pm$4} & \small{1.6$\pm$0.4} & \small{0000}\\
\small{852} & \small{$1.12\pm0.01$} & \small{$21.0$} &  \small{$0.36$} & \small{$0.9\pm0.1$} & \small{0000} & \small{$5.77\pm0.07$} & \small{1.04} & \small{1.01} & \small{12$\pm$5} & \small{0.9$\pm$0.3} & \small{0000}\\
\small{854} & \small{$1.18\pm0.03$} & \small{$20.7$} &  \small{$0.42$} & \small{$0.9\pm0.1$} & \small{0000} & \small{$5.32\pm0.90$} & \small{1.05} & \small{0.98} & \small{24$\pm$6} & \small{1.1$\pm$0.3} & \small{0000}\\
\small{856} & \small{$1.05\pm0.02$} & \small{$21.2$} &  \small{$0.17$} & \small{...} & \small{0000} & \small{$5.53\pm0.14$} & \small{1.06} & \small{0.98} & \small{28$\pm$5} & \small{0.9$\pm$0.2} & \small{0000}\\
\small{857} & \small{$1.08\pm0.00$} & \small{$21.5$} &  \small{$0.16$} & \small{$0.9\pm0.1$} & \small{0000} & \small{$6.00\pm0.13$} & \small{1.14} & \small{1.03} & \small{40$\pm$7} & \small{0.7$\pm$0.2} & \small{0000}\\
\small{858} & \small{$1.10\pm0.03$} & \small{$21.2$} &  \small{$0.15$} & \small{$1.3\pm0.2$} & \small{0000} & \small{$6.07\pm0.32$} & \small{1.06} & \small{0.97} & \small{36$\pm$6} & \small{1.1$\pm$0.4} & \small{0000}\\
\small{859} & \small{$1.16\pm0.02$} & \small{$21.2$} &  \small{$0.15$} & \small{$0.7\pm0.1$} & \small{0000} & \small{$5.98\pm0.47$} & \small{1.04} & \small{0.99} & \small{46$\pm$7} & \small{1.1$\pm$0.3} & \small{0000}\\
\small{860} & \small{$1.04\pm0.02$} & \small{$21.2$} &  \small{$0.29$} & \small{$1.0\pm0.1$} & \small{0000} & \small{$6.21\pm0.79$} & \small{1.00} & \small{0.84} & \small{44$\pm$5} & \small{1.0$\pm$0.2} & \small{0000}\\
\small{863} & \small{$1.09\pm0.01$} & \small{$21.4$} &  \small{$0.16$} & \small{$0.7\pm0.1$} & \small{0000} & \small{$5.78\pm0.15$} & \small{0.99} & \small{0.94} & \small{33$\pm$6} & \small{0.8$\pm$0.2} & \small{0000}\\
\small{864} & \small{$1.06\pm0.01$} & \small{$20.9$} &  \small{$0.28$} & \small{...} & \small{0100} & \small{$5.08\pm0.44$} & \small{1.05} & \small{0.96} & \small{25$\pm$6} & \small{1.4$\pm$0.3} & \small{0000}\\
\small{865} & \small{$1.01\pm0.01$} & \small{$21.2$} &  \small{$0.25$} & \small{...} & \small{0100} & \small{$5.75\pm0.44$} & \small{0.96} & \small{0.90} & \small{38$\pm$6} & \small{0.8$\pm$0.2} & \small{0000}\\
\small{866} & \small{$1.04\pm0.02$} & \small{$21.3$} &  \small{$0.21$} & \small{$1.1\pm0.1$} & \small{0111} & \small{$5.36\pm0.40$} & \small{0.98} & \small{0.88} & \small{16$\pm$4} & \small{0.7$\pm$0.2} & \small{0000}\\
\small{867} & \small{$1.13\pm0.01$} & \small{$21.2$} &  \small{$0.21$} & \small{$1.1\pm0.1$} & \small{0000} & \small{$5.69\pm0.20$} & \small{0.92} & \small{0.90} & \small{43$\pm$5} & \small{0.8$\pm$0.3} & \small{0000}\\
\small{872} & \small{$1.28\pm0.01$} & \small{$20.9$} &  \small{$0.15$} & \small{$0.6\pm0.1$} & \small{0000} & \small{$4.93\pm0.75$} & \small{1.03} & \small{1.00} & \small{22$\pm$4} & \small{1.1$\pm$0.5} & \small{0000}\\
\small{874} & \small{$1.10\pm0.02$} & \small{$21.0$} &  \small{$0.22$} & \small{...} & \small{0000} & \small{$5.74\pm0.51$} & \small{0.93} & \small{0.88} & \small{33$\pm$6} & \small{2.4$\pm$0.1} & \small{0000}\\
\small{877} & \small{$1.24\pm0.05$} & \small{$21.0$} &  \small{$0.22$} & \small{...} & \small{0100} & \small{$5.13\pm0.41$} & \small{0.77} & \small{0.80} & \small{26$\pm$7} & \small{1.1$\pm$0.3} & \small{0000}\\
\small{878} & \small{$1.33\pm0.01$} & \small{$20.9$} &  \small{$0.22$} & \small{$0.9\pm0.1$} & \small{0000} & \small{$4.00\pm0.90$} & \small{1.00} & \small{0.95} & \small{10$\pm$12} & \small{0.9$\pm$0.3} & \small{0000}\\
\small{881} & \small{$1.10\pm0.00$} & \small{$21.1$} &  \small{$0.15$} & \small{$0.8\pm0.1$} & \small{0000} & \small{$5.58\pm0.32$} & \small{1.11} & \small{1.03} & \small{45$\pm$8} & \small{1.0$\pm$0.2} & \small{0000}\\
\small{883} & \small{$1.10\pm0.01$} & \small{$21.2$} &  \small{$0.17$} & \small{$1.0\pm0.2$} & \small{0000} & \small{$6.11\pm0.21$} & \small{0.78} & \small{0.70} & \small{12$\pm$3} & \small{1.6$\pm$0.5} & \small{0000}\\
\small{887} & \small{$1.08\pm0.02$} & \small{$21.3$} &  \small{$0.17$} & \small{$0.9\pm0.1$} & \small{0000} & \small{$6.43\pm0.59$} & \small{1.06} & \small{0.98} & \small{38$\pm$6} & \small{0.8$\pm$0.3} & \small{0000}\\
\small{888} & \small{$1.01\pm0.01$} & \small{$20.9$} &  \small{$0.22$} & \small{...} & \small{0000} & \small{$5.25\pm0.41$} & \small{1.10} & \small{1.03} & \small{29$\pm$7} & \small{1.3$\pm$0.3} & \small{0000}\\
\small{890} & \small{$1.08\pm0.02$} & \small{$21.0$} &  \small{$0.22$} & \small{...} & \small{0000} & \small{$4.76\pm0.71$} & \small{1.04} & \small{1.00} & \small{35$\pm$6} & \small{0.9$\pm$0.3} & \small{0000}\\
\small{893} & \small{$1.35\pm0.07$} & \small{$20.5$} &  \small{$0.20$} & \small{$1.0\pm0.1$} & \small{1100} & \small{$5.96\pm0.28$} & \small{1.18} & \small{1.04} & \small{20$\pm$8} & \small{2.0$\pm$0.5} & \small{0000}\\
\small{896} & \small{$1.36\pm0.08$} & \small{$20.8$} &  \small{$0.18$} & \small{$0.9\pm0.1$} & \small{0000} & \small{$6.02\pm0.14$} & \small{1.00} & \small{0.96} & \small{27$\pm$6} & \small{1.0$\pm$0.3} & \small{0000}\\
\small{901} & \small{$1.10\pm0.03$} & \small{$21.3$} &  \small{$0.14$} & \small{$1.0\pm0.1$} & \small{0100} & \small{$5.54\pm0.26$} & \small{1.02} & \small{0.95} & \small{35$\pm$9} & \small{1.1$\pm$0.3} & \small{0000}\\
\small{902} & \small{$1.10\pm0.00$} & \small{$21.2$} &  \small{$0.17$} & \small{$0.8\pm0.1$} & \small{0000} & \small{$5.96\pm0.13$} & \small{1.01} & \small{0.93} & \small{41$\pm$6} & \small{1.0$\pm$0.3} & \small{0000}\\
\small{904} & \small{$1.20\pm0.04$} & \small{$21.0$} &  \small{$0.17$} & \small{$0.9\pm0.1$} & \small{0000} & \small{$6.10\pm0.10$} & \small{1.06} & \small{0.99} & \small{34$\pm$9} & \small{1.1$\pm$0.3} & \small{0000}\\
\small{935} & \small{$1.04\pm0.02$} & \small{$21.0$} &  \small{$0.17$} & \small{$1.1\pm0.1$} & \small{0000} & \small{$5.79\pm0.32$} & \small{1.02} & \small{1.02} & \small{16$\pm$6} & \small{0.8$\pm$0.3} & \small{0000}\\
\small{939} & \small{$1.11\pm0.04$} & \small{$20.8$} &  \small{$0.15$} & \small{...} & \small{0000} & \small{$5.96\pm0.22$} & \small{1.00} & \small{0.92} & \small{17$\pm$6} & \small{1.3$\pm$0.4} & \small{0000}\\
\end{longtable}
\tablefoot{We describe the meaning of each column including the identifier of each column in the electronic table available on the DR1 web page.
\tablefoottext{a}{Average airmass (OBS\_AIR\_MEAN) and rms (OBS\_AIR\_RMS) of the observations for the frames used to create the considered datacube.}
\tablefoottext{b}{Average night-sky surface brightness (OBS\_SKY\_MAG) in the $V$ band during the observations in units of mag\,arcsec$^{-2}$.}
\tablefoottext{c}{Average night-sky attenuation (OBS\_EXT\_MEAN) in the V band during the observations in magnitudes.}
\tablefoottext{d}{Average natural seeing (OBS\_SEE\_RD\_AVE) in the V-band during the observations in arcsec (FWHM).}
\tablefoottext{e}{Reduction quality flags, combining the four individual column flags (FLAG\_RED\_O, FLAG\_RED\_R, FLAG\_RED\_W, FLAG\_RED\_V) as described in Sect.~\ref{sect:QC}.}
\tablefoottext{f}{Average spectral resolution (CAL\_SP\_RES) and error (CAL\_SP\_RES\_ERR) in \AA\ (FWHM), measured by fitting the night-sky emission lines with single Gaussian functions.}
\tablefoottext{g}{Ratio between the SDSS $g$ band flux derived from the datacube and the one derived from the SDSS images for a 30$\arcsec$-diameter aperture (CAL\_DR8\_RATIO\_G).}
\tablefoottext{h}{Ratio between the SDSS $r$ band flux derived from the datacube and the one derived from the SDSS images for a 30$\arcsec$-diameter aperture (CAL\_DR8\_RATIO\_R).}
\tablefoottext{i}{Average signal-to-noise ratio (CAL\_SN\_MEAN\_WIN) and error (CAL\_SN\_ERR\_WIN) estimated for the full wavelength range at one half light radius from the center.}
\tablefoottext{j}{Average flux (CAL\_LS\_WIN) and error (CAL\_LS\_ERR\_WIN) at the 3$\sigma$ continuum detection limit in units of $10^{-18}\,\mathrm{erg}\,\mathrm{s}^{-1}\,\mathrm{cm}^{-2}\mathrm{\AA}^{-1}$.}
\tablefoottext{k}{Quality control flags, combining the four individual column flags (FLAG\_AIR, FLAG\_ASTR, FLAG\_FLUXCAL, FLAG\_DEPTH) as described in Sect.~\ref{sect:QC}.}
}
\twocolumn
\end{longtab}
\begin{longtab}
\onecolumn
\begin{longtable}{lccccccccc}
\caption{CALIFA DR1 quality control parameters for the V1200 data }\\\hline\hline
\label{tab:QC_par_V1200}
\small{ID} & \small{airmass} & \small{$\mu_{B,\mathrm{sky}}$} & \small{$A_V$} & \small{seeing} & \small{flags(R)} & \small{$\delta_\lambda$} &  \small{S/N($R_{50}$)} & \small{$I_{3\sigma}$} & \small{flags(QC)}\\\hline
\endfirsthead
\caption{continued.}\\\hline\hline
\small{ID} & \small{airmass} & \small{$\mu_{B,\mathrm{sky}}$} & \small{$A_V$} & \small{seeing} & \small{flags(R)} & \small{$\delta\lambda$} & \small{S/N($R_{50}$)} & \small{$I_{3\sigma}$} & \small{flags(QC)}\\\hline
\endhead
\hline
\endfoot
\small{001} & \small{$1.01\pm0.02$} & \small{$22.4$} & \small{$0.22$} & \small{...} & \small{0000} & \small{$3.05\pm0.44$} & \small{13$\pm$4} & \small{2.1$\pm$0.8} & \small{000}\\
\small{003} & \small{$1.07\pm0.06$} & \small{$22.0$} & \small{$0.17$} & \small{...} & \small{0010} & \small{$2.77\pm0.17$} & \small{9$\pm$3} & \small{2.0$\pm$0.7} & \small{000}\\
\small{007} & \small{$1.18\pm0.03$} & \small{$21.7$} & \small{$0.14$} & \small{...} & \small{0000} & \small{$2.52\pm0.13$} & \small{12$\pm$3} & \small{3.4$\pm$1.9} & \small{000}\\
\small{010} & \small{$1.27\pm0.09$} & \small{$22.2$} & \small{$0.17$} & \small{...} & \small{0000} & \small{$2.56\pm0.15$} & \small{8$\pm$3} & \small{3.2$\pm$0.7} & \small{000}\\
\small{014} & \small{$1.25\pm0.08$} & \small{$22.2$} & \small{$0.19$} & \small{...} & \small{0010} & \small{$2.88\pm0.28$} & \small{11$\pm$3} & \small{1.5$\pm$0.6} & \small{000}\\
\small{039} & \small{$1.08\pm0.05$} & \small{$22.1$} & \small{$0.21$} & \small{$1.1\pm0.2$} & \small{0000} & \small{$3.10\pm0.14$} & \small{12$\pm$3} & \small{1.8$\pm$0.4} & \small{000}\\
\small{042} & \small{$1.21\pm0.10$} & \small{$22.3$} & \small{$0.15$} & \small{$0.9\pm0.1$} & \small{0100} & \small{$3.91\pm1.11$} & \small{11$\pm$3} & \small{1.8$\pm$0.5} & \small{000}\\
\small{043} & \small{$1.01\pm0.01$} & \small{$22.3$} & \small{$0.27$} & \small{...} & \small{0000} & \small{$2.89\pm0.34$} & \small{14$\pm$4} & \small{1.8$\pm$0.6} & \small{000}\\
\small{053} & \small{$1.36\pm0.20$} & \small{$21.9$} & \small{$0.16$} & \small{...} & \small{0100} & \small{$2.62\pm0.32$} & \small{11$\pm$2} & \small{3.0$\pm$0.8} & \small{000}\\
\small{073} & \small{$1.07\pm0.04$} & \small{$22.2$} & \small{$0.26$} & \small{...} & \small{0110} & \small{$2.98\pm0.24$} & \small{8$\pm$2} & \small{3.4$\pm$1.1} & \small{000}\\
\small{088} & \small{$1.34\pm0.15$} & \small{$21.4$} & \small{$0.21$} & \small{...} & \small{1000} & \small{$3.51\pm0.44$} & \small{10$\pm$3} & \small{4.4$\pm$2.1} & \small{000}\\
\small{100} & \small{$1.40\pm0.16$} & \small{$21.8$} & \small{$0.15$} & \small{$1.1\pm0.1$} & \small{0100} & \small{$3.09\pm0.69$} & \small{20$\pm$5} & \small{2.8$\pm$0.6} & \small{000}\\
\small{127} & \small{$1.39\pm0.34$} & \small{$21.7$} & \small{$0.15$} & \small{...} & \small{1000} & \small{$3.04\pm0.14$} & \small{7$\pm$2} & \small{3.9$\pm$1.5} & \small{000}\\
\small{146} & \small{$1.48\pm0.01$} & \small{$22.2$} & \small{$0.14$} & \small{$1.3\pm0.2$} & \small{0100} & \small{$3.04\pm0.13$} & \small{13$\pm$3} & \small{2.1$\pm$0.6} & \small{000}\\
\small{151} & \small{$1.57\pm0.22$} & \small{$21.4$} & \small{$0.17$} & \small{...} & \small{0000} & \small{$2.68\pm0.25$} & \small{13$\pm$4} & \small{2.8$\pm$0.8} & \small{000}\\
\small{155} & \small{$1.08\pm0.05$} & \small{$22.1$} & \small{$0.13$} & \small{$0.9\pm0.1$} & \small{0000} & \small{$2.97\pm0.23$} & \small{6$\pm$1} & \small{4.1$\pm$1.0} & \small{000}\\
\small{156} & \small{$1.07\pm0.05$} & \small{$22.3$} & \small{$0.14$} & \small{...} & \small{0000} & \small{$3.68\pm0.75$} & \small{9$\pm$6} & \small{3.5$\pm$0.6} & \small{000}\\
\small{273} & \small{$1.06\pm0.02$} & \small{$22.2$} & \small{$0.14$} & \small{$1.1\pm0.1$} & \small{0100} & \small{$2.96\pm0.24$} & \small{15$\pm$3} & \small{1.6$\pm$0.5} & \small{000}\\
\small{274} & \small{$1.33\pm0.11$} & \small{$22.0$} & \small{$0.17$} & \small{...} & \small{0000} & \small{$2.66\pm0.16$} & \small{12$\pm$4} & \small{3.1$\pm$0.8} & \small{000}\\
\small{277} & \small{$1.08\pm0.04$} & \small{$21.7$} & \small{$0.15$} & \small{...} & \small{0100} & \small{$3.05\pm0.37$} & \small{15$\pm$2} & \small{3.4$\pm$1.0} & \small{000}\\
\small{306} & \small{$1.12\pm0.01$} & \small{$22.1$} & \small{$0.17$} & \small{$1.6\pm0.4$} & \small{0010} & \small{$2.93\pm0.17$} & \small{7$\pm$2} & \small{1.8$\pm$0.5} & \small{000}\\
\small{307} & \small{$1.18\pm0.08$} & \small{$22.2$} & \small{$0.15$} & \small{$0.8\pm0.1$} & \small{0000} & \small{$2.44\pm0.18$} & \small{6$\pm$2} & \small{3.8$\pm$1.3} & \small{000}\\
\small{309} & \small{$1.23\pm0.08$} & \small{$22.1$} & \small{$0.14$} & \small{...} & \small{0000} & \small{$3.25\pm0.20$} & \small{8$\pm$2} & \small{3.1$\pm$0.5} & \small{000}\\
\small{319} & \small{$1.07\pm0.05$} & \small{$22.2$} & \small{$0.15$} & \small{$1.1\pm0.1$} & \small{0000} & \small{$2.41\pm0.09$} & \small{14$\pm$4} & \small{1.7$\pm$0.6} & \small{000}\\
\small{326} & \small{$1.29\pm0.06$} & \small{$21.9$} & \small{$0.17$} & \small{...} & \small{0100} & \small{$2.69\pm0.22$} & \small{7$\pm$3} & \small{3.9$\pm$1.5} & \small{000}\\
\small{341} & \small{$1.23\pm0.10$} & \small{$22.0$} & \small{$0.16$} & \small{$1.2\pm0.1$} & \small{0000} & \small{$2.76\pm0.19$} & \small{13$\pm$4} & \small{1.8$\pm$0.6} & \small{000}\\
\small{364} & \small{$1.02\pm0.02$} & \small{$22.1$} & \small{$0.17$} & \small{$0.9\pm0.1$} & \small{0000} & \small{$2.84\pm0.18$} & \small{18$\pm$4} & \small{1.8$\pm$0.6} & \small{000}\\
\small{475} & \small{$1.11\pm0.07$} & \small{$22.1$} & \small{$0.17$} & \small{...} & \small{0000} & \small{$2.95\pm0.30$} & \small{20$\pm$14} & \small{2.0$\pm$0.6} & \small{000}\\
\small{479} & \small{$1.16\pm0.04$} & \small{$22.3$} & \small{$0.17$} & \small{...} & \small{0110} & \small{$2.58\pm0.10$} & \small{8$\pm$3} & \small{3.9$\pm$1.3} & \small{000}\\
\small{486} & \small{$1.19\pm0.09$} & \small{$22.1$} & \small{$0.15$} & \small{$0.8\pm0.1$} & \small{0000} & \small{$3.13\pm0.22$} & \small{10$\pm$5} & \small{2.2$\pm$0.6} & \small{000}\\
\small{515} & \small{$1.18\pm0.09$} & \small{$22.3$} & \small{$0.15$} & \small{$0.9\pm0.1$} & \small{0000} & \small{$2.82\pm0.29$} & \small{6$\pm$1} & \small{3.7$\pm$0.9} & \small{000}\\
\small{518} & \small{$1.18\pm0.03$} & \small{$22.6$} & \small{$0.13$} & \small{...} & \small{0000} & \small{$3.22\pm0.11$} & \small{15$\pm$4} & \small{2.1$\pm$0.6} & \small{000}\\
\small{528} & \small{$1.48\pm0.08$} & \small{$22.0$} & \small{$0.17$} & \small{...} & \small{0010} & \small{$2.37\pm0.02$} & \small{3$\pm$1} & \small{3.4$\pm$1.1} & \small{000}\\
\small{548} & \small{$1.19\pm0.05$} & \small{$21.9$} & \small{$0.18$} & \small{$0.8\pm0.1$} & \small{0000} & \small{$3.08\pm0.31$} & \small{19$\pm$7} & \small{2.9$\pm$0.7} & \small{020}\\
\small{577} & \small{$1.06\pm0.04$} & \small{$22.2$} & \small{$0.22$} & \small{...} & \small{0000} & \small{$3.05\pm0.13$} & \small{8$\pm$7} & \small{1.5$\pm$0.4} & \small{020}\\
\small{607} & \small{$1.24\pm0.07$} & \small{$22.3$} & \small{$0.14$} & \small{...} & \small{0000} & \small{$2.89\pm0.34$} & \small{23$\pm$4} & \small{2.5$\pm$0.6} & \small{000}\\
\small{608} & \small{$1.22\pm0.11$} & \small{$21.9$} & \small{$0.16$} & \small{...} & \small{0000} & \small{$2.60\pm0.27$} & \small{9$\pm$3} & \small{2.1$\pm$0.8} & \small{000}\\
\small{609} & \small{$1.07\pm0.05$} & \small{$22.4$} & \small{$0.15$} & \small{$0.8\pm0.1$} & \small{0000} & \small{$2.75\pm0.20$} & \small{8$\pm$2} & \small{2.8$\pm$0.5} & \small{000}\\
\small{610} & \small{$1.19\pm0.08$} & \small{$22.5$} & \small{$0.17$} & \small{...} & \small{0000} & \small{$2.38\pm0.05$} & \small{10$\pm$4} & \small{2.6$\pm$0.8} & \small{000}\\
\small{657} & \small{$1.07\pm0.05$} & \small{$22.4$} & \small{$0.19$} & \small{$1.5\pm0.2$} & \small{0000} & \small{$2.84\pm0.11$} & \small{8$\pm$2} & \small{1.6$\pm$0.4} & \small{000}\\
\small{663} & \small{$1.10\pm0.01$} & \small{$22.0$} & \small{$0.17$} & \small{...} & \small{0000} & \small{$3.56\pm0.27$} & \small{12$\pm$3} & \small{2.2$\pm$0.6} & \small{000}\\
\small{664} & \small{$1.08\pm0.05$} & \small{$22.0$} & \small{$0.26$} & \small{$1.2\pm0.1$} & \small{0000} & \small{$2.82\pm0.21$} & \small{19$\pm$3} & \small{2.0$\pm$0.6} & \small{000}\\
\small{665} & \small{$1.18\pm0.11$} & \small{$22.1$} & \small{$0.10$} & \small{...} & \small{0000} & \small{$2.69\pm0.13$} & \small{6$\pm$3} & \small{4.0$\pm$1.1} & \small{000}\\
\small{676} & \small{$1.01\pm0.01$} & \small{$22.5$} & \small{$0.17$} & \small{$0.8\pm0.1$} & \small{0000} & \small{$3.85\pm1.30$} & \small{6$\pm$2} & \small{3.0$\pm$0.5} & \small{000}\\
\small{680} & \small{$1.08\pm0.05$} & \small{$22.4$} & \small{$0.10$} & \small{...} & \small{0010} & \small{$2.63\pm0.06$} & \small{9$\pm$5} & \small{1.4$\pm$0.4} & \small{000}\\
\small{684} & \small{$1.07\pm0.05$} & \small{$22.3$} & \small{$0.13$} & \small{$0.9\pm0.1$} & \small{0000} & \small{$2.98\pm0.16$} & \small{17$\pm$3} & \small{2.0$\pm$0.5} & \small{000}\\
\small{758} & \small{$1.20\pm0.05$} & \small{$22.2$} & \small{$0.26$} & \small{$1.0\pm0.1$} & \small{0000} & \small{$3.06\pm0.11$} & \small{10$\pm$3} & \small{2.2$\pm$0.7} & \small{000}\\
\small{764} & \small{$0.74\pm0.64$} & \small{...} & \small{$0.13$} & \small{...} & \small{0000} & \small{$2.94\pm0.20$} & \small{9$\pm$3} & \small{1.7$\pm$0.5} & \small{000}\\
\small{769} & \small{$1.03\pm0.02$} & \small{$22.2$} & \small{$0.17$} & \small{...} & \small{0000} & \small{$3.05\pm0.21$} & \small{11$\pm$4} & \small{2.1$\pm$0.6} & \small{000}\\
\small{783} & \small{$1.18\pm0.08$} & \small{$22.3$} & \small{$0.17$} & \small{...} & \small{0010} & \small{$2.80\pm0.23$} & \small{15$\pm$3} & \small{2.8$\pm$0.5} & \small{000}\\
\small{797} & \small{$1.07\pm0.05$} & \small{$22.5$} & \small{$0.17$} & \small{$1.3\pm0.2$} & \small{0010} & \small{$2.89\pm0.28$} & \small{11$\pm$3} & \small{1.6$\pm$0.5} & \small{000}\\
\small{798} & \small{$1.13\pm0.07$} & \small{$22.2$} & \small{$0.30$} & \small{$0.8\pm0.1$} & \small{0000} & \small{$2.87\pm0.11$} & \small{10$\pm$3} & \small{2.0$\pm$0.8} & \small{000}\\
\small{806} & \small{$1.05\pm0.05$} & \small{$22.2$} & \small{$0.21$} & \small{$1.1\pm0.1$} & \small{0000} & \small{$2.99\pm0.19$} & \small{8$\pm$2} & \small{1.8$\pm$0.5} & \small{000}\\
\small{820} & \small{$1.13\pm0.09$} & \small{$22.3$} & \small{$0.15$} & \small{$0.7\pm0.1$} & \small{0011} & \small{$2.99\pm0.23$} & \small{6$\pm$1} & \small{3.6$\pm$0.7} & \small{000}\\
\small{822} & \small{$1.10\pm0.06$} & \small{$22.1$} & \small{$0.16$} & \small{$0.8\pm0.1$} & \small{0000} & \small{$3.02\pm0.08$} & \small{13$\pm$4} & \small{1.6$\pm$0.5} & \small{000}\\
\small{823} & \small{$1.18\pm0.04$} & \small{$22.0$} & \small{$0.26$} & \small{$0.9\pm0.1$} & \small{0000} & \small{$2.84\pm0.12$} & \small{12$\pm$2} & \small{2.1$\pm$0.7} & \small{000}\\
\small{824} & \small{$1.19\pm0.05$} & \small{$21.6$} & \small{$0.24$} & \small{$1.1\pm0.2$} & \small{0000} & \small{$2.77\pm0.08$} & \small{11$\pm$2} & \small{4.0$\pm$1.1} & \small{000}\\
\small{826} & \small{$1.17\pm0.05$} & \small{$22.0$} & \small{$0.17$} & \small{...} & \small{0000} & \small{$2.58\pm0.32$} & \small{14$\pm$4} & \small{3.0$\pm$0.9} & \small{000}\\
\small{828} & \small{$1.18\pm0.05$} & \small{$22.6$} & \small{$0.16$} & \small{$1.0\pm0.1$} & \small{0010} & \small{$2.78\pm0.07$} & \small{15$\pm$4} & \small{1.3$\pm$0.4} & \small{000}\\
\small{829} & \small{$1.25\pm0.08$} & \small{$22.0$} & \small{$0.17$} & \small{$1.4\pm0.2$} & \small{0010} & \small{$4.00\pm0.69$} & \small{13$\pm$3} & \small{2.2$\pm$0.3} & \small{000}\\
\small{832} & \small{$1.03\pm0.02$} & \small{$22.1$} & \small{$0.17$} & \small{...} & \small{0000} & \small{$2.92\pm0.27$} & \small{18$\pm$5} & \small{2.1$\pm$0.6} & \small{000}\\
\small{833} & \small{$1.07\pm0.03$} & \small{$22.2$} & \small{$0.10$} & \small{...} & \small{0000} & \small{$2.99\pm0.69$} & \small{5$\pm$2} & \small{3.4$\pm$1.2} & \small{000}\\
\small{835} & \small{$1.04\pm0.04$} & \small{$22.3$} & \small{$0.16$} & \small{$1.0\pm0.1$} & \small{0000} & \small{$2.66\pm0.11$} & \small{20$\pm$5} & \small{1.4$\pm$0.5} & \small{000}\\
\small{837} & \small{$1.16\pm0.05$} & \small{$22.2$} & \small{$0.14$} & \small{$0.8\pm0.1$} & \small{0010} & \small{$2.91\pm0.13$} & \small{25$\pm$6} & \small{1.6$\pm$0.5} & \small{000}\\
\small{840} & \small{$1.26\pm0.11$} & \small{$22.1$} & \small{$0.10$} & \small{...} & \small{0000} & \small{$3.71\pm1.47$} & \small{12$\pm$3} & \small{2.1$\pm$0.9} & \small{000}\\
\small{843} & \small{$1.12\pm0.06$} & \small{$22.4$} & \small{$0.26$} & \small{...} & \small{0000} & \small{$2.74\pm0.15$} & \small{10$\pm$4} & \small{1.9$\pm$0.4} & \small{000}\\
\small{845} & \small{$1.19\pm0.17$} & \small{$21.7$} & \small{$0.28$} & \small{$0.8\pm0.1$} & \small{0000} & \small{$2.92\pm0.10$} & \small{11$\pm$4} & \small{2.3$\pm$1.0} & \small{000}\\
\small{846} & \small{$1.18\pm0.08$} & \small{$22.0$} & \small{$0.25$} & \small{...} & \small{0000} & \small{$2.92\pm0.14$} & \small{9$\pm$2} & \small{2.0$\pm$0.9} & \small{000}\\
\small{847} & \small{$1.03\pm0.03$} & \small{$21.9$} & \small{$0.32$} & \small{$0.8\pm0.1$} & \small{1000} & \small{$2.95\pm0.27$} & \small{11$\pm$4} & \small{2.3$\pm$0.8} & \small{000}\\
\small{848} & \small{$0.83\pm0.73$} & \small{$14.7$} & \small{$0.23$} & \small{$1.0\pm0.1$} & \small{1000} & \small{$2.66\pm0.11$} & \small{16$\pm$3} & \small{2.3$\pm$0.6} & \small{000}\\
\small{850} & \small{$1.05\pm0.02$} & \small{$22.0$} & \small{$0.19$} & \small{$0.8\pm0.1$} & \small{0000} & \small{$3.00\pm0.10$} & \small{13$\pm$3} & \small{2.2$\pm$0.5} & \small{000}\\
\small{851} & \small{$1.13\pm0.04$} & \small{$22.1$} & \small{$0.28$} & \small{$0.9\pm0.1$} & \small{0000} & \small{$3.00\pm0.74$} & \small{8$\pm$2} & \small{2.0$\pm$0.5} & \small{000}\\
\small{852} & \small{$1.40\pm0.10$} & \small{$21.6$} & \small{$0.19$} & \small{$1.4\pm0.2$} & \small{0010} & \small{$3.34\pm0.39$} & \small{6$\pm$2} & \small{2.1$\pm$0.6} & \small{000}\\
\small{854} & \small{$1.22\pm0.06$} & \small{$22.0$} & \small{$0.37$} & \small{$0.9\pm0.1$} & \small{0001} & \small{$2.69\pm0.16$} & \small{10$\pm$4} & \small{2.2$\pm$0.8} & \small{000}\\
\small{856} & \small{$1.06\pm0.04$} & \small{$22.7$} & \small{$0.24$} & \small{$0.9\pm0.1$} & \small{1100} & \small{$2.69\pm0.22$} & \small{7$\pm$2} & \small{3.2$\pm$0.8} & \small{000}\\
\small{857} & \small{$1.10\pm0.02$} & \small{$22.4$} & \small{$0.22$} & \small{$0.9\pm0.1$} & \small{0000} & \small{$2.83\pm0.13$} & \small{16$\pm$4} & \small{1.9$\pm$1.1} & \small{000}\\
\small{858} & \small{$1.28\pm0.17$} & \small{$22.1$} & \small{$0.23$} & \small{$0.9\pm0.1$} & \small{0000} & \small{$2.97\pm0.27$} & \small{13$\pm$5} & \small{2.4$\pm$0.8} & \small{000}\\
\small{859} & \small{$1.18\pm0.05$} & \small{$22.2$} & \small{$0.32$} & \small{$0.7\pm0.1$} & \small{0000} & \small{$2.65\pm0.10$} & \small{14$\pm$3} & \small{2.6$\pm$0.6} & \small{000}\\
\small{860} & \small{$1.05\pm0.03$} & \small{$22.3$} & \small{$0.27$} & \small{$0.9\pm0.3$} & \small{0000} & \small{$2.78\pm0.31$} & \small{23$\pm$3} & \small{2.4$\pm$0.6} & \small{000}\\
\small{863} & \small{$1.20\pm0.06$} & \small{$22.3$} & \small{$0.19$} & \small{$0.9\pm0.1$} & \small{0000} & \small{$2.68\pm0.07$} & \small{13$\pm$4} & \small{2.2$\pm$0.8} & \small{000}\\
\small{864} & \small{$1.05\pm0.02$} & \small{$22.4$} & \small{$0.20$} & \small{...} & \small{0000} & \small{$3.56\pm0.32$} & \small{14$\pm$3} & \small{1.7$\pm$0.5} & \small{000}\\
\small{865} & \small{$1.02\pm0.02$} & \small{$22.3$} & \small{$0.26$} & \small{$0.9\pm0.1$} & \small{0000} & \small{$2.94\pm0.14$} & \small{17$\pm$4} & \small{1.7$\pm$0.6} & \small{000}\\
\small{866} & \small{$1.08\pm0.05$} & \small{$22.1$} & \small{$0.34$} & \small{...} & \small{0000} & \small{$2.95\pm0.06$} & \small{6$\pm$2} & \small{1.7$\pm$0.4} & \small{000}\\
\small{867} & \small{$1.13\pm0.01$} & \small{$22.1$} & \small{$0.38$} & \small{...} & \small{0000} & \small{$3.61\pm2.32$} & \small{17$\pm$2} & \small{2.4$\pm$0.6} & \small{000}\\
\small{872} & \small{$1.37\pm0.06$} & \small{$21.7$} & \small{$0.22$} & \small{$1.0\pm0.1$} & \small{0000} & \small{$2.66\pm0.14$} & \small{7$\pm$2} & \small{2.8$\pm$0.8} & \small{000}\\
\small{874} & \small{$1.13\pm0.05$} & \small{$21.8$} & \small{$0.30$} & \small{$0.8\pm0.1$} & \small{0000} & \small{$3.29\pm0.47$} & \small{11$\pm$3} & \small{2.7$\pm$0.4} & \small{000}\\
\small{877} & \small{$1.10\pm0.06$} & \small{$21.7$} & \small{$0.22$} & \small{$1.0\pm0.1$} & \small{0000} & \small{$3.01\pm0.22$} & \small{8$\pm$3} & \small{3.5$\pm$1.0} & \small{000}\\
\small{878} & \small{$1.36\pm0.06$} & \small{$22.1$} & \small{$0.18$} & \small{$1.1\pm0.1$} & \small{0010} & \small{$2.86\pm0.11$} & \small{5$\pm$7} & \small{2.1$\pm$0.7} & \small{000}\\
\small{881} & \small{$1.17\pm0.12$} & \small{$22.0$} & \small{$0.30$} & \small{$0.9\pm0.2$} & \small{0000} & \small{$3.04\pm0.19$} & \small{16$\pm$4} & \small{2.0$\pm$0.7} & \small{000}\\
\small{883} & \small{$1.12\pm0.04$} & \small{$22.1$} & \small{$0.31$} & \small{$0.9\pm0.1$} & \small{0000} & \small{$2.71\pm0.19$} & \small{3$\pm$1} & \small{4.6$\pm$1.7} & \small{000}\\
\small{887} & \small{$1.05\pm0.01$} & \small{$22.2$} & \small{$0.29$} & \small{$0.8\pm0.1$} & \small{0100} & \small{$3.03\pm0.42$} & \small{13$\pm$4} & \small{2.8$\pm$1.0} & \small{000}\\
\small{888} & \small{$1.08\pm0.06$} & \small{$21.9$} & \small{$0.41$} & \small{...} & \small{0000} & \small{$2.87\pm0.14$} & \small{8$\pm$3} & \small{2.5$\pm$0.8} & \small{000}\\
\small{890} & \small{$1.02\pm0.01$} & \small{$21.8$} & \small{$0.23$} & \small{$0.9\pm0.1$} & \small{0000} & \small{$3.03\pm0.22$} & \small{13$\pm$3} & \small{2.5$\pm$0.6} & \small{000}\\
\small{893} & \small{$1.37\pm0.21$} & \small{$21.8$} & \small{$0.15$} & \small{$1.0\pm0.1$} & \small{0010} & \small{$2.87\pm0.14$} & \small{8$\pm$3} & \small{2.4$\pm$0.4} & \small{000}\\
\small{896} & \small{$1.04\pm0.03$} & \small{$22.3$} & \small{$0.14$} & \small{$1.4\pm0.3$} & \small{0110} & \small{$2.80\pm0.30$} & \small{11$\pm$3} & \small{2.5$\pm$0.6} & \small{000}\\
\small{901} & \small{$1.20\pm0.11$} & \small{$22.1$} & \small{$0.25$} & \small{$1.0\pm0.2$} & \small{0100} & \small{$3.09\pm0.15$} & \small{11$\pm$6} & \small{3.5$\pm$1.7} & \small{000}\\
\small{902} & \small{$1.34\pm0.14$} & \small{$21.7$} & \small{$0.21$} & \small{$1.1\pm0.1$} & \small{0000} & \small{$3.04\pm0.15$} & \small{18$\pm$3} & \small{1.9$\pm$0.5} & \small{000}\\
\small{904} & \small{$1.29\pm0.09$} & \small{$21.7$} & \small{$0.22$} & \small{$1.3\pm0.3$} & \small{0001} & \small{$3.03\pm0.25$} & \small{14$\pm$4} & \small{2.5$\pm$0.8} & \small{000}\\
\small{935} & \small{$1.10\pm0.07$} & \small{$22.3$} & \small{$0.18$} & \small{...} & \small{0110} & \small{$2.43\pm0.04$} & \small{6$\pm$2} & \small{2.5$\pm$0.7} & \small{000}\\
\small{939} & \small{$1.35\pm0.08$} & \small{$21.9$} & \small{$0.28$} & \small{$0.9\pm0.1$} & \small{0100} & \small{$3.23\pm0.28$} & \small{9$\pm$3} & \small{1.9$\pm$0.7} & \small{000}\\
\end{longtable}
\tablefoot{The meaning of the columns is the same as for Table~\ref{tab:QC_par_V500} except that the sky surface brightness is measured in the $B$ band and the QC flag (FLAG\_FLUXCAL) is not defined for the V1200 data.}
\twocolumn
\end{longtab}

\section{Data Quality}\label{sect:QC}
This first CALIFA data release provides science grade data of 100 galaxies to the community.
Hence, in order to define the set of galaxies to be released, a detailed
characterization of the data quality for the existing data is needed to select
and define a minimal useful suite of quality control (QC) parameters.

Overall, the final data quality for CALIFA depends on a number of independent
factors such as (i) general instrument reliability and temperature stability,
(ii) ambient conditions during observations, and (iii) the robustness of the
data reduction pipeline. We define a set of parameters and QC flags in the
areas of astrometric, spectroscopic and photometric characterization
that are measured for all galaxies. We amend these
  parameters with four quality flags for each of the two spectral setups
  (V500/V1200; eight in total) that summarize the reduction process for each
  galaxy and indicate data quality.

We strongly urge users to consider these QC parameters with regard to
  their science application. For example, if a science goal focuses primarily
  on kinematics, then a substantial offset in absolute astrometric
  registration is not a limitation for the analysis of any one galaxy.
  However, if, under the same circumstance, the science goal requires deriving
  joint spatially-extended information from matching with the SDSS, then
  caution is advised.

  The values and uncertainties listed for the individual QC parameters are
  intended to provide the user with a summary of the information content of
  the CALIFA data on a specific galaxy in order to allow for the initial
  assessment of whether or not a specific datacube can be used for a planned
  analysis. In the following we display and comment on the distribution of
  the key flags and QC parameters while demonstrating the overall sample
  properties and the nature of outliers.

The QC parameters and QC flags can be found in Tables~\ref{tab:QC_par_V500}
and \ref{tab:QC_par_V1200} for the V500 and V1200 data, respectively, which
are also available in electronic form on the DR1 web page and accessible through 
the Virtual Observatory (see Sect.~\ref{sect:VO_serve} for details).

\subsection{Initial reduction pipeline flags}\label{sect:flags}

As an initial description to evaluate the data and reduction quality, we
provide each of the four flags per setup that come directly from the reduction
process. Some of the extended parameters in the sections below will add to
further validate and expand on these initial flags.

Potential minor issues are noted with a value=1, while the standard case
receives value=0. Galaxies with major issues, i.e.\ flag values=2 have been
observed as part of the CALIFA survey, but were excluded from this data
release. The reduction flags are:

\begin{itemize} 

\item FLAG\_RED\_O: observing conditions quality. 

  Set to 1 if the night sky is brighter than $\mu_V = 20.5$
  mag\,$\mathrm{arcsec}^{-2}$ (V500) or $\mu_B = 20.0$
  mag\,$\mathrm{arcsec}^{-2}$ (V1200), respectively. There exists a correlation
  between limiting continuum sensitivity of the datacubes and the night-sky
  brightness. For nights brighter than this limit the datacubes may not be as
  deep as targeted.
 
  Flag is set to 2 if observing conditions were non-photometric or if
  background of unknown origin is seen in the data.

\item FLAG\_RED\_R: reduction/calibration quality.

  Set to 1 if reduction deviates from the standard procedure, e.g.\ due to a
  lack of directly associated continuum or arc-lamp exposures or a saturation
  in the available calibration data.

  Flag is set to 2 if manual intervention was required to recover a minimum of
  possible data from the raw data.

\item FLAG\_RED\_W: wavelength calibration quality.

  Set to 1 if the standard deviation of the pipeline-estimated systemic
  velocity in different wavelength regions is larger than $25\,\mathrm{km}\,\mathrm{s}^{-1}$.

  Flag set to 2 if the deviation is more than 2 standard deviations, or $34\,\mathrm{km}\,\mathrm{s}^{-1}$.

\item FLAG\_RED\_V: visual inspection quality.

  Summary of a visual inspection of the data. If there is an obvious defect
  affecting a small fraction of the FoV (like the presence of a strong line or
  a imperfectly traced fiber), flag is set to 1. It is set to 2
  when it is evident that the data have low quality for any unknown reason.

  Finally this flag is set to 1, even if all previous cases have
    value=0, when a master continuum or master arc frame, instead of an
    individual frame, has been used during the reduction of any of the frames used
  to create the final datacube.
\end{itemize}

\subsection{Astrometric accuracy and spatial resolution}\label{sect:spatial_res}
\subsubsection{Astrometric registration accuracy}
The first area of testing and evaluation of the pipeline output is the
absolute astrometric registration of the datacube coordinate systems to the
International Coordinate Reference System (ICRS). Astrometric registration is
of central importance when CALIFA data are to be combined with e.g.\ imaging
data from other surveys to extract spatially resolved information at certain
locations in a galaxy. For this purpose the CALIFA pipeline uses a simple
scheme whereby the tabulated coordinates of the galaxy
  $V$ band photometric center, as given in Table~\ref{tab:DR1_sample}, are
assigned to the measured barycenter of the reconstructed image in the CALIFA
datacubes.

We tested independently the robustness and accuracy of this approach
  by both visual inspection of the assigned galaxy centers in both setups, as
  well as in V500 matching the galaxy location of a reconstructed $g$ band
  from CALIFA to, where available, corresponding SDSS images. The deviations
  of the CALIFA astrometry from the ICRS are typically small.

 In the V500 setup the offset for most galaxies lies below 1\farcs4. However,
  three galaxies show offsets between 1\farcs4 and 3\arcsec, and further two
  substantially larger offsets of $\sim$12\arcsec and $\sim$22\arcsec. In
  these cases, either the centers of the galaxies are not well defined due to
  dust lanes in strongly inclined system, or the galaxies have a bright field
  star near their center:

\begin{itemize}
\item UGC~10650: Edge-on spiral with fuzzy center. Registration should be
  better than 3\arcsec.

\item NGC~4676A: One galaxy of The Mice with unclear center and a possible
  offset of $\sim$1\farcs5 in the astrometric registration.

\item NGC~6032: Highly inclined galaxy with prominent dust lane and difficult
  to define center. $\sim$2\arcsec\ offset in the coordinates.

\item NGC~0477: Astrometric center set to nearby star; resulting offset is
  $\sim$12\arcsec.

\item NGC~3991: Center of this inclined spiral is not the brightest
  component. The correct center and hence coordinate system is offset by
  $\sim$22\arcsec.
\end{itemize}

In the V1200 setup we tested the astrometric co-registration with
  respect to the V500 reference frame. Generally the registration is accurate
  to better than 2\arcsec. Only for two galaxies we find a more substantial
  offset:

\begin{itemize}
\item NGC~4470: Galaxy with difficult to define center. Offset of
  $\sim$9\arcsec\ between the two setups.

\item NGC~4676A: As stated for V500 the center is not well defined for this
  galaxy and an offset of $\sim$6\arcsec\ exists between the two setups.
\end{itemize}

For this DR1 we provide the pipeline registration as described above and do
not correct offsets found through these separate external checks. In future
data releases it is planned to implement more sophisticated registration
methods directly in the pipeline thereby reducing the number of
outliers and improving the overall astrometric registration
accuracy. We describe the astrometric offsets for DR1 with the flag
  FLAG\_ASTR, given for the V500 and V1200 in Tables~\ref{tab:QC_par_V500}
  and \ref{tab:QC_par_V1200}, respectively. For V500 the values 0, 1, and 2
  describe offsets $<$1\farcs4, 1\farcs4 to 3\farcs0, and $>$3\farcs0. In the
  V1200 table values for FLAG\_ASTR of 0 and 2 refer to relative offsets between
  V500 and V1200 smaller or larger than 3\arcsec.

\subsubsection{Seeing and spatial resolution}\label{sect:seeing}
\begin{figure}
\resizebox{\hsize}{!}{\includegraphics{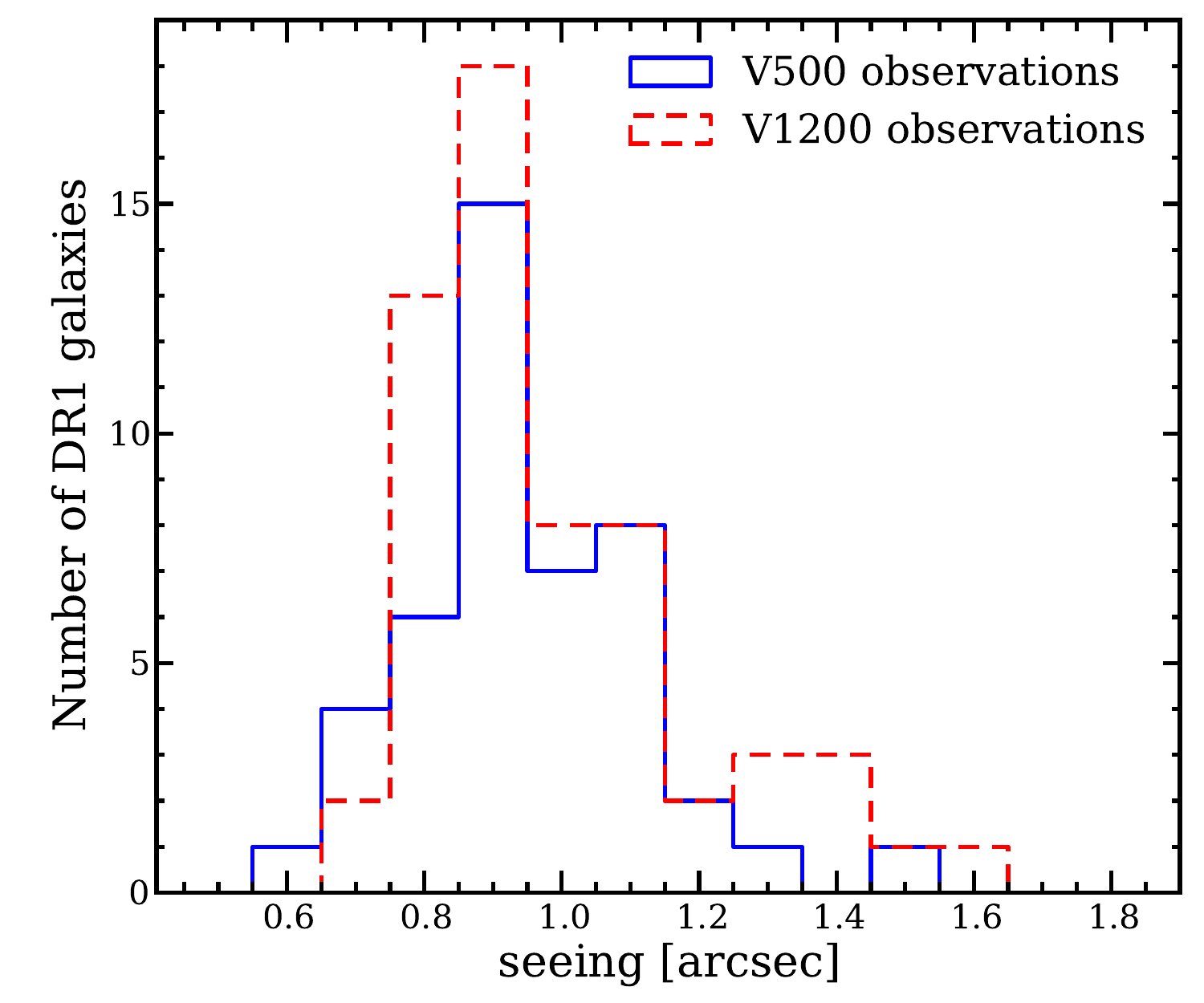}}\\
\caption{Distribution of the seeing during the CALIFA observations as measured
  by the automatic Differential Image Motion Monitor
  \citep[DIMM,][]{Aceituno:2004}.}
  \label{fig:seeing}
\end{figure}

The achievable spatial resolution of imaging data is usually determined by the
telescope aperture together with the atmospheric and instrumental seeing
during ground-based observations. In the case of CALIFA, the coarse sampling
of the large PPak fibers modified by the adopted 3-fold dithering pattern as
described in S12 imposes additional limitations on the final spatial
resolution.

\begin{figure*}
\sidecaption
\includegraphics[width=12cm]{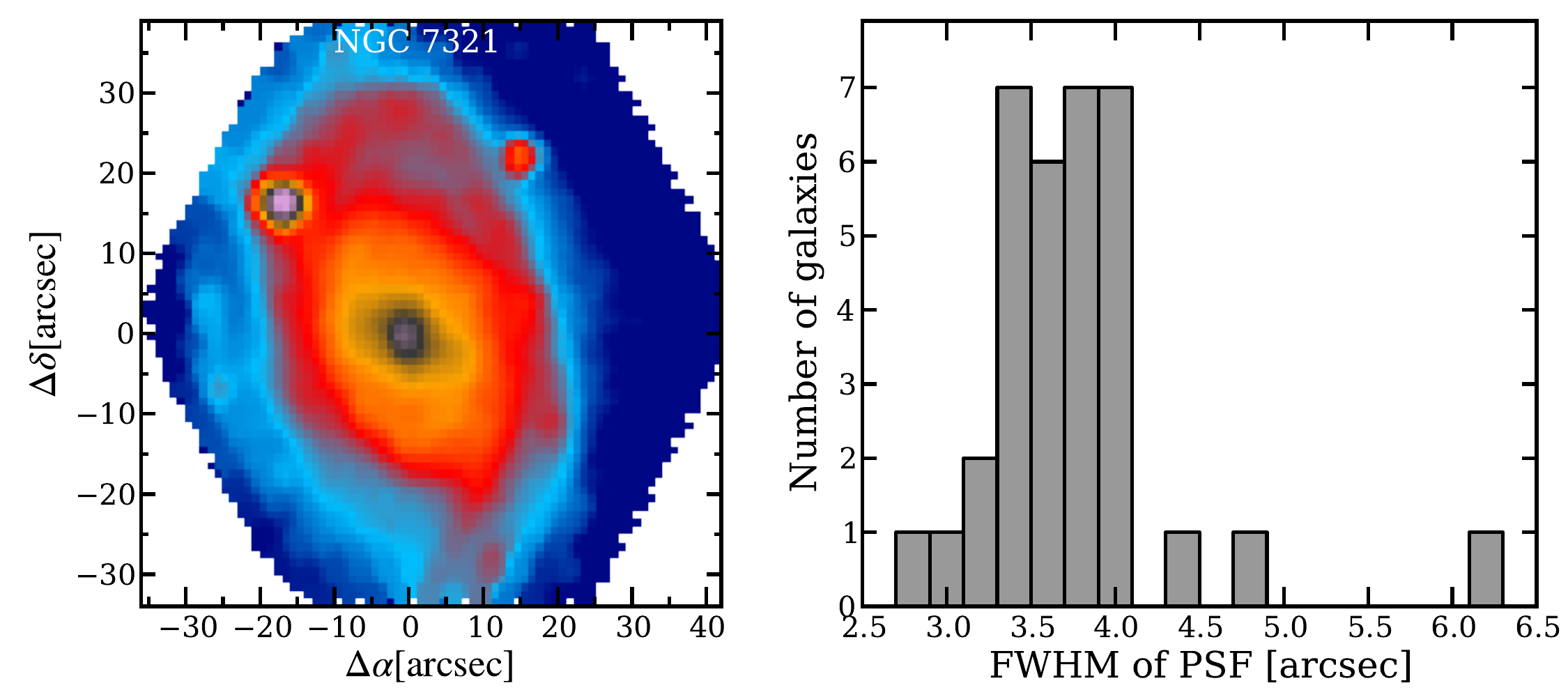}
\caption{\emph{Left panel:} $V$ band image of NGC~7321, an example
  of a CALIFA galaxy with a star in the FoV to measure the PSF of the
  data. \emph{Right panel:} Distribution of the FWHM of the CALIFA PSF as
  measured from 34 cubes with a sufficiently bright star in the field. The
  median of the distribution is 3\farcs7; the spread of values is
    due to the flux of the underlying galaxy structure and undersampling of
    the stellar image and hence high-value outliers are upper limits in these
  cases.}
 \label{fig:psf_hist}
\end{figure*}

We measure the seeing in the CALIFA data from two independent
  sources. The first one, the differential image motion monitor
  \citep[DIMM,][]{Aceituno:2004}, operates fully automatically at the Calar
  Alto observatory during the night\footnote{Since the DIMM has tighter
    operational constraints (humidity lower than 80\% and wind speed less than
    $12\,\mathrm{m}\,\mathrm{s}^{-1}$) than the 3.5m telescope, seeing
    information is not available for every CALIFA observation. Hence DIMM
    seeing values can be missing from Tables~\ref{tab:QC_par_V500} and
    \ref{tab:QC_par_V1200}, but the overall seeing distribution is not
    expected to be very different.}. The DIMM seeing for the DR1 sample is
  shown in Fig.~\ref{fig:seeing} and has a median value of $0\farcs9$ (FWHM). We
  confirm this value with measurements of the width of the guide star on
  images taken by the Acquisition and Guiding (AG) Camera of the PMAS
  instrument \citep{Roth:2005}, which, as expected, has a slightly larger
  median value of 1\farcs15 (FWHM). This confirms that atmospheric and
  instrumental seeing are not the limiting factors in the spatial resolution
  of the CALIFA cubes.

\medskip

Instead, the large aperture of each fiber ($2\farcs7$
diameter) in the PPak fiber bundle does not permit to take full
advantage of the good seeing conditions. The final spatial resolution
of the CALIFA data is set by fiber size and the dither scheme together with
the adopted image reconstruction algorithm. The width of the point spread
function (PSF) is directly measured from the V500 CALIFA datacubes in 35 cases
where a bright field star is in the FoV (see left panel of
Fig.~\ref{fig:psf_hist} for an example). The measured FWHM distribution of the
PSF (Fig.~\ref{fig:psf_hist} right panel) has a well-defined median value of
$3\farcs7$ (FWHM). To verify these measurements we simulated the expected CALIFA PSF
for 1\arcsec\ Gaussian seeing while adopting the fiber sampling, dither
pattern and image reconstruction algorithm used for the DR1. We obtained a
FWHM for the PSF of $\geq3\arcsec$ as a lower limit, which is in
agreement with our empirical measurements.

This spatial resolution is for a large part set by the adopted kernel for the
inverse-distance weighting scheme for the image reconstruction (see S12 for
details). It was chosen to produce smooth images without obvious structure
caused by the dither pattern. Other image combination approaches such as
drizzling \citep{Fruchter:2002} can be used to reach closer to the intrinsic
spatial resolution of CALIFA data, which is significantly better than
$3\arcsec$ due to the dithering scheme used. The choice of image combination
procedure depends on the actual goal and will be an area of improvement in
future CALIFA data releases.

\subsection{Wavelength calibration accuracy}
During the data reduction the spectral resolution was homogenized to reach a
target FWHM of 6\AA\ (V500) and 2.3\AA\ (V1200), respectively, over the
  whole wavelength range, as explained in S12. To go beyond the simple
flag-cut from the pipeline diagnostics we tested the spectral resolution and
the wavelength calibration, and its spectral and spatial consistency.

\begin{figure}
\resizebox{\hsize}{!}{\includegraphics{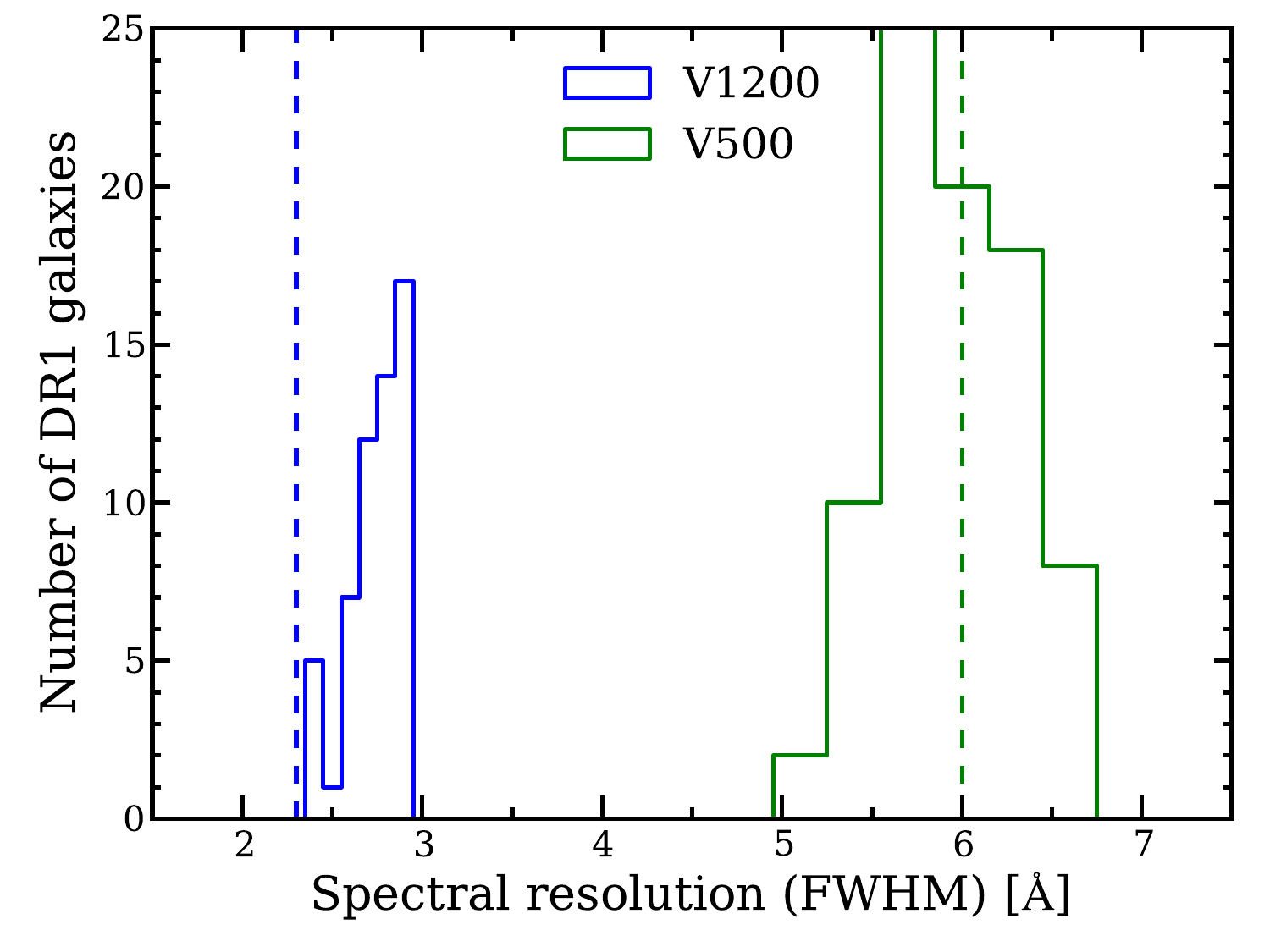}}
\caption{Distribution of the final spectral resolution (FWHM) in the combined
  cubes as measured from the sky lines in the error cube. The vertical dashed
  lines indicate the spectral resolution to which the spectra were aimed to be
  homogenized during the data reduction. Measurements for V1200 (blue, left)
  are upper limits since the night sky lines used are intrinsically
  resolved at this level. The spread is due to uncertainties in
    calculating the spectral resolution from only a few available lines.}
  \label{fig:spectral_res}
\end{figure}

The distribution of spectral resolutions for all galaxies is shown in
Fig.~\ref{fig:spectral_res} for both instrument setups. It was estimated by
measuring the width of night sky lines in the error cube, which contains a
reliable trace of the sky lines in the input data. For the V500 setup we find
that it is centered on the target spectral resolution (6\AA). This is not the
case for the V1200 setup. For this setup only Hg pollution lines
(\ion{Hg}{i}\,$\lambda$4046\AA\ and $\lambda$4358\AA) are
available in the corresponding wavelength range.  Since these lines are
resolved at the resolution of this setup the derived empirical resolution is a
conservative upper limit.

We tested the overall wavelength calibration in two ways. First, we measured the centroids of the same night sky lines and determined their scatter across the FoV. In all
  cases, the centroids are fully consistent with zero offset from the nominal
  wavelength, while the scatter is consistent with pure measurement errors and
  the absence of any detectable systematic spatial variation.

Secondly, we tested whether there exists any systematic offsets of the
wavelength calibration at different wavelengths of the spectra. For this
purpose we modeled the recession velocity in the spectra of every galaxy's
coadded central 10\arcsec\ apertures in different spectral regions. We did
this using the penalized Pixel-Fitting method \citep[pPXF,][]{Cappellari:2004}
to piecewise fit stellar models, hence measuring central wavelengths in
several (3--5) independent spectral bins showing strong
absorption features. The resulting velocity differences between these bins
were again always consistent with pure measurement noise and the absence of
systematics.

In summary, no spatial or spectral variation of the wavelength calibration is
found beyond the uncertainty imposed by the spectral resolutions of 6\AA\
and 2.3\AA\ (FWHM) for the different setups -- hence the random shot noise of the 
data fully dominates the accuracy of wavelength calibration.

\subsection{Spectrophotometric quality and accuracy}
\subsubsection{Sky subtraction quality}
\begin{figure}
\resizebox{\hsize}{!}{\includegraphics{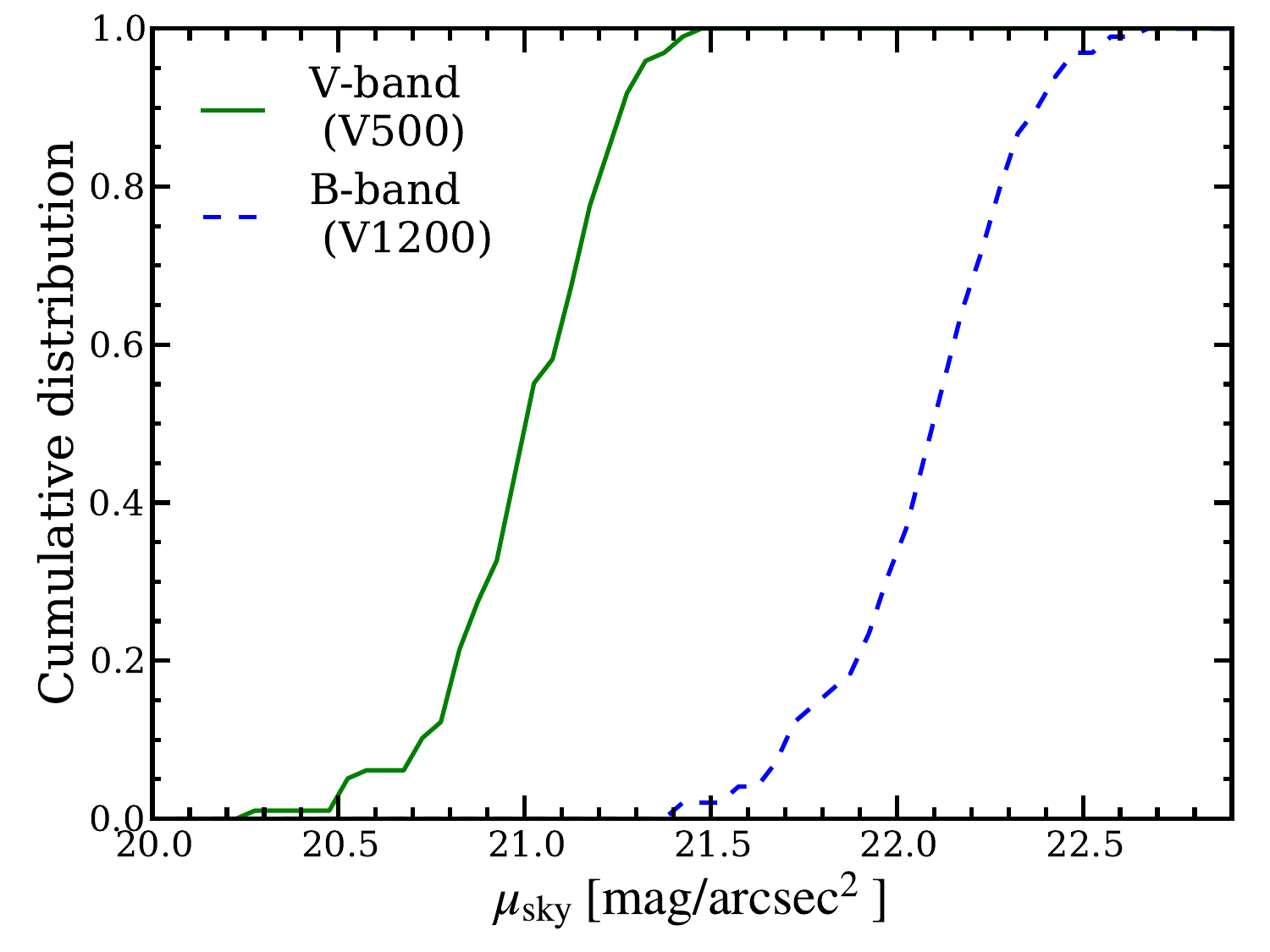}}
\caption{Cumulative distribution of the $V$ band (V500, left curve) and $B$ band (V1200, right curve)
  night-sky surface brightness during the observations of the V500 and
  V1200 CALIFA data, respectively.}
  \label{fig:sky_brightness}
\end{figure}
\begin{figure*}
 \includegraphics[width=\textwidth]{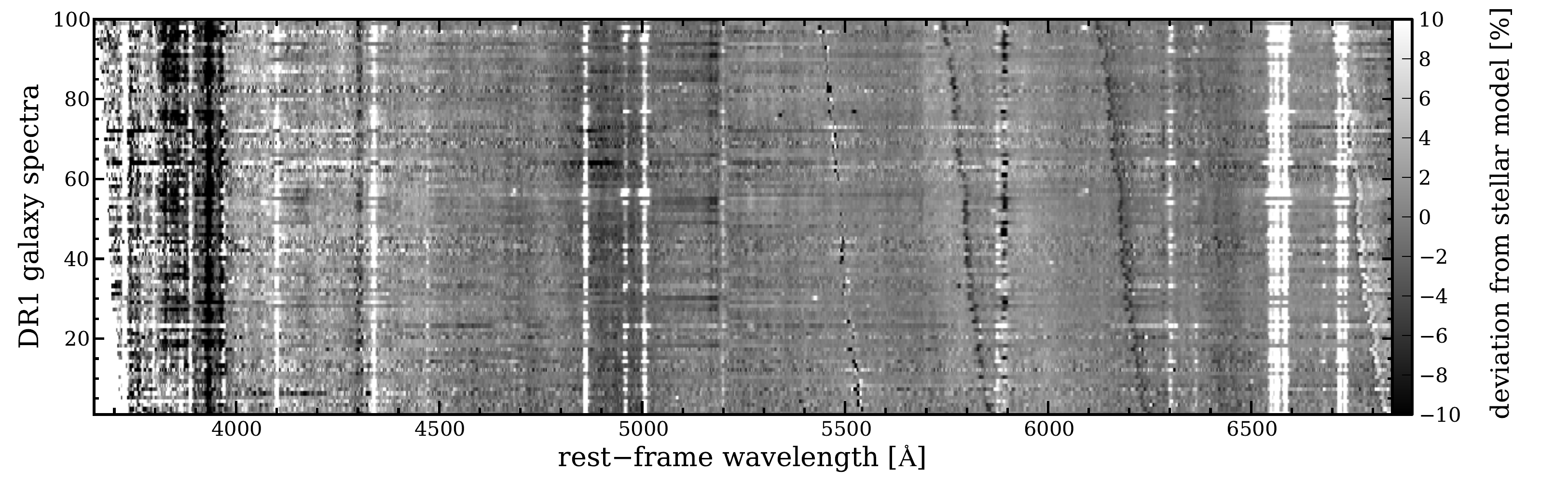}
  \caption{Stack of 100 DR1 central-region spectra in the rest-frame, sorted by
   redshift. Shown are the residuals after subtraction of the best-fitting
   stellar population model with {\sc Starlight}. Systematic deviations from a
   idealized model can be seen as vertical stripes (rest-frame
   mismatches, e.g.\ imperfect stellar model or emission lines), slanted stripes (observed-frame
   mismatches, e.g.\ imperfect sky model) or general blue-to-red features
   (imperfect relative flux calibration with respect to SDSS). }
\label{fig:sky_subtraction}
\end{figure*}

Cumulative distributions of the $V$ and $B$ band night-sky surface brightness
during observations are shown in Fig.~\ref{fig:sky_brightness} for the V500
and V1200 CALIFA data, respectively. They were derived from the sky-fibers
following the analysis described in S12. There is considerable scatter in
night-sky brightness of about 1.5 magnitudes in both cases, indicating the
highly variable conditions at the observatory. The typical night-sky
brightness is $\mu_V\sim21.0\,\mathrm{mag}\,\mathrm{arcsec}^{-2}$ for the V500,
and $\mu_B\sim22.1\,\mathrm{mag}\,\mathrm{arcsec}^{-2}$ for the V1200
data. While the former is $\sim$1 mag brighter than the mean dark night sky at
the observatory
($\overline{\mu_V}$\,$\sim$\,$22\,\mathrm{mag}\,\mathrm{arcsec}^{-2}$), the
latter is much more similar to the mean
($\overline{\mu_B}$\,$\sim$\,$22.5\,\mathrm{mag}\,\mathrm{arcsec}^{-2}$),
according to \citet{Sanchez:2007b}. This is expected because we have reserved
the darkest nights (new moon $\pm1$ night) for the V1200 setup, while gray
nights are used for the more efficient V500 setup.

We have conducted a series of tests to assess the sky subtraction accuracy in
the continuum as delivered by the data reduction pipeline. As the
  initial test we again used the {\sc Starlight} code
  \citep{CidFernandes:2005} to model the central pipeline-reduced spectra with
  stellar templates and allowing for the presence of a dust
  component. Figure~\ref{fig:sky_subtraction} shows the residuals of these
fits as a fixed rest-frame wavelength stack of spectra for the 100 DR1
galaxies. Here spectra were always integrated over a central 10\arcsec\
aperture.

In this view, all rest-frame features form vertical stripes while
atmospheric and instrumental features are slanted away from the vertical. The
main effects from imperfections in sky subtraction can be seen around
6000\AA\ and 6250\AA, where sky-line residuals at a few percent of the stellar
template flux exist.

In order to be more quantitative regarding how much the sky subtraction could
be systematically improved, if at all, we have taken advantage of the ability
of the pPXF package to simultaneously fit sky and stellar templates to
optimize the sky subtraction of our data\footnote{A caveat is that
  while ideally the following method should be implemented at the pipeline
  level, at this point it is too complex and needs too much intervention to be
  implemented there, or to be carried out for all the sample galaxies -- at
  the same time it does not provide optimum sky fitting for all the
  fibers.}.

\begin{figure*}
\sidecaption
\includegraphics[width=12cm,height=5cm]{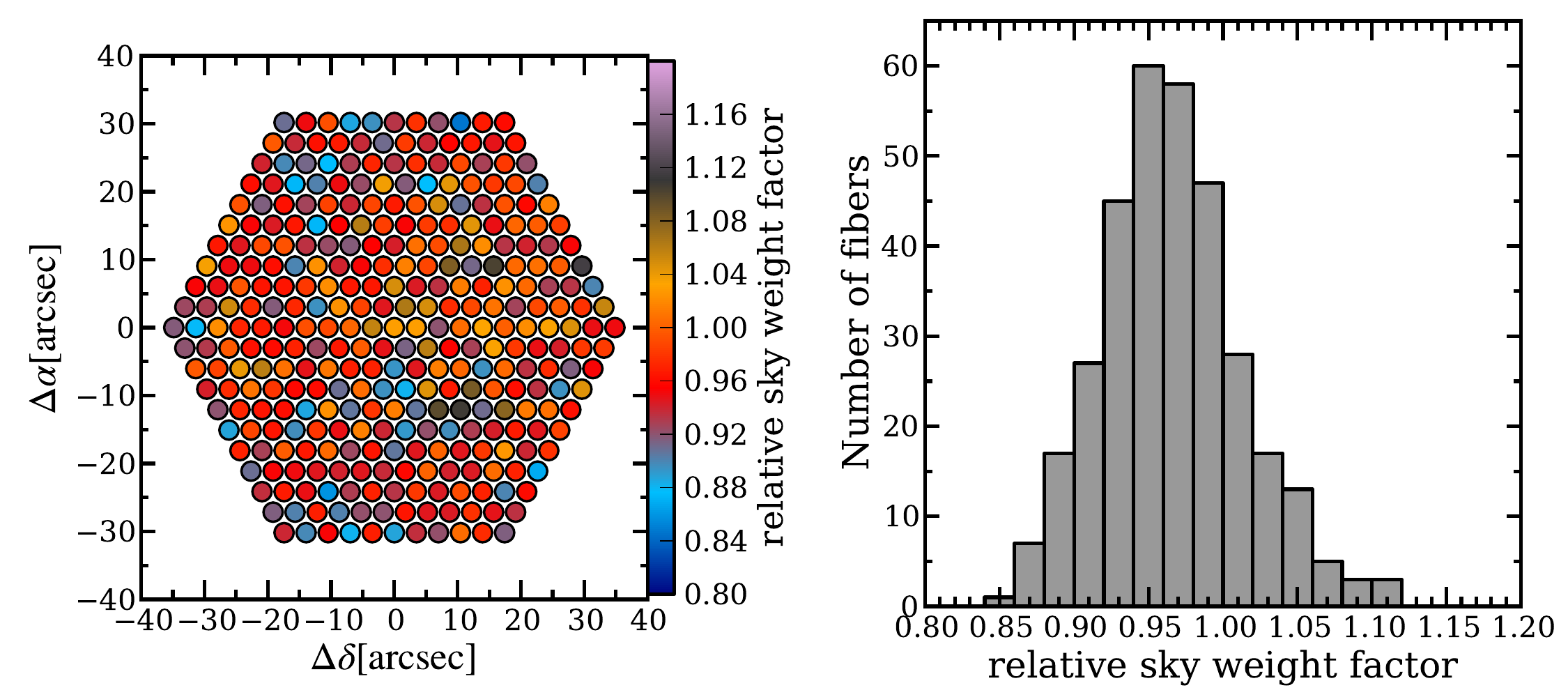}
\caption{\emph{Left panel:} Sky weight factor of the sky spectrum as inferred
  with {\tt pPXF} across the PPak field for the second pointing of
  UGC~03944. \emph{Right panel:} Distribution of sky weight factors for that
  particular pointing.}
\label{fig:sky_accuracy}
\end{figure*}

We selected all sky fibers from each of the three pointings for a particular
object and, as in the pipeline, discarded the six brightest fibers to ensure
that we eliminate cases that are clearly not dominated by the sky. The
remaining 30 sky fibers were used as sky templates for pPXF.  The
  input galaxy spectra were selected from fibers located in a ring in the
  inner disk of the galaxy where light intrinsic to the galaxy dominates. We
noticed that the use of fibers from the outer parts of the galaxy (i.e.\ more
sky dominated) can lead to errors in the derived kinematic parameters, giving
the wrong sky-template plus stars combination as the output of the pPXF
fitting. After masking emission lines to avoid nebular contamination we ran
pPXF, which gives as output the best stellar combination plus the best sky
combination. This procedure therefore provides the combination of sky fibers
that best matches the data. For further optimization, we ran pPXF on a
fiber-per-fiber basis to adjust this best sky solution to each of the fibers
including the sky-dominated outer fibers. The resulting sky
weights are rather constant over the FoV of the instrument with a scatter of
4\%, as is illustrated in Fig.~\ref{fig:sky_accuracy}.

This method was applied to a set of objects in the DR1 sample. We then
compared the sky-subtracted spectra of a pointing coming from the pipeline and
our method, before cube reconstruction. We carried out this comparison by
calculating the S/N in the continuum: for those fibers for which the method
described above improves the sky subtraction, the average difference in the S/N
is 3.7 $\pm$ 2.8\%.

Hence, the resulting sky subtraction is, globally, not significantly improved
compared to the pipeline sky subtraction. The average improvement
achieved in some of the fibers is of the order of a few percent, thus nearly as 
the pipeline sky subtraction is within a few percent in S/N as
accurate as the sky subtraction carried out using a sophisticated template
fitting algorithm.

For completeness, we note that among all lines the strongest residual
from the sky occurs in the \ion{Hg}{i}\,$\lambda$4358\AA,
\ion{Hg}{i}\,$\lambda$5461\AA, \ion{O}{i}\,$\lambda$5577\AA, \ion{Na}{i}\,D
(around 5890\AA), \ion{O}{i}\,$\lambda$6300\AA\, and
\ion{O}{i}\,$\lambda$6364\AA\ emission lines as well as at the telluric $B$ band
absorption.

\subsubsection{Spectrophotometric accuracy}\label{sect:specphot_cal_depth}

\begin{figure*}
 \includegraphics[width=0.5\textwidth]{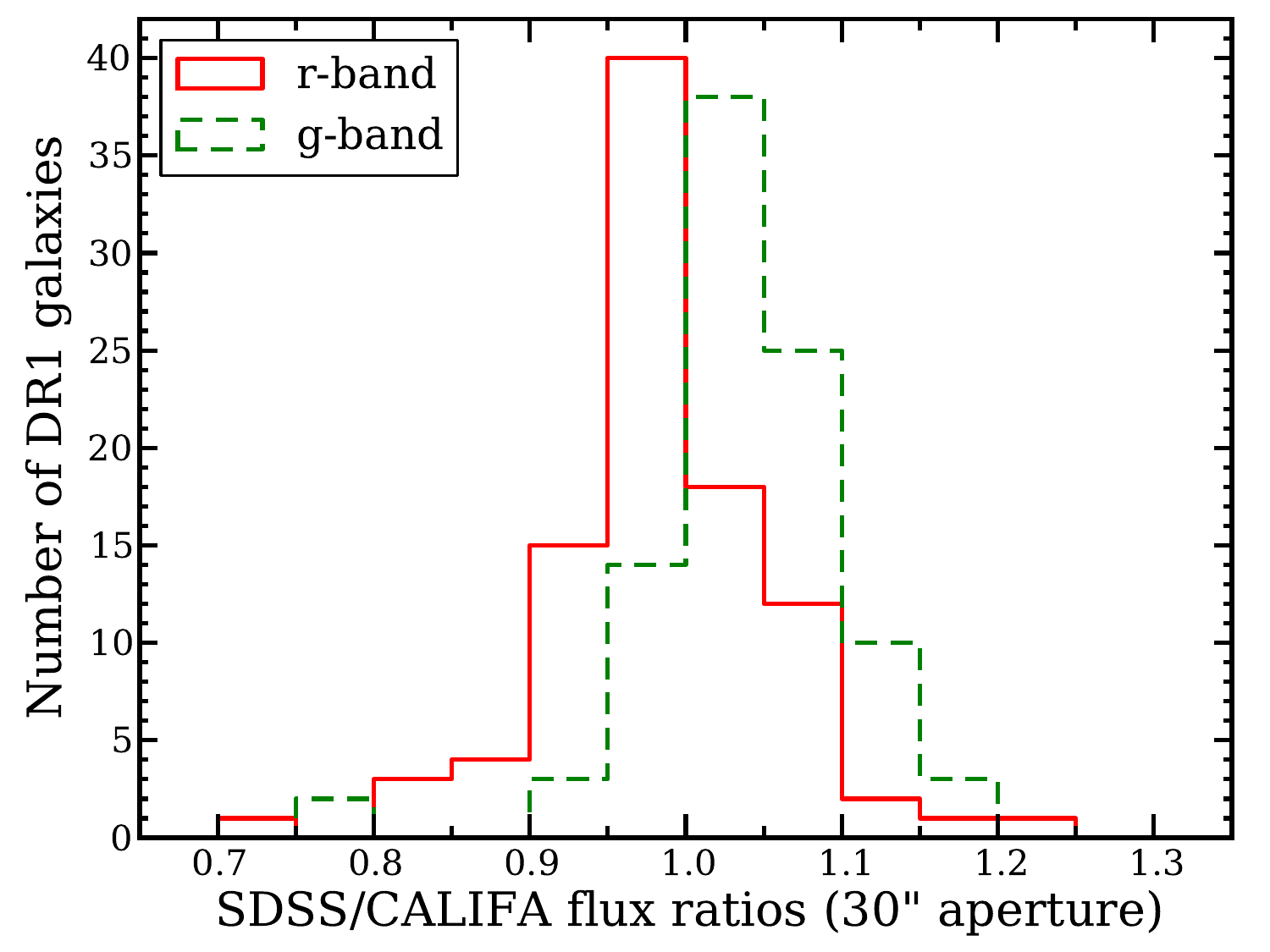}
 \includegraphics[width=0.5\textwidth]{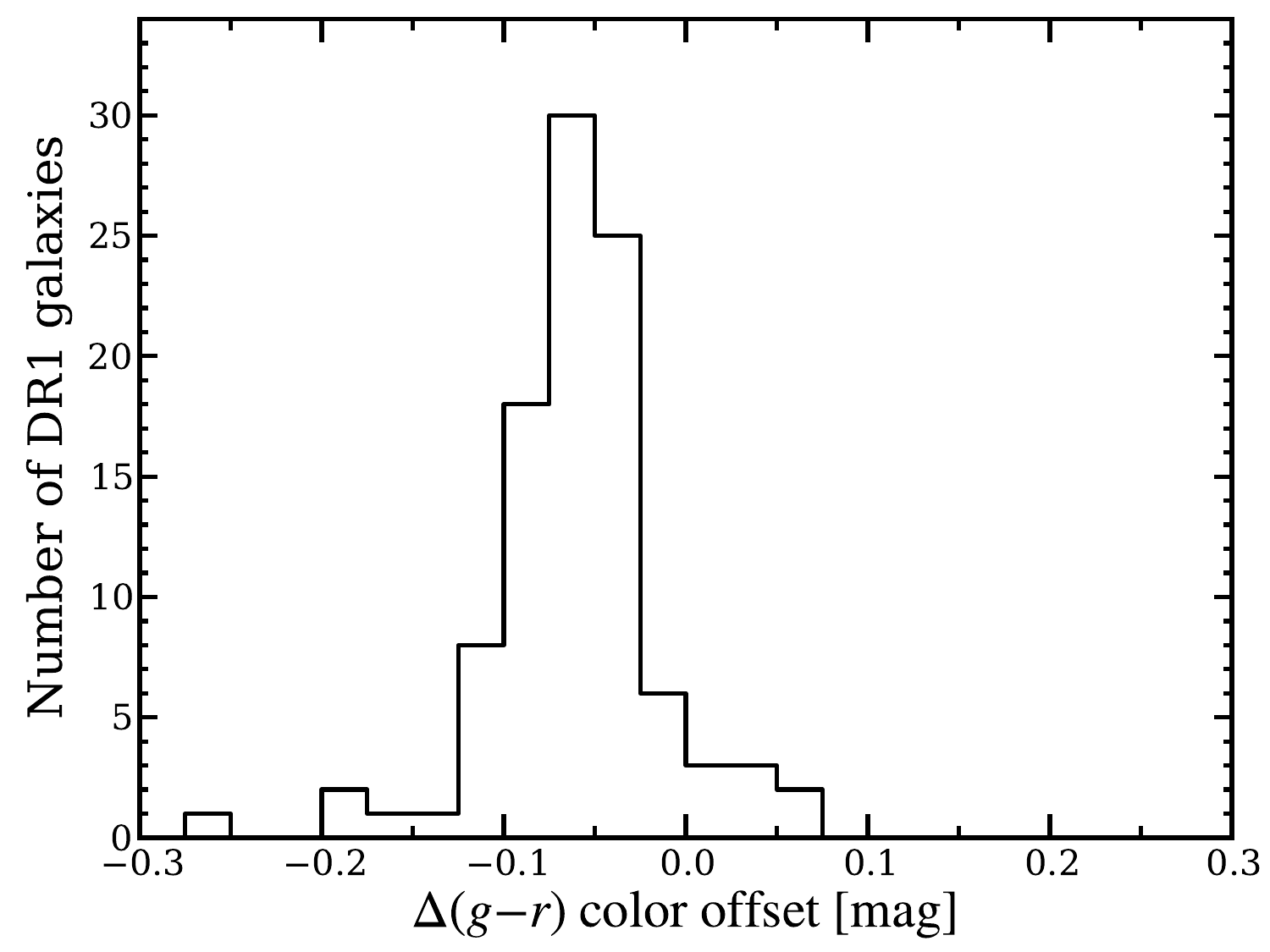}

 \caption{\emph{Left panel:} Distribution of the 30\arcsec\ aperture
   photometry scale factor between the SDSS DR8 images and re-calibrated
   CALIFA data. We compare the photometry only for the $g$ and $r$ bands,
   which are both entirely covered by the V500 wavelength range. \emph{Right
     panel:} Distribution of the corresponding color offset between the SDSS
   DR8 images and the synthetic CALIFA broad-band images.  }
  \label{fig:DR1_scale_check}
\end{figure*}

As part of the standard CALIFA pipeline each V500 datacube was rescaled in the
absolute flux level to match the SDSS DR7 broad-band photometry within an
aperture of 30\arcsec\ diameter, while the V1200 data were matched to
the V500 data (S12). A mosaiced SDSS image was created for each galaxy
using \texttt{SDSSmosaic}, an IRAF\footnote{IRAF is distributed by the
  National Optical Astronomy Observatories, which are operated by the
  Association of Universities for Research in Astronomy, Inc., under
  cooperative agreement with the National Science Foundation.} package
developed by S.~Zibetti \citep{Zibetti:2009}, which also subtracts the
background from each SDSS scan by fitting a plane surface that allows for linear
gradient along the scan direction. All stripes of the mosaic were combined
using the program \texttt{SWarp} \citep{Bertin:2002}. Finally, we obtained the
Galactic-extinction corrected absolute magnitudes with SExtractor
\citep{Bertin:1996} using the photometric zero-points as provided by the SDSS
DR7 photometric pipeline.

This process to rescale fluxes for absolute re-calibration will not be perfect
as can be seen from the slight systematic residuals in the blue and red edges of the spectrum
in Figure~\ref{fig:sky_subtraction}. To quantitatively cross-check the
achieved accuracy in the absolute photometry of the CALIFA data, we use the
improved photometry provided by the 8th data release of SDSS
\citep{Aihara:2011} to independently measure any systematic offset. The
distribution of photometric scale factors for the $g$ and $r$ bands is shown
in Fig.~\ref{fig:DR1_scale_check} (left panel). The mean SDSS/CALIFA $g$ and
$r$ band ratios in the release are $1.06\pm0.13$ and $1.00\pm0.10$,
respectively, which shows that the absolute photometric calibration of CALIFA
data is better than 15\%.  We attribute the mild systematic offset to the
improvements in the SDSS sky subtraction accuracy around the low-redshift
galaxies with size of $>$40\arcsec\ selected for CALIFA. However, there are
also several objects with photometric scale factors below 0.8 and above 1.2
for which we identify bright field stars as being the most likely cause of a
bias in the background estimates either for the SDSS or CALIFA data. For DR1
we actually drew a line at values of 0.5 and 2.0 for the scale factors and
removed a few objects that showed values beyond these limits. In the right
panel of Fig.~\ref{fig:DR1_scale_check} we show the distribution in
$\Delta(g-r)$ color difference between the SDSS and CALIFA data. We find a
systematic offset of $\Delta(g-r)$=$-$0.06\,mag (median) with a scatter of only
0.05\,mag. Thus, the spectrophotometric accuracy across most of the covered
wavelength range is 6\% for the CALIFA data.  

For the V500 setup we add a flag for the quality of photometric
  calibration, FLAG\_FLUXCAL. If for both SDSS $g$ and $r$ band the ratio
  between SDSS and CALIFA lies between 0.7 and 1.4, we assign a flag value of
  0. For at least one band behaving worse than this, with ratios up to 2.0 or
  down to 0.5, we assign FLAG\_FLUXCAL=1, and we reject the data for even worse
  values.

\medskip

\begin{figure}
\resizebox{\hsize}{!}{\includegraphics{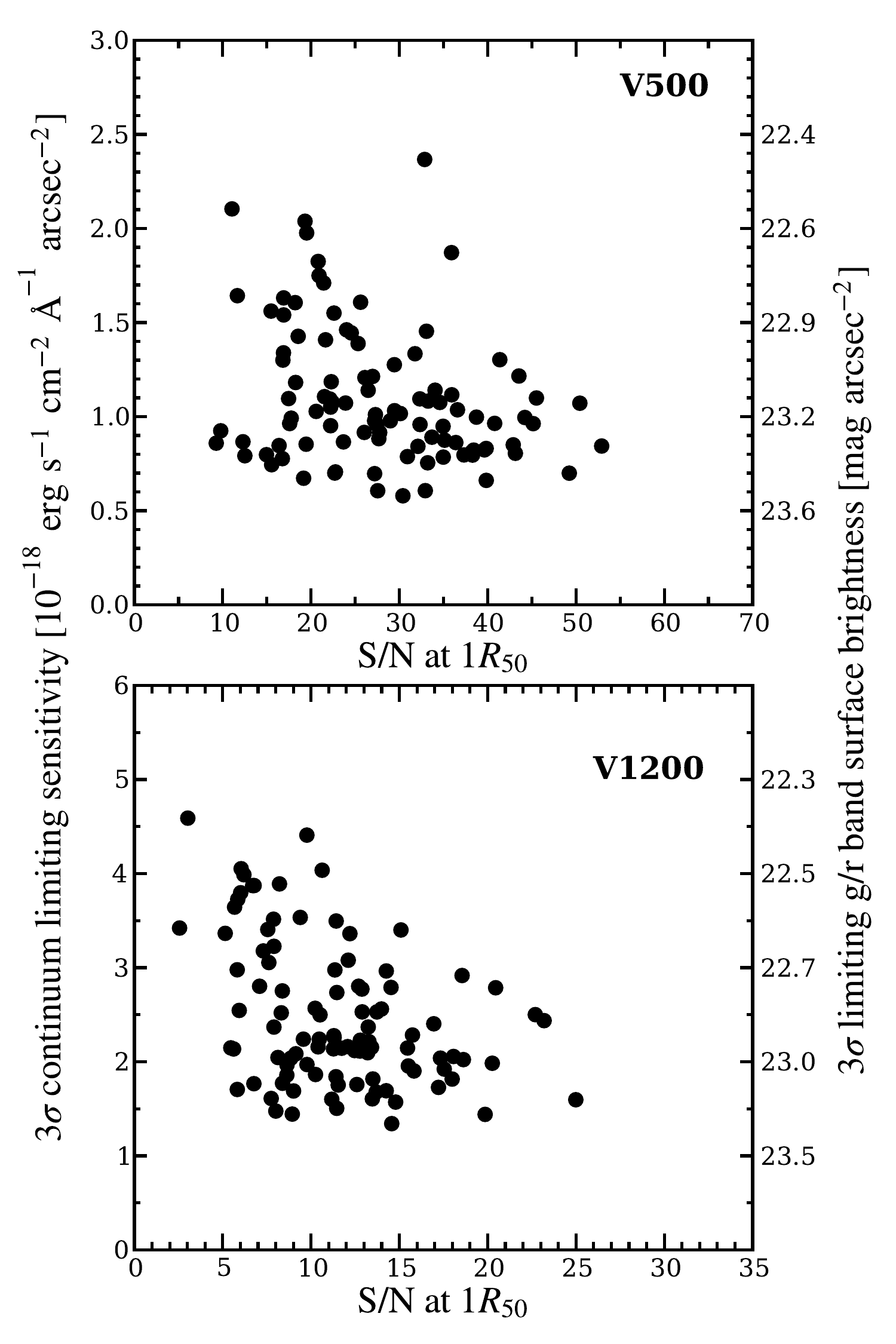}}
\caption{Limiting $3\sigma$ continuum sensitivity measured at the outskirts of the galaxy as a function of the average continuum S/N at the
  half light radius ($R_{50}$). The corresponding broad-band surface brightness limits in $r$ (V500) and $g$ (V1200) are indicated on the right $y$-axis. 
  The limiting continuum sensitivity and the S/N was computed from the median signal and noise in the wavelength region 4480--4520\AA\ and 5590--5680\AA\ for the V1200 and V500
  data, respectively.}
  \label{fig:DR1_depth}
\end{figure}

\subsection{Limiting sensitivity and signal-to-noise}\label{sect:depth}
The limiting continuum sensitivity of the spectrophotometrically recalibrated SDSS
spectra are presented in Fig.~\ref{fig:DR1_depth}. We compare the
3$\sigma$ continuum flux density detection limit per interpolated
$1\,\mathrm{arcsec}^{2}$-spaxel for the faintest regions against the average S/N per spectral resolution element within in an elliptical annulus of $\pm1\arcsec$ around the galaxies 
$r$ band half-light semi-major axis ($R_{50}$). Both values were estimated from the noise within the narrow
wavelength ranges 4480--4520\AA\ and 5590--5680\AA\ for the V1200 and V500 setup, respectively, that are nearly free of stellar absorption features. 
As expected, the 3$\sigma$ continuum flux density detection
limit\footnote{Please note that this is a continuum flux
    density. For e.g.\ any emission line the spectral resolution has to be taken into account, i.e. 6\AA\ and 2.3\AA\ FWHM for the V500 and V1200
    setups. In order to detect an unresolved emission line with a 3$\sigma$ significance in their peak flux density requires an integrated emission line flux of
    $\sim$\,$1\times10^{-17},\mathrm{erg}\,\mathrm{s}^{-1}\,\mathrm{cm}^{-2}\,\mathrm{arcsec}^{-2}$ and $\sim$\,$0.6\times10^{-17},\mathrm{erg}\,\mathrm{s}^{-1}\,\mathrm{cm}^{-2}\,\mathrm{arcsec}^{-2}$ in the V500 and V1200 data, respectively.}  for the V500 data
(as low as $I_{3\sigma}=1.0\times10^{-18}\,\mathrm{erg}\,\mathrm{s}^{-1}\,\mathrm{cm}^{-2}\,\mathrm{\AA}^{-1}\,\mathrm{arcsec}^{-2}$
in the median at 5635\AA) is a factor of $\sim$2 fainter than for the V1200
data
($I_{3\sigma}=2.2\times10^{-18}\,\mathrm{erg}\,\mathrm{s}^{-1}\,\mathrm{cm}^{-2}\,\mathrm{\AA}^{-1}\,\mathrm{arcsec}^{-2}$
in the median at 4500\AA) mainly because of the difference in spectral
resolution. These continuum sensitivities can be directly transformed into equivalent limiting broad band surface brightnesses
of $23.6\,\mathrm{mag}\,\mathrm{arcsec}^{-2}$ in the $r$ band for the V500 and $23.4\,\mathrm{mag}\,\mathrm{arcsec}^{-2}$ in the $g$ band for the V1200 data.
The dispersion in the limiting continuum sensitivity can partially be
explained by the variance in the sky brightness of the corresponding night,
because both are significantly correlated, as expected. However, other
atmospheric conditions, such as dust attenuation, transparency of the night,
etc.\ are important characteristics that alter the achievable depth at fixed
exposure times. We provide a flag for the limiting sensitivity,
  FLAG\_DEPTH, which indicates with 0, 1, 2, whether the 3$\sigma$ continuum
  limiting sensitivity is $<$5, between 5 and 20, or
  $>$$20\times10^{-18}\,\mathrm{erg}\,\mathrm{s}^{-1}\,\mathrm{cm}^{-2}\,\mathrm{\AA}^{-1}\,\mathrm{arcsec}^{-2}$
  .

Another interesting quantity for practical purposes is the S/N of
the data. Since the galaxy brightness enters as the signal, S/N
  correlates only partially with the limiting sensitivity, which is a measure
  of the noise.  The mean S/N in the continuum per spaxel at the half-light semi-major axis 
($1R_\mathrm{50}$) of all objects is $\sim$14 for the V1200 setup, while it is $\sim$34.5 for the
V500 data.  Thus, we achieve a S/N$>$10 for the majority of the objects even
for the V1200 setup. This means that using adaptive binning scheme, a decent
S/N can be achieved far beyond $1R_\mathrm{50}$.

\subsection{Overall quality judgment}
The quality of a given dataset is defined by a combination of the
  quality flags introduced above. We used a subset of these in order to guarantee that 
  we deliver scientifically first-grade data in this data release. Minimum (maximum) 
  threshold values have been defined as follows: Airmass, sec($z$)$<$2; relative spectrophotometric calibration
relative to SDSS, $0.50< (f_\mathrm{CALIFA}/f_\mathrm{SDSS})<2$ for both $g$
and $r$ band; and a limiting sensitivity better than
$2\times10^{-17}\,\mathrm{erg}\,\mathrm{s}^{-1}\,\mathrm{cm}^{-2}\,\mathrm{arcsec}^{-2}$
in the abovementioned spectral windows (Sect.~\ref{sect:depth}). We also add
the described checks on seeing, wavelength calibration and spectral resolution
and initial reduction flags.

For each object we list the most important QC parameters discussed in
  this section in Tables~\ref{tab:QC_par_V500} and \ref{tab:QC_par_V1200}. We carried out very extensive vetting and
description in order to deliver 100 galaxies with prime quality data. These
listed parameters can be used to check the data quality of the
released cubes per object in more detail.

\section{Access to the CALIFA DR1 data}\label{sect:DR1_access}
\subsection{The CALIFA DR1 search and retrieval tool}
The public data is distributed through the CALIFA DR1 web page (\url{http://califa.caha.es/DR1}). A simple web form interface, developed especially for this first data release, allows to select data of a particular target galaxy, or a subsample of objects  within some constraints on galaxy properties and observing conditions. Among the selection parameters we include the instrument setup, galaxy coordinates, redshift, $g$ band magnitudes, observing date, Hubble type, bar strength, inclination estimated from axis ratio, $V$ band atmospheric attenuation, airmass, and relative accuracy of the SDSS/CALIFA photometric calibration. See W12 for a more detailed characterization of the CALIFA galaxies. 

If CALIFA datasets are available within the given constraints, they are listed in the succeeding web page to make a final selection for download. The download process requests a target directory on the local machine to store the data, after the downloading option was selected. The CALIFA data are delivered as fully reduced datacubes in FITS format, described in Sect.~\ref{sect:data}, separately for each of the two CALIFA spectral settings, i.e. the V500 and V1200 setup. Each DR1 dataset is uniquely identified by their file name,  \textit{GALNAME}.V1200.rscube.fits.gz and \textit{GALNAME}.V500.rscube.fits.gz for the V1200 and V500 setup respectively, where \textit{GALNAME} is the name of the CALIFA galaxy listed in Table~\ref{tab:DR1_sample}.

\subsection{CALIFA Galaxy Explorer page}

\begin{figure*}
\includegraphics[width=4.48cm]{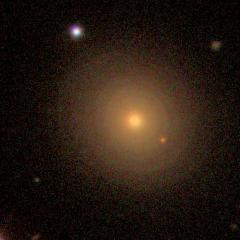}
\hfill
\includegraphics[width=4.48cm]{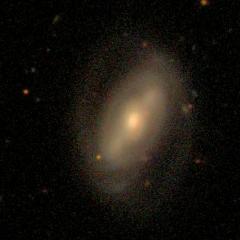}
\hfill
\includegraphics[width=4.48cm]{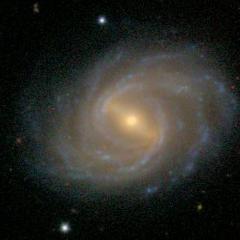}
\hfill
\includegraphics[width=4.48cm]{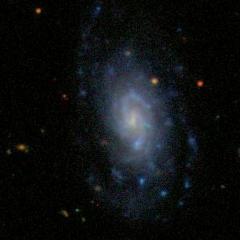}
\includegraphics[width=4.5cm]{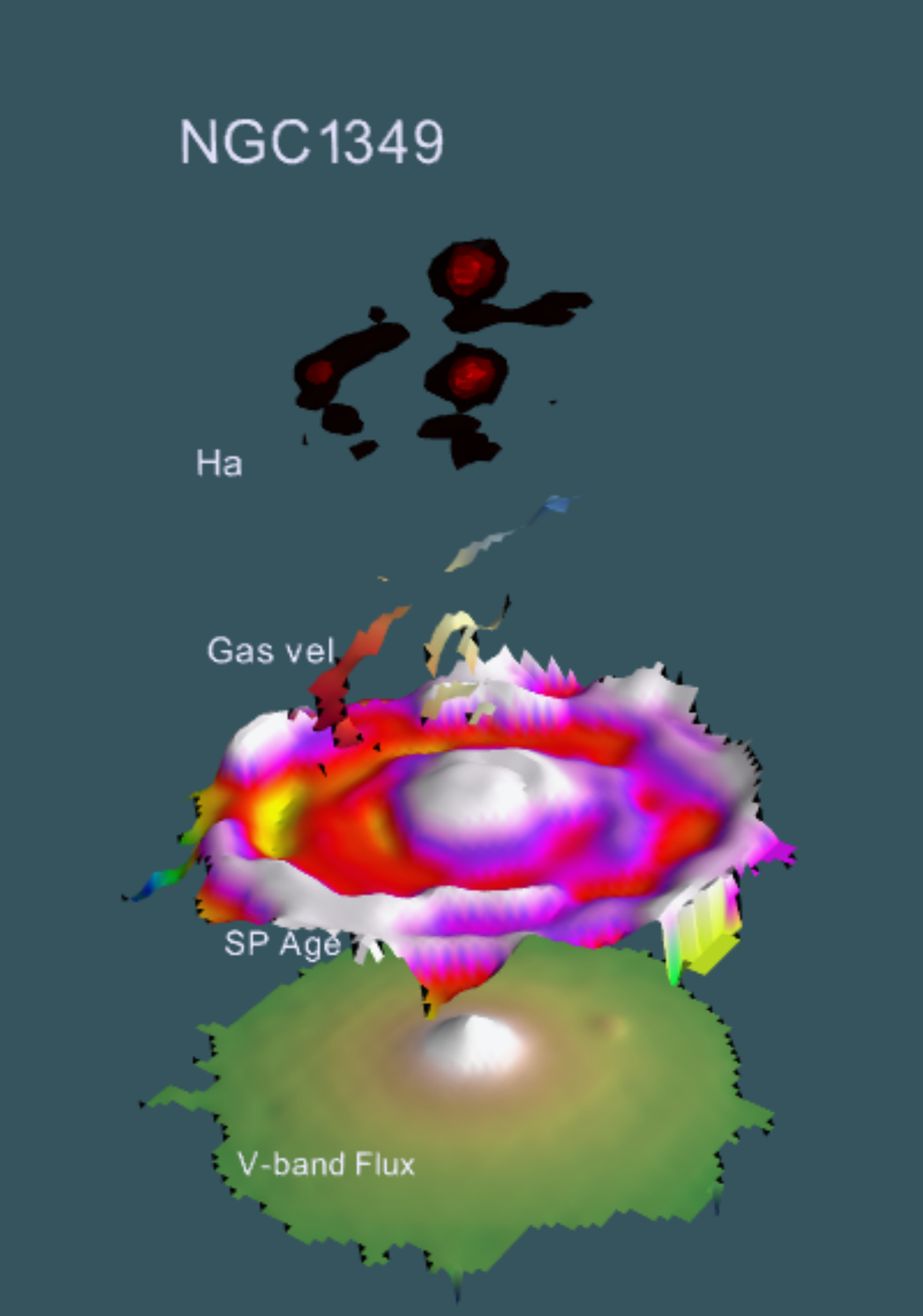}
\hfill
\includegraphics[width=4.5cm]{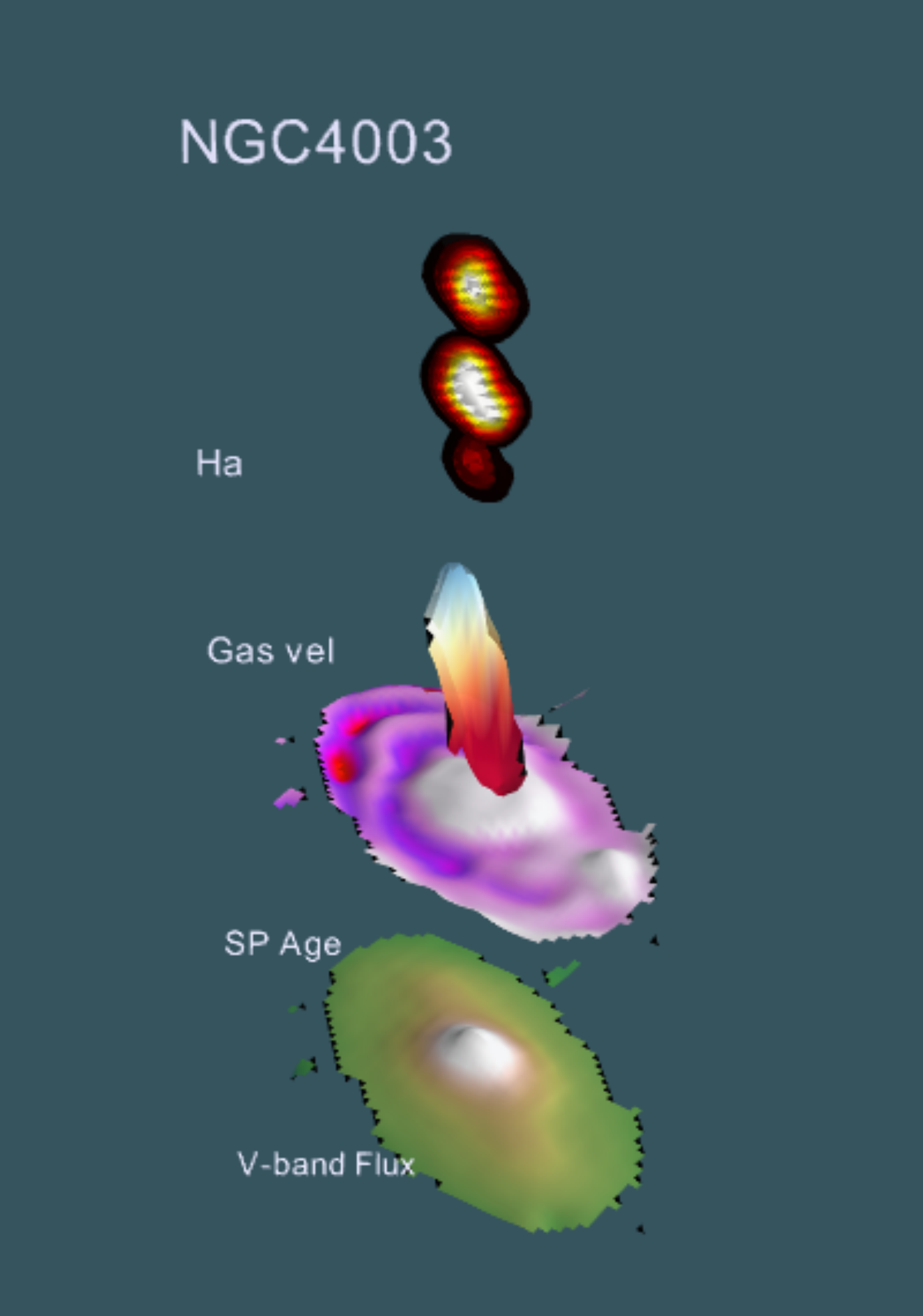}
\hfill
\includegraphics[width=4.5cm]{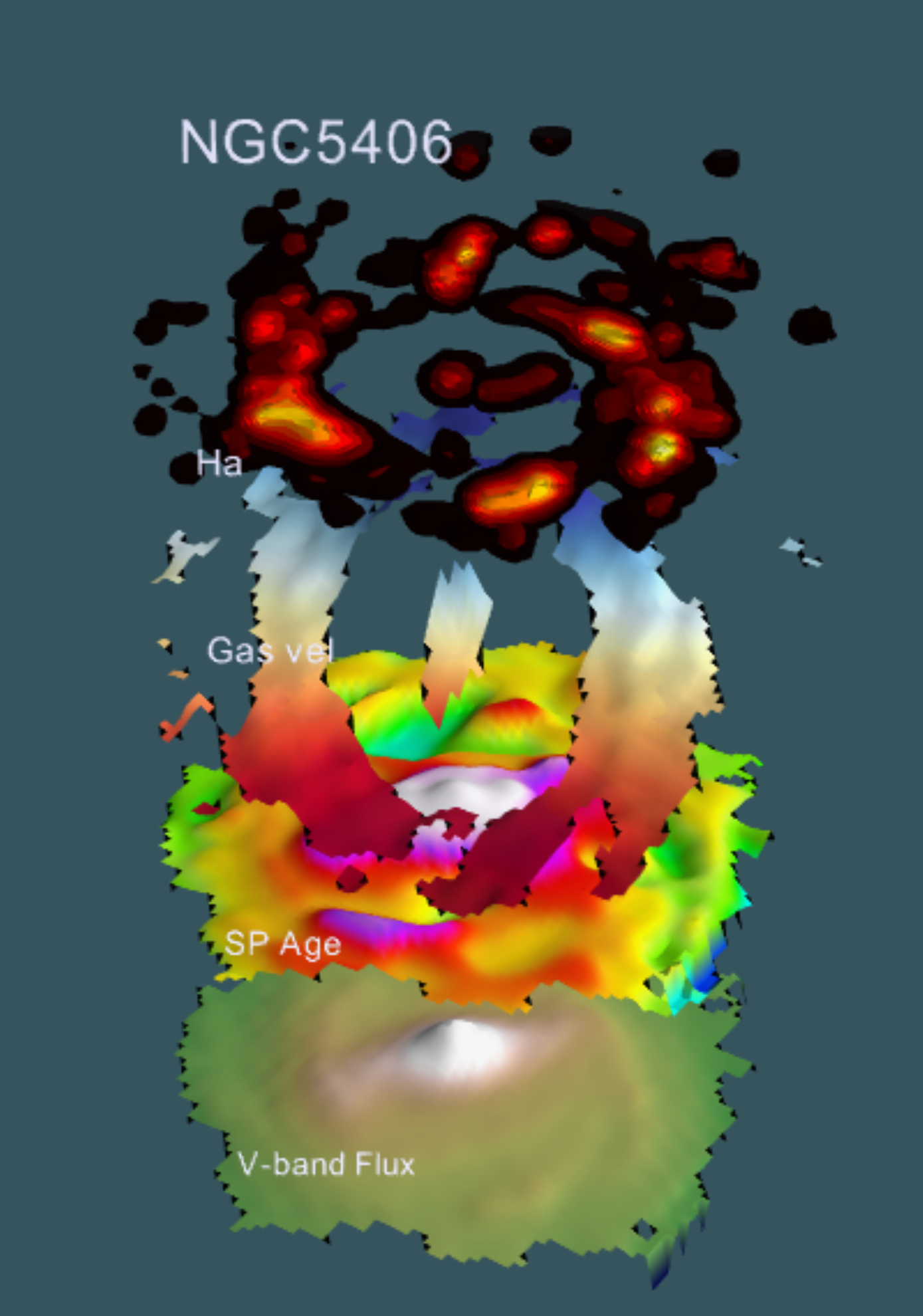}
\hfill
\includegraphics[width=4.5cm]{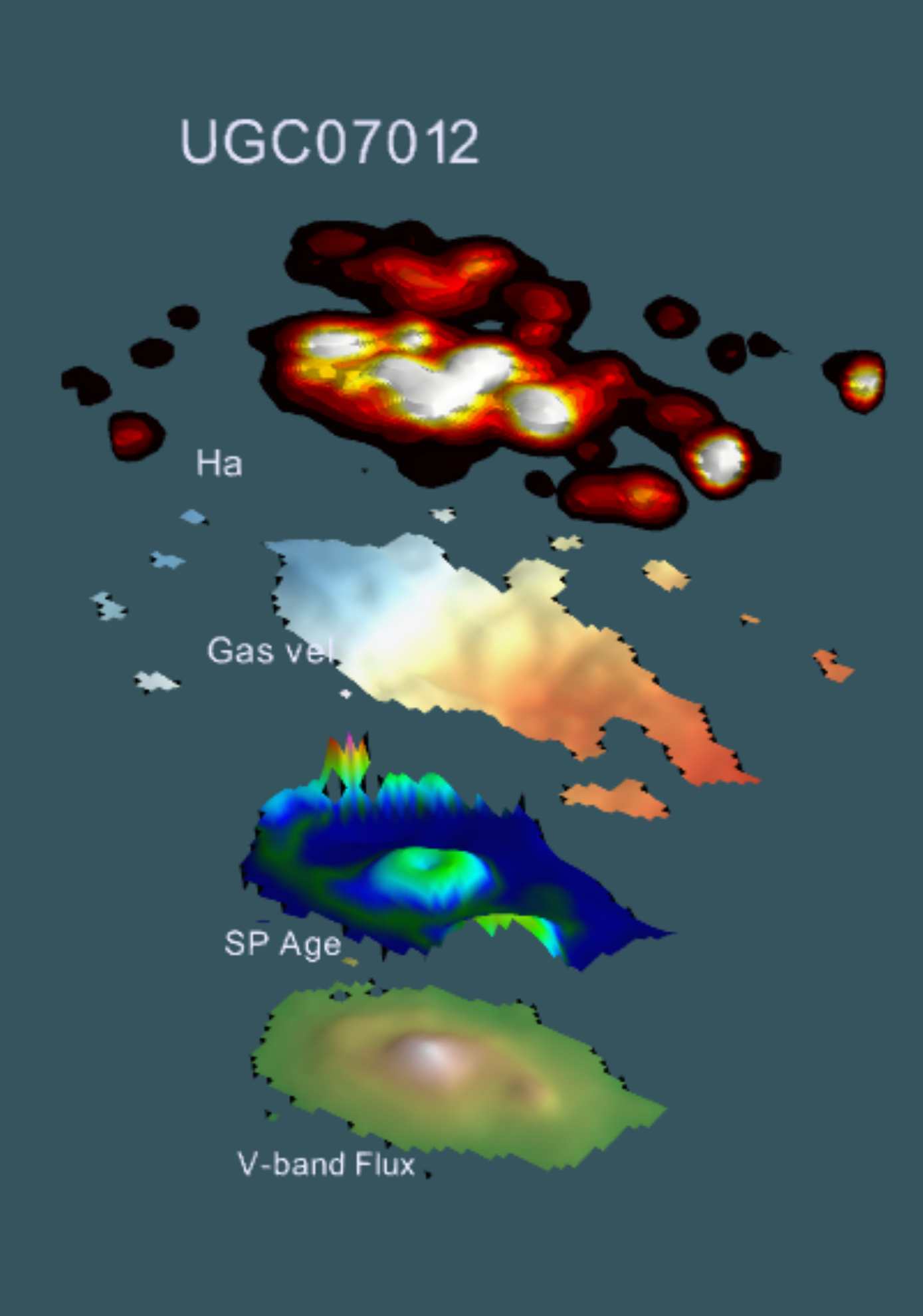}
\includegraphics[width=4.53cm]{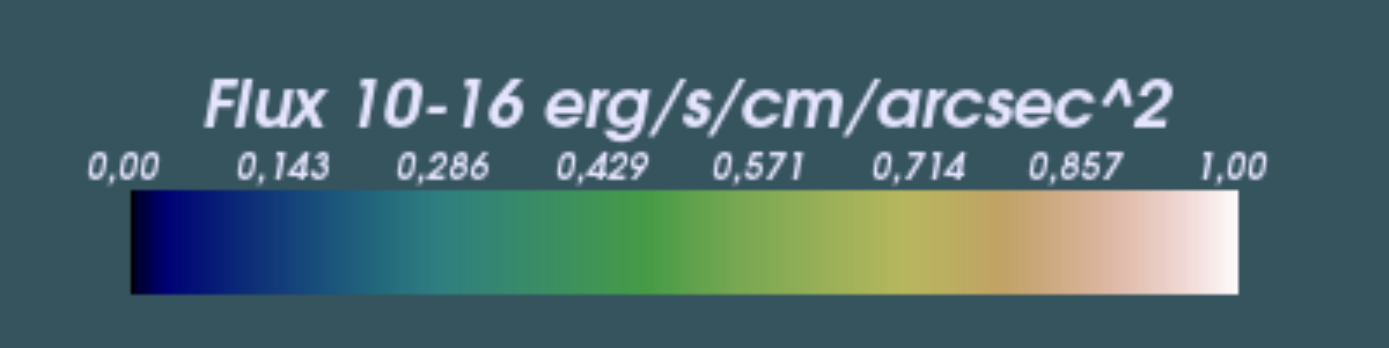}
\hfill
\includegraphics[width=4.53cm]{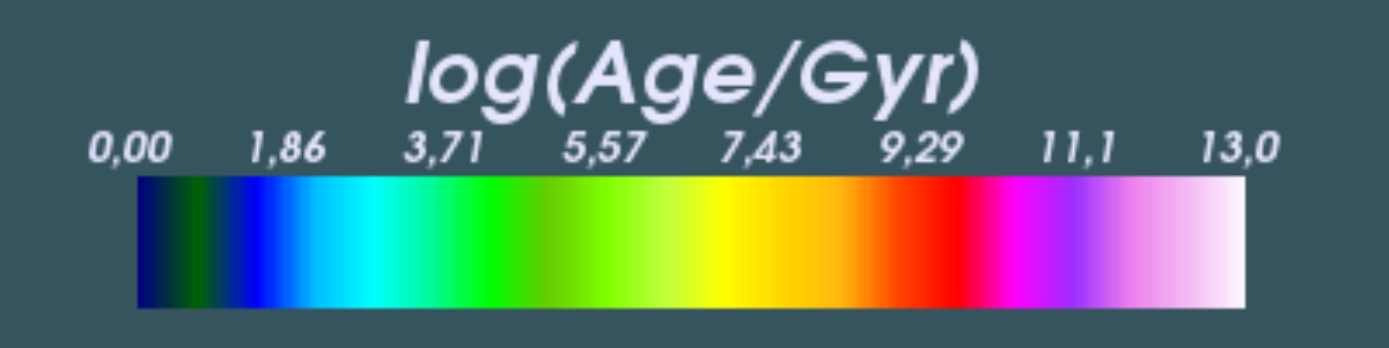}
\hfill
\includegraphics[width=4.53cm]{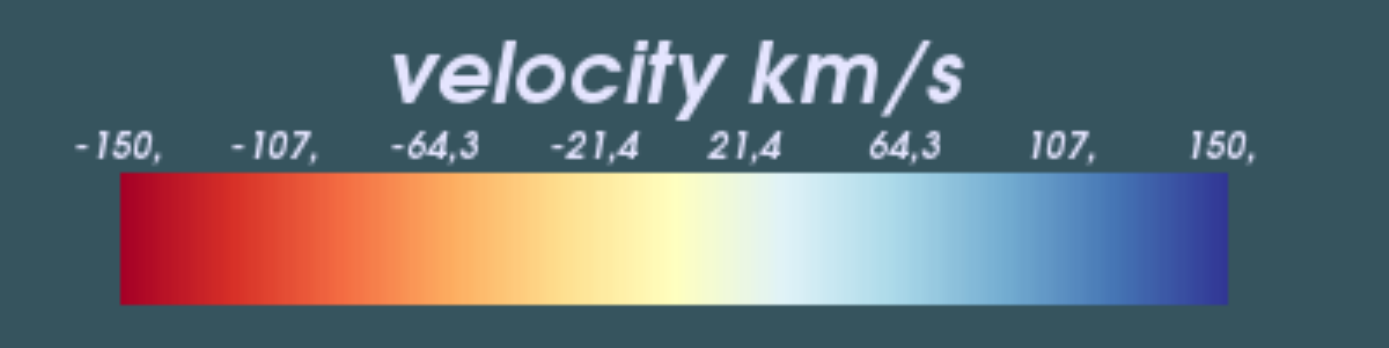}
\hfill
\includegraphics[width=4.53cm]{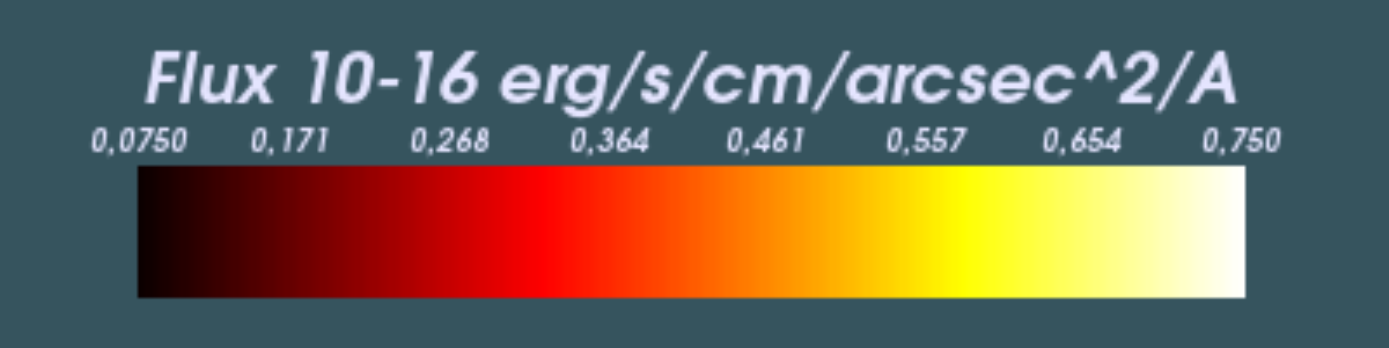}

\caption{\emph{Top panels}: SDSS postage stamp images of CALIFA target galaxies, with a FoV of $90\arcsec \times90\arcsec$. \emph{Bottom panels:} Corresponding examples of quick-look CALIFA 3D visualization images shown on the CALIFA Galaxy Explorer page for four representative galaxies. From bottom to top it shows the reconstructed $V$ band light distribution, the stellar population luminosity-weighted age distribution, the H$\alpha$ gas velocity field, and the H$\alpha$+[\ion{N}{ii}]\,$\lambda\lambda 6543,6584$ emission-line 3D distribution inferred from the CALIFA V500 datacube. The associated color bars are shown below from left to right.} 
\label{fig:CGE_diagram}
\end{figure*}
As it can be a challenge for the general user to judge a-priori which CALIFA dataset could be useful for their specific scientific interest, we prepared a CALIFA Galaxy Explorer page (hereafter CGE) that provides a presentation card for each individual CALIFA DR1 galaxy to the scientific community. It is a powerful tool for a quick-look reference of the galaxy properties and the actual content of the CALIFA DR1 data. The CGE is linked to each object in the output table of the CALIFA retrieval tool and  a column with a link is added to a dedicated DR1 sample table available on the DR1 web page. 

The CGE page is structured in three different levels. In the first level, a false-color ($gri$ bands) SDSS postage stamp image ($1.5\arcmin\times1.5\arcmin$ in size) is presented together with a short table containing the galaxy name, galaxy equatorial coordinates (in J2000 format) and the designated CALIFA ID. Alternative names of the galaxies can be found through a direct link to NED together with the redshift, the SDSS $u$, $g$, $r$, $i$, and $z$ band Petrosian magnitudes, and the corresponding Galactic extinction in each band.
 
In the second level, the user finds the CALIFA data products and their content. A compact visualization of the CALIFA data (see Fig.~\ref{fig:CGE_diagram}) shows the reconstructed $V$ band image, the stellar population age distribution, the gas velocity field, and the H$\alpha$+[\ion{N}{ii}]\,$\lambda\lambda 6543,6584$ emission-line 3D distribution at first glance.   Those maps were created by a preliminary analysis of the V500 datacubes with the {\sc FIT3D} software tool \citep{Sanchez:2007d}. Additionally,  the $B$ and $V$ band images reconstructed from the CALIFA datacubes are shown for both the V1200 and V500 setups, respectively, together with the corresponding 30\arcsec\ aperture integrated spectra.

In the third level, we report the result of our visual morphological classification of the galaxy. More detailed galaxy information is shown in a separate table which includes additional parameters available through the Hyperleda database.

\subsection{Virtual Observatory services}\label{sect:VO_serve}
CALIFA data can be accessed using the Virtual Observatory (VO) Table
Access Protocol \citep[TAP,][]{Dowler:2011} at the TAP URL \url{http://dc.g-vo.org/tap}
(service ivo://org.gavo.dc/\_\_system\_\_/tap/run). 
We provide tables describing each cube as a
single data product (califadr1.cubes) including the QC parameters reported in Tables~\ref{tab:QC_par_V500} and \ref{tab:QC_par_V1200} as well as tables containing all
fluxes of the entire DR1 by position and wavelength (califadr1.fluxv500 and califadr1.fluxv1200) for the two setups separately.  
The table schemata can be obtained using the usual TAP means and tables can be easily retrieved and queried through the VO client TOPCAT\footnote{\url{http://www.star.bris.ac.uk/~mbt/topcat/}}. 
Example queries and links will be provided on the DR1 web page to guide the user.
The cubes can also be found in the GAVO data center's ivoa.obscore
tables with a complete ObsCore metadata set \citep{Tody:2011} and the corresponding links are also stored in the califadr1.cubes table.

Individual spectra can also be accessed using IVOA's Simple Spectral
Access Protocol using the service ivo://org.gavo.dc/califa/q/s.  This
allows for easy access from within analysis programs like {\tt Splat} and an example will be provided on the DR1 web page.

An overview of VO-accessible resources generated from CALIFA data is
available at http://dc.g-vo.org/browse/califa/q.

\subsection{Visualization of CALIFA data}
Visualization of CALIFA data is essential for  its correct treatment, analysis and eventual scientific interpretation. IFS visualization tools are very useful for this purpose, as they provide a 2D view at any wavelength slice or enable exploration of the spectra at any spaxel. There are various visualization tools available in the community.  Some are designed for a quick-look to the data, others allow for more interactivity, while yet other are developed for a particular project or data format.

For the visualization of CALIFA data, stored in a FITS datacube format, we list here several tools as examples:
1) {\tt ds9}\footnote{\url{http://hea-www.harvard.edu/RD/ds9}}, from version 7, can load and view 3D datacubes in multiple dimensions;
2) {\tt QFitsView}\footnote{\url{http://www.mpe.mpg.de/~ott/QFitsView}} by Thomas Ott, is a generic FITS viewer program capable of handling datacubes and performing basic analysis operations;
3) {\tt E3D}\footnote{\url{http://www.aip.de/Euro3D/E3D}}  \citep{Sanchez:2004c}, is a high-level IFS package which allows interactive visualization, spatial resampling, and basic analysis;
4) {\tt IFSview}\footnote{\url{http://www.caha.es/sanchez/IFSview/}} is a light and stand-alone {\tt Python} script using {\tt TkInter}, {\tt pyfits} and {\tt matplotlib}, to visualize FITS datacubes. Its ultimate goal is to transfer all the functionality of {\tt E3D} into a more flexible and easy to install tool;
5) {\tt CASA Viewer} as part of the  {\tt CASA} software package\footnote{\url{http://casa.nrao.edu}}, it was initially designed for the handling and analysis of 3D radio data, but will be extended to display optical/NIR datacubes in the upcoming version 4.0 (Kuntchner et al. in prep.). It can also read error vectors that can be properly used in the analysis of the data;
6) {\tt PINGSoft}\footnote{\url{http://califa.caha.es/pingsoft}}  \citep{Rosales-Ortega:2011}, is a set of IDL routines designed to visualize, manipulate, and analyze IFS data regardless of the original instrument and spaxel size/shape.  Given that it was originally developed for data from the PPak instrument \citep{Rosales-Ortega:2010}, it was recently updated to handle the CALIFA data format (see below).
All those tools are also available through the CALIFA DR1 web page.

In the case of {\tt PINGSoft}, the package is adapted to work natively with the CALIFA data, all routines automatically identify the different FITS extensions (HDUs) of the CALIFA format, allowing the user to spatially and spectrally explore the CALIFA data. In addition, {\tt PINGSoft} includes routines to extract regions of interest by hand, within a given geometric aperture or based on a user-given mask, to integrate the spectra within a given region or based on the S/N derived from  continuum/emission line features; to convolve the data with a full set of narrow and broad-band filters for visualization or analysis purposes; to perform radial binnings with either fixed spatial bins or S/N floor; to derive cross-correlations velocity fields; to perform spatial adaptive binning (e.g. Voronoi tessellation), as well as some miscellaneous codes for generic tasks in astronomy and spectroscopy. {\tt PINGSoft} also includes a routine to split the different CALIFA FITS extensions into individual FITS files\footnote{A similar {\tt Python} routine that can split the FITS extensions into separate files and convert the CALIFA data to  {\tt CASA} data format can be found in the CALIFA web page.}.

In summary, there have been important efforts in the development of IFS visualization and manipulation tools. Given the number of different software options available in the literature, spatial and spectral visualization of the CALIFA data should not represent a limitation for potential users.

\section{Conclusions}
In this article we describe the data distributed with the first public data release of the Calar Alto Legacy Integral Field Area (CALIFA) survey. The CALIFA DR1 provides science-grade and quality-checked integral-field spectroscopy data for 100 nearby galaxies publicly distributed to the community at \url{http://califa.caha.es/DR1}. The current DR1 sample comprises one sixth of the total foreseen sample, but it already covers a similar range of colors, magnitudes (see Fig.~\ref{fig:DR1_CM_diag}), and morphological types (see Fig.~\ref{fig:DR1_morph}). Thus, the released IFS data are representative of the typical spectroscopic properties of galaxies at each color/magnitude bin, although it lacks the proper statistical depth of the final sample. With this sample it will be possible to probe the evolutionary sequence across the color-magnitude diagram, even if it may not yet be possible to disentangle the effects of different galaxy properties and their evolution at each bin (like active nuclei, morphology, presence of bars...).

An improved data reduction pipeline (V1.3c) has allowed us to increase the quality of the data beyond the original specifications of the survey. The modifications with respect to the previous pipeline (V1.2) are described in  Appendix~\ref{apx:data_reduction}. The final FITS data format, including science data, propagated error vectors, masks, and error weighting factors was outlined in Sect~\ref{sect:data_format}. The quality of each datacube was cross-checked by a set of defined figures of merit, allowing us to select and distribute the data with the highest quality. Among the parameters cross-checked are the depth, wavelength calibration accuracy, spectral resolution, sky subtraction, and stability of the observations and reduction process. 

Despite its limited statistical scope, the CALIFA DR1 sample already comprises the largest (so far) IFS survey of nearby galaxies of all morphological types, sampling the full color-magnitude diagram, and covering the full optical extension of galaxies with a large wavelength coverage. The scientific use of this dataset is verified by several  publications currently in preparation by the CALIFA collaboration on different science topics, including the ISM in early-type galaxies, dynamical and gas ionization state in merging galaxies, the stellar mass-growth in galaxies, dynamical state of galaxies, and oxygen abundance gradients. All those topics cover most of the science goals described in the official proposal for CALIFA survey, the CALIFA Red Book\footnote{available at \url{http://www.caha.es/CALIFA/Accepted_Proposal.pdf}}, but there is certainly much more to explore. We encourage the community to make use of the CALIFA data. In return we request to cite the technical CALIFA papers about the survey design \citep{Sanchez:2012a}, sample characterization (Walcher et al., in preparation), and this paper together with some recommended acknowledgements\footnote{This study makes use of the data provided by the Calar Alto Legacy Integral Field Area (CALIFA) survey (http://califa.caha.es/). Based on observations collected at the Centro Astronómico Hispano Alemán (CAHA) at Calar Alto, operated jointly by the Max-Planck-Institut fűr Astronomie (MPIA) and the Instituto de Astrofisica de Andalucia (CSIC).}.

The amount of data distributed in DR1 already comprises $\sim$200\,000 independent spectra. While this is already a large sample in itself, it represents only 17\% of the total number of spectra foreseen at the completion of the CALIFA survey. Up to now we have acquired data for $\sim$200 galaxies in both setups, for which we will continue to perform the quality control tests implemented for this DR. It is expected that the second CALIFA data release will take place after 2013, when we have collected high-quality science-grade data of 300 galaxies in both setups.

\begin{acknowledgements}
CALIFA is the first legacy survey being performed at Calar Alto. The CALIFA collaboration would like to thank the IAA-CSIC and MPIA-MPG as major partners of the observatory, and CAHA itself, for the unique access
to telescope time and support in manpower and infrastructures.  The CALIFA collaboration thanks also the CAHA staff for the dedication to this project. We thank the {\it Viabilidad  Dise\~no  Acceso y Mejora } funding program, ICTS-2009-10, for funding the data acquisition of this project.\\

We thank the referee Mark Westmoquette for a prompt report with many valuable comments that improved the clarity of the article.\\

BH gratefully acknowledges the support by the DFG via grant Wi 1369/29-1. KJ was supported by the Emmy Noether-Programme of the german DFG under grant Ja 1114/3-1. SFS, FFRO, and DM thank the {\it Plan Nacional de Investigaci\'on y Desarrollo} funding programs, AYA2010-22111-C03-03 and AYA2010-10904E, of the Spanish {\it Ministerio de   Ciencia e Innovaci{\'o}n}, for the support given to this project. SFS also thanks the the {\it Ram\'on y Cajal} project RyC-2011-07590 of the spanish {\it Ministerio de Economia y Competitividad}, for the support giving to this project. RCF acknowledges the support of the brazilian agencies CAPES and CNPq.
JF-B acknowledges support from the {\it Ram\'on y Cajal} Program financed by the Spanish Ministry of Economy and Competitiveness (MINECO). This research has been supported by the Spanish Ministry of Economy and Competitiveness (MINECO) under grants AYA2010-21322-C03-02 and AIB-2010-DE-00227. 
RG-B has been supported by  the Spanish {\it Ministerio de   Ciencia e Innovaci{\'o}n} under grant AYA2010-15081.
RAM was also funded by the spanish programme of International Campus of Excellence Moncloa (CEI). 
DM and AMI are supported by the Spanish Research Council within the program JAE-Doc, Junta para la Ampliación de Estudios, co-funded by the FSE. 
FFR-O acknowledge financial support for the ESTALLIDOS collaboration by the Spanish {\it Ministerio de Ciencia e Innovaci{\'o}n} under grant  AYA2010-21887-C04-03. 
JMG is supported by grant SFRH/BPD/66958/2009 from FCT. 
BJ has been supported by the Czech program for the long-term development of the research institution No. RVO67985815 and by the grant PIPPMS No. M100031201 of the Czech Academy of Sciences.
CK has been funded by the project AYA2010-21887 from the Spanish PNAYA. 
AdO and IM has been partially funded by the projects AYA2010-15196 from the Spanish {\it Ministerio de Ciencia e Innovaci{\'o}n} and TIC 114 and PO08-TIC-3531 from Junta de Andaluc\'{\i}a.
BG-L acknowledges the support of the {\it Ramón y Caja}l programme (RYC-2006-002037) and the grant AYA2009-12903 of the Spanish MICINN.
PP is supported by a Ciencia 2008 contract, funded by FCT/MCTES (Portugal) and POPH/FSE (EC). 
PS-B is supported by the {\it Ministerio de Ciencia e Innovaci\'on} of Spain through the {\it Ram\'on y Cajal} programme. PSB also acknowledges a Marie Curie Intra-European Reintegration grant within the 6th European framework program and financial support from the Spanish Plan Nacional del Espacio del Ministerio de Educaci\'on y Ciencia (AYA2007-67752-C03-01).
KS acknowledges support from the National Sciences and Engineering Research Council of Canada.
VS acknowledges financial support from Funda\c{c}\~{a}o para a Ci\^{e}ncia e a Tecnologia (FCT) under the program Ci\^{e}ncia 2008 and the research grant PTDC/CTE-AST/112582/2009. 
VW Acknowledges support from an European Research Council starting grand and a Marie Curie reintegration grant.
JMV and AMI have been partially funded by the projects AYA2010-21887 from the Spanish PNAYA, CSD2006 - 00070 “1st Science with GTC” from the CONSOLIDER 2010 programme of the Spanish MICINN, and TIC114 Galaxias y Cosmología of the Junta de Andalucía (Spain). The Dark Cosmology Centre is funded by the Danish National Research Foundation. \\

We also made use of data from the SDSS, SDSS-II and SDSS-III for which funding was provided by the Alfred P. Sloan Foundation, the Participating Institutions, the National Science Foundation, the U.S. Department of Energy, the National Aeronautics and Space Administration, the Japanese Monbukagakusho, the Max Planck Society, and the Higher Education Funding Council for England. The SDSS was managed by the Astrophysical Research Consortium for the Participating Institutions.\\

This research has made use of the NASA/IPAC Extragalactic Database (NED) which is operated by the Jet Propulsion Laboratory, California Institute of Technology, under contract with the National Aeronautics and Space Administration.\\

PMAS is operated as a common user instrument at CAHA under contract with the Leibniz-Insitute for Astrophysics Potsdam (AIP).
\end{acknowledgements}

\bibliographystyle{aa}
\bibliography{references}

\begin{thebibliography}{47}
\expandafter\ifx\csname natexlab\endcsname\relax\def\natexlab#1{#1}\fi

\bibitem[{{Abazajian} {et~al.}(2009){Abazajian}, {Adelman-McCarthy},
  {Ag{\"u}eros}, {Allam}, {Allende Prieto}, {An}, {Anderson}, {Anderson},
  {Annis}, {Bahcall}, \& et~al.}]{Abazajian:2009}
{Abazajian}, K.~N., {Adelman-McCarthy}, J.~K., {Ag{\"u}eros}, M.~A., {et~al.}
  2009, \apjs, 182, 543

\bibitem[{{Aceituno}(2004)}]{Aceituno:2004}
{Aceituno}, J. 2004, Calar Alto Newsletter No.8,
  http://www.caha.es/newsletter/news04b/Aceituno/Newsletter.html

\bibitem[{{Aihara} {et~al.}(2011){Aihara}, {Allende Prieto}, {An}, {Anderson},
  {Aubourg}, {Balbinot}, {Beers}, {Berlind}, {Bickerton}, {Bizyaev}, {Blanton},
  {Bochanski}, {Bolton}, {Bovy}, {Brandt}, {Brinkmann}, {Brown}, {Brownstein},
  {Busca}, {Campbell}, {Carr}, {Chen}, {Chiappini}, {Comparat}, {Connolly},
  {Cortes}, {Croft}, {Cuesta}, {da Costa}, {Davenport}, {Dawson}, {Dhital},
  {Ealet}, {Ebelke}, {Edmondson}, {Eisenstein}, {Escoffier}, {Esposito},
  {Evans}, {Fan}, {Femen{\'{\i}}a Castell{\'a}}, {Font-Ribera}, {Frinchaboy},
  {Ge}, {Gillespie}, {Gilmore}, {Gonz{\'a}lez Hern{\'a}ndez}, {Gott}, {Gould},
  {Grebel}, {Gunn}, {Hamilton}, {Harding}, {Harris}, {Hawley}, {Hearty}, {Ho},
  {Hogg}, {Holtzman}, {Honscheid}, {Inada}, {Ivans}, {Jiang}, {Johnson},
  {Jordan}, {Jordan}, {Kazin}, {Kirkby}, {Klaene}, {Knapp}, {Kneib},
  {Kochanek}, {Koesterke}, {Kollmeier}, {Kron}, {Lampeitl}, {Lang}, {Le Goff},
  {Lee}, {Lin}, {Long}, {Loomis}, {Lucatello}, {Lundgren}, {Lupton}, {Ma},
  {MacDonald}, {Mahadevan}, {Maia}, {Makler}, {Malanushenko}, {Malanushenko},
  {Mandelbaum}, {Maraston}, {Margala}, {Masters}, {McBride}, {McGehee},
  {McGreer}, {M{\'e}nard}, {Miralda-Escud{\'e}}, {Morrison}, {Mullally},
  {Muna}, {Munn}, {Murayama}, {Myers}, {Naugle}, {Neto}, {Nguyen}, {Nichol},
  {O'Connell}, {Ogando}, {Olmstead}, {Oravetz}, {Padmanabhan},
  {Palanque-Delabrouille}, {Pan}, {Pandey}, {P{\^a}ris}, {Percival},
  {Petitjean}, {Pfaffenberger}, {Pforr}, {Phleps}, {Pichon}, {Pieri}, {Prada},
  {Price-Whelan}, {Raddick}, {Ramos}, {Reyl{\'e}}, {Rich}, {Richards}, {Rix},
  {Robin}, {Rocha-Pinto}, {Rockosi}, {Roe}, {Rollinde}, {Ross}, {Ross},
  {Rossetto}, {S{\'a}nchez}, {Sayres}, {Schlegel}, {Schlesinger}, {Schmidt},
  {Schneider}, {Sheldon}, {Shu}, {Simmerer}, {Simmons}, {Sivarani}, {Snedden},
  {Sobeck}, {Steinmetz}, {Strauss}, {Szalay}, {Tanaka}, {Thakar}, {Thomas},
  {Tinker}, {Tofflemire}, {Tojeiro}, {Tremonti}, {Vandenberg}, {Vargas
  Maga{\~n}a}, {Verde}, {Vogt}, {Wake}, {Wang}, {Weaver}, {Weinberg}, {White},
  {White}, {Yanny}, {Yasuda}, {Yeche}, \& {Zehavi}}]{Aihara:2011}
{Aihara}, H., {Allende Prieto}, C., {An}, D., {et~al.} 2011, \apjs, 193, 29

\bibitem[{{Alonso-Herrero} {et~al.}(2012){Alonso-Herrero}, {Rosales-Ortega},
  {S{\'a}nchez}, {Kennicutt}, {Pereira-Santaella}, \&
  {D{\'{\i}}az}}]{Alonso-Herrero:2012}
{Alonso-Herrero}, A., {Rosales-Ortega}, F.~F., {S{\'a}nchez}, S.~F., {et~al.}
  2012, \mnras, 425, L46

\bibitem[{{Baldwin} {et~al.}(1981){Baldwin}, {Phillips}, \&
  {Terlevich}}]{Baldwin:1981}
{Baldwin}, J.~A., {Phillips}, M.~M., \& {Terlevich}, R. 1981, \pasp, 93, 5

\bibitem[{{Bertin} \& {Arnouts}(1996)}]{Bertin:1996}
{Bertin}, E. \& {Arnouts}, S. 1996, \aaps, 117, 393

\bibitem[{{Bertin} {et~al.}(2002){Bertin}, {Mellier}, {Radovich}, {Missonnier},
  {Didelon}, \& {Morin}}]{Bertin:2002}
{Bertin}, E., {Mellier}, Y., {Radovich}, M., {et~al.} 2002, in Astronomical
  Society of the Pacific Conference Series, Vol. 281, Astronomical Data
  Analysis Software and Systems XI, ed. D.~A. {Bohlender}, D.~{Durand}, \&
  T.~H. {Handley}, 228

\bibitem[{{Blanton} {et~al.}(2005){Blanton}, {Schlegel}, {Strauss},
  {Brinkmann}, {Finkbeiner}, {Fukugita}, {Gunn}, {Hogg}, {Ivezi{\'c}}, {Knapp},
  {Lupton}, {Munn}, {Schneider}, {Tegmark}, \& {Zehavi}}]{Blanton:2005}
{Blanton}, M.~R., {Schlegel}, D.~J., {Strauss}, M.~A., {et~al.} 2005, \aj, 129,
  2562

\bibitem[{{Cappellari} \& {Copin}(2003)}]{Cappellari:2003}
{Cappellari}, M. \& {Copin}, Y. 2003, \mnras, 342, 345

\bibitem[{{Cappellari} \& {Emsellem}(2004)}]{Cappellari:2004}
{Cappellari}, M. \& {Emsellem}, E. 2004, \pasp, 116, 138

\bibitem[{{Cardelli} {et~al.}(1989){Cardelli}, {Clayton}, \&
  {Mathis}}]{Cardelli:1989}
{Cardelli}, J.~A., {Clayton}, G.~C., \& {Mathis}, J.~S. 1989, \apj, 345, 245

\bibitem[{{Cid Fernandes} {et~al.}(2005){Cid Fernandes}, {Mateus}, {Sodr{\'e}},
  {Stasi{\'n}ska}, \& {Gomes}}]{CidFernandes:2005}
{Cid Fernandes}, R., {Mateus}, A., {Sodr{\'e}}, L., {Stasi{\'n}ska}, G., \&
  {Gomes}, J.~M. 2005, \mnras, 358, 363

\bibitem[{{Cid Fernandes} {et~al.}(2010){Cid Fernandes}, {Stasi{\'n}ska},
  {Schlickmann}, {Mateus}, {Vale Asari}, {Schoenell}, \&
  {Sodr{\'e}}}]{CidFernandes:2010}
{Cid Fernandes}, R., {Stasi{\'n}ska}, G., {Schlickmann}, M.~S., {et~al.} 2010,
  \mnras, 403, 1036

\bibitem[{{Dowler} {et~al.}(2011){Dowler}, {Rixon}, \& {Tody}}]{Dowler:2011}
{Dowler}, P., {Rixon}, G., \& {Tody}, D. 2011, ArXiv e-prints, arXiv:1110.0497

\bibitem[{{Fruchter} \& {Hook}(2002)}]{Fruchter:2002}
{Fruchter}, A.~S. \& {Hook}, R.~N. 2002, \pasp, 114, 144

\bibitem[{{Garc{\'{\i}}a-Benito}(2009)}]{Garcia-Benito:2009}
{Garc{\'{\i}}a-Benito}, R. 2009, PhD thesis, Universidad Autonoma de Madrid

\bibitem[{{Garc{\'{\i}}a-Benito} {et~al.}(2010){Garc{\'{\i}}a-Benito},
  {D{\'{\i}}az}, {H{\"a}gele}, {P{\'e}rez-Montero}, {L{\'o}pez},
  {V{\'{\i}}lchez}, {P{\'e}rez}, {Terlevich}, {Terlevich}, \&
  {Rosa-Gonz{\'a}lez}}]{Garcia-Benito:2010}
{Garc{\'{\i}}a-Benito}, R., {D{\'{\i}}az}, A., {H{\"a}gele}, G.~F., {et~al.}
  2010, \mnras, 408, 2234

\bibitem[{{Gonz{\'a}lez Delgado} {et~al.}(2005){Gonz{\'a}lez Delgado},
  {Cervi{\~n}o}, {Martins}, {Leitherer}, \&
  {Hauschildt}}]{Gonzalez-Delgado:2005}
{Gonz{\'a}lez Delgado}, R.~M., {Cervi{\~n}o}, M., {Martins}, L.~P.,
  {Leitherer}, C., \& {Hauschildt}, P.~H. 2005, \mnras, 357, 945

\bibitem[{{Greisen} \& {Calabretta}(2002)}]{Greisen:2002}
{Greisen}, E.~W. \& {Calabretta}, M.~R. 2002, \aap, 395, 1061

\bibitem[{{Heckman}(1980)}]{Heckman:1980}
{Heckman}, T.~M. 1980, \aap, 87, 152

\bibitem[{{Horne}(1986)}]{Horne:1986}
{Horne}, K. 1986, \pasp, 98, 609

\bibitem[{{Husemann} {et~al.}(2012){Husemann}, {Kamann}, {Sandin},
  {S{\'a}nchez}, {Garc{\'{\i}}a-Benito}, \& {Mast}}]{Husemann:2012a}
{Husemann}, B., {Kamann}, S., {Sandin}, C., {et~al.} 2012, \aap, 545, A137

\bibitem[{{Kauffmann} {et~al.}(2003){Kauffmann}, {Heckman}, {Tremonti},
  {Brinchmann}, {Charlot}, {White}, {Ridgway}, {Brinkmann}, {Fukugita}, {Hall},
  {Ivezi{\'c}}, {Richards}, \& {Schneider}}]{Kauffmann:2003}
{Kauffmann}, G., {Heckman}, T.~M., {Tremonti}, C., {et~al.} 2003, \mnras, 346,
  1055

\bibitem[{{Kehrig} {et~al.}(2012){Kehrig}, {Monreal-Ibero}, {Papaderos},
  {V{\'{\i}}lchez}, {Gomes}, {Masegosa}, {S{\'a}nchez}, {Lehnert}, {Cid
  Fernandes}, {Bland-Hawthorn}, {Bomans}, {Marquez}, {Mast}, {Aguerri},
  {L{\'o}pez-S{\'a}nchez}, {Marino}, {Pasquali}, {Perez}, {Roth},
  {S{\'a}nchez-Bl{\'a}zquez}, \& {Ziegler}}]{Kehrig:2012}
{Kehrig}, C., {Monreal-Ibero}, A., {Papaderos}, P., {et~al.} 2012, \aap, 540,
  A11

\bibitem[{{Kelz} {et~al.}(2006){Kelz}, {Verheijen}, {Roth}, {Bauer}, {Becker},
  {Paschke}, {Popow}, {S{\'a}nchez}, \& {Laux}}]{Kelz:2006}
{Kelz}, A., {Verheijen}, M.~A.~W., {Roth}, M.~M., {et~al.} 2006, \pasp, 118,
  129

\bibitem[{{Kewley} {et~al.}(2001){Kewley}, {Dopita}, {Sutherland}, {Heisler},
  \& {Trevena}}]{Kewley:2001}
{Kewley}, L.~J., {Dopita}, M.~A., {Sutherland}, R.~S., {Heisler}, C.~A., \&
  {Trevena}, J. 2001, \apj, 556, 121

\bibitem[{{M{\'a}rmol-Queralt{\'o}} {et~al.}(2011){M{\'a}rmol-Queralt{\'o}},
  {S{\'a}nchez}, {Marino}, {Mast}, {Viironen}, {Gil de Paz},
  {Iglesias-P{\'a}ramo}, {Rosales-Ortega}, \& {Vilchez}}]{Marmol-Queralto:2011}
{M{\'a}rmol-Queralt{\'o}}, E., {S{\'a}nchez}, S.~F., {Marino}, R.~A., {et~al.}
  2011, \aap, 534, A8

\bibitem[{{Rosales-Ortega}(2009)}]{Rosales-Ortega:2009}
{Rosales-Ortega}, F.~F. 2009, PhD thesis, PhD Thesis, University of Cambridge,
  2010.

\bibitem[{{Rosales-Ortega}(2011)}]{Rosales-Ortega:2011}
{Rosales-Ortega}, F.~F. 2011, \na, 16, 220

\bibitem[{{Rosales-Ortega} {et~al.}(2011){Rosales-Ortega}, {D{\'{\i}}az},
  {Kennicutt}, \& {S{\'a}nchez}}]{Rosales-Ortega:2011b}
{Rosales-Ortega}, F.~F., {D{\'{\i}}az}, A.~I., {Kennicutt}, R.~C., \&
  {S{\'a}nchez}, S.~F. 2011, \mnras, 415, 2439

\bibitem[{{Rosales-Ortega} {et~al.}(2010){Rosales-Ortega}, {Kennicutt},
  {S{\'a}nchez}, {D{\'{\i}}az}, {Pasquali}, {Johnson}, \&
  {Hao}}]{Rosales-Ortega:2010}
{Rosales-Ortega}, F.~F., {Kennicutt}, R.~C., {S{\'a}nchez}, S.~F., {et~al.}
  2010, \mnras, 405, 735

\bibitem[{{Rosales-Ortega} {et~al.}(2012){Rosales-Ortega}, {S{\'a}nchez},
  {Iglesias-P{\'a}ramo}, {D{\'{\i}}az}, {V{\'{\i}}lchez}, {Bland-Hawthorn},
  {Husemann}, \& {Mast}}]{Rosales-Ortega:2012}
{Rosales-Ortega}, F.~F., {S{\'a}nchez}, S.~F., {Iglesias-P{\'a}ramo}, J.,
  {et~al.} 2012, \apjl, 756, L31

\bibitem[{{Roth} {et~al.}(2005){Roth}, {Kelz}, {Fechner}, {Hahn}, {Bauer},
  {Becker}, {B{\"o}hm}, {Christensen}, {Dionies}, {Paschke}, {Popow}, {Wolter},
  {Schmoll}, {Laux}, \& {Altmann}}]{Roth:2005}
{Roth}, M.~M., {Kelz}, A., {Fechner}, T., {et~al.} 2005, \pasp, 117, 620

\bibitem[{{S{\'a}nchez}(2004)}]{Sanchez:2004c}
{S{\'a}nchez}, S.~F. 2004, Astronomische Nachrichten, 325, 167

\bibitem[{{S{\'a}nchez}(2006)}]{Sanchez:2006a}
{S{\'a}nchez}, S.~F. 2006, AN, 327, 850

\bibitem[{{S{\'a}nchez} {et~al.}(2007{\natexlab{a}}){S{\'a}nchez}, {Aceituno},
  {Thiele}, {P{\'e}rez-Ram{\'{\i}}rez}, \& {Alves}}]{Sanchez:2007b}
{S{\'a}nchez}, S.~F., {Aceituno}, J., {Thiele}, U., {P{\'e}rez-Ram{\'{\i}}rez},
  D., \& {Alves}, J. 2007{\natexlab{a}}, \pasp, 119, 1186

\bibitem[{{S{\'a}nchez} {et~al.}(2007{\natexlab{b}}){S{\'a}nchez}, {Cardiel},
  {Verheijen}, {Pedraz}, \& {Covone}}]{Sanchez:2007d}
{S{\'a}nchez}, S.~F., {Cardiel}, N., {Verheijen}, M.~A.~W., {Pedraz}, S., \&
  {Covone}, G. 2007{\natexlab{b}}, \mnras, 376, 125

\bibitem[{{S{\'a}nchez} {et~al.}(2012{\natexlab{a}}){S{\'a}nchez}, {Kennicutt},
  {Gil de Paz}, {van de Ven}, {V{\'{\i}}lchez}, {Wisotzki}, {Walcher}, {Mast},
  {Aguerri}, {Albiol-P{\'e}rez}, {Alonso-Herrero}, {Alves}, {Bakos},
  {Bart{\'a}kov{\'a}}, {Bland-Hawthorn}, {Boselli}, {Bomans},
  {Castillo-Morales}, {Cortijo-Ferrero}, {de Lorenzo-C{\'a}ceres}, {Del Olmo},
  {Dettmar}, {D{\'{\i}}az}, {Ellis}, {Falc{\'o}n-Barroso}, {Flores},
  {Gallazzi}, {Garc{\'{\i}}a-Lorenzo}, {Gonz{\'a}lez Delgado}, {Gruel},
  {Haines}, {Hao}, {Husemann}, {Igl{\'e}sias-P{\'a}ramo}, {Jahnke}, {Johnson},
  {Jungwiert}, {Kalinova}, {Kehrig}, {Kupko}, {L{\'o}pez-S{\'a}nchez},
  {Lyubenova}, {Marino}, {M{\'a}rmol-Queralt{\'o}}, {M{\'a}rquez}, {Masegosa},
  {Meidt}, {Mendez-Abreu}, {Monreal-Ibero}, {Montijo}, {Mour{\~a}o},
  {Palacios-Navarro}, {Papaderos}, {Pasquali}, {Peletier}, {P{\'e}rez},
  {P{\'e}rez}, {Quirrenbach}, {Rela{\~n}o}, {Rosales-Ortega}, {Roth},
  {Ruiz-Lara}, {S{\'a}nchez-Bl{\'a}zquez}, {Sengupta}, {Singh}, {Stanishev},
  {Trager}, {Vazdekis}, {Viironen}, {Wild}, {Zibetti}, \&
  {Ziegler}}]{Sanchez:2012a}
{S{\'a}nchez}, S.~F., {Kennicutt}, R.~C., {Gil de Paz}, A., {et~al.}
  2012{\natexlab{a}}, \aap, 538, A8

\bibitem[{{S{\'a}nchez} {et~al.}(2011){S{\'a}nchez}, {Rosales-Ortega},
  {Kennicutt}, {Johnson}, {Diaz}, {Pasquali}, \& {Hao}}]{Sanchez:2011}
{S{\'a}nchez}, S.~F., {Rosales-Ortega}, F.~F., {Kennicutt}, R.~C., {et~al.}
  2011, \mnras, 410, 313

\bibitem[{{S{\'a}nchez} {et~al.}(2012{\natexlab{b}}){S{\'a}nchez},
  {Rosales-Ortega}, {Marino}, {Iglesias-P{\'a}ramo}, {V{\'{\i}}lchez},
  {Kennicutt}, {D{\'{\i}}az}, {Mast}, {Monreal-Ibero}, {Garc{\'{\i}}a-Benito},
  {Bland-Hawthorn}, {P{\'e}rez}, {Gonz{\'a}lez Delgado}, {Husemann},
  {L{\'o}pez-S{\'a}nchez}, {Cid Fernandes}, {Kehrig}, {Walcher}, {Gil de Paz},
  \& {Ellis}}]{Sanchez:2012b}
{S{\'a}nchez}, S.~F., {Rosales-Ortega}, F.~F., {Marino}, R.~A., {et~al.}
  2012{\natexlab{b}}, \aap, 546, A2

\bibitem[{{Tody} {et~al.}(2011){Tody}, {Plante}, \& {Harrison}}]{Tody:2011}
{Tody}, D., {Plante}, R., \& {Harrison}, P. 2011, ArXiv e-prints,
  arXiv:1110.0499

\bibitem[{{van Dokkum}(2001)}]{Dokkum:2001}
{van Dokkum}, P.~G. 2001, \pasp, 113, 1420

\bibitem[{{Vazdekis} {et~al.}(2010){Vazdekis}, {S{\'a}nchez-Bl{\'a}zquez},
  {Falc{\'o}n-Barroso}, {Cenarro}, {Beasley}, {Cardiel}, {Gorgas}, \&
  {Peletier}}]{Vazdekis:2010}
{Vazdekis}, A., {S{\'a}nchez-Bl{\'a}zquez}, P., {Falc{\'o}n-Barroso}, J.,
  {et~al.} 2010, \mnras, 404, 1639

\bibitem[{{Veilleux} \& {Osterbrock}(1987)}]{Veilleux:1987}
{Veilleux}, S. \& {Osterbrock}, D.~E. 1987, \apjs, 63, 295

\bibitem[{{Verheijen} {et~al.}(2004){Verheijen}, {Bershady}, {Andersen},
  {Swaters}, {Westfall}, {Kelz}, \& {Roth}}]{Verheijen:2004}
{Verheijen}, M.~A.~W., {Bershady}, M.~A., {Andersen}, D.~R., {et~al.} 2004,
  Astronomische Nachrichten, 325, 151

\bibitem[{{Viironen} {et~al.}(2012){Viironen}, {S{\'a}nchez},
  {Marmol-Queralt{\'o}}, {Iglesias-P{\'a}ramo}, {Mast}, {Marino},
  {Crist{\'o}bal-Hornillos}, {Gil de Paz}, {van de Ven}, {Vilchez}, \&
  {Wisotzki}}]{Viironen:2012}
{Viironen}, K., {S{\'a}nchez}, S.~F., {Marmol-Queralt{\'o}}, E., {et~al.} 2012,
  \aap, 538, A144

\bibitem[{{Zibetti} {et~al.}(2009){Zibetti}, {Charlot}, \&
  {Rix}}]{Zibetti:2009}
{Zibetti}, S., {Charlot}, S., \& {Rix}, H.-W. 2009, \mnras, 400, 1181

\end{thebibliography}
\appendix
\section{CALIFA data reduction pipeline improvements}\label{apx:data_reduction}
\subsection{Detection of cosmics in single exposures} 
Although the individual exposure times for the CALIFA data are relatively short, numerous pixels on the CCD may still be hit by cosmic rays. These pixels must be identified and properly handled to avoid artifacts in the final data products. In pipeline V1.2, cosmic rays in single exposures were detected using a simplified Laplacian edge detection scheme based on \citet{Dokkum:2001}. However, the performance was limited, with many cosmics remaining undetected.  For this reason we developed a new detection algorithm \citep[{\tt PyCosmic},][]{Husemann:2012a} that achieves a much higher detection efficiency by combining a Laplacian edge detection scheme with a point spread function convolution approach. {\tt PyCosmic} is applied to each  raw frame just after it has been bias-subtracted and converted to electron counts. All detected cosmic rays are replaced by the median value of their surrounding pixels in {\tt PyCosmic}. Nevertheless, we use the associated bad pixel mask to properly mask those pixels throughout the V1.3c pipeline as suggested in \citet{Husemann:2012a}.

\subsection{Estimation and correction of instrument flexures} 
The PMAS instrument suffers from significant flexures \citep{Roth:2005}, which introduces small shifts of the signal on the CCD in the dispersion and cross-dispersion direction depending on the hour angle of the telescope.  Pipeline V1.2 invoked a pre-defined position list of $\sim$30 prominent night-sky emission line across the CCD to estimate the \emph{absolute} shifts in both directions. Subsequently, each raw image was resampled to a common reference frame if the offset exceeded 0.1\,pixel. This method was unstable because the absolute flexure offsets can exceed several pixels in which case the sky line positions might be miss-identified with lines in the neighboring fibers. 

\begin{figure}
 \resizebox{\hsize}{!}{\includegraphics{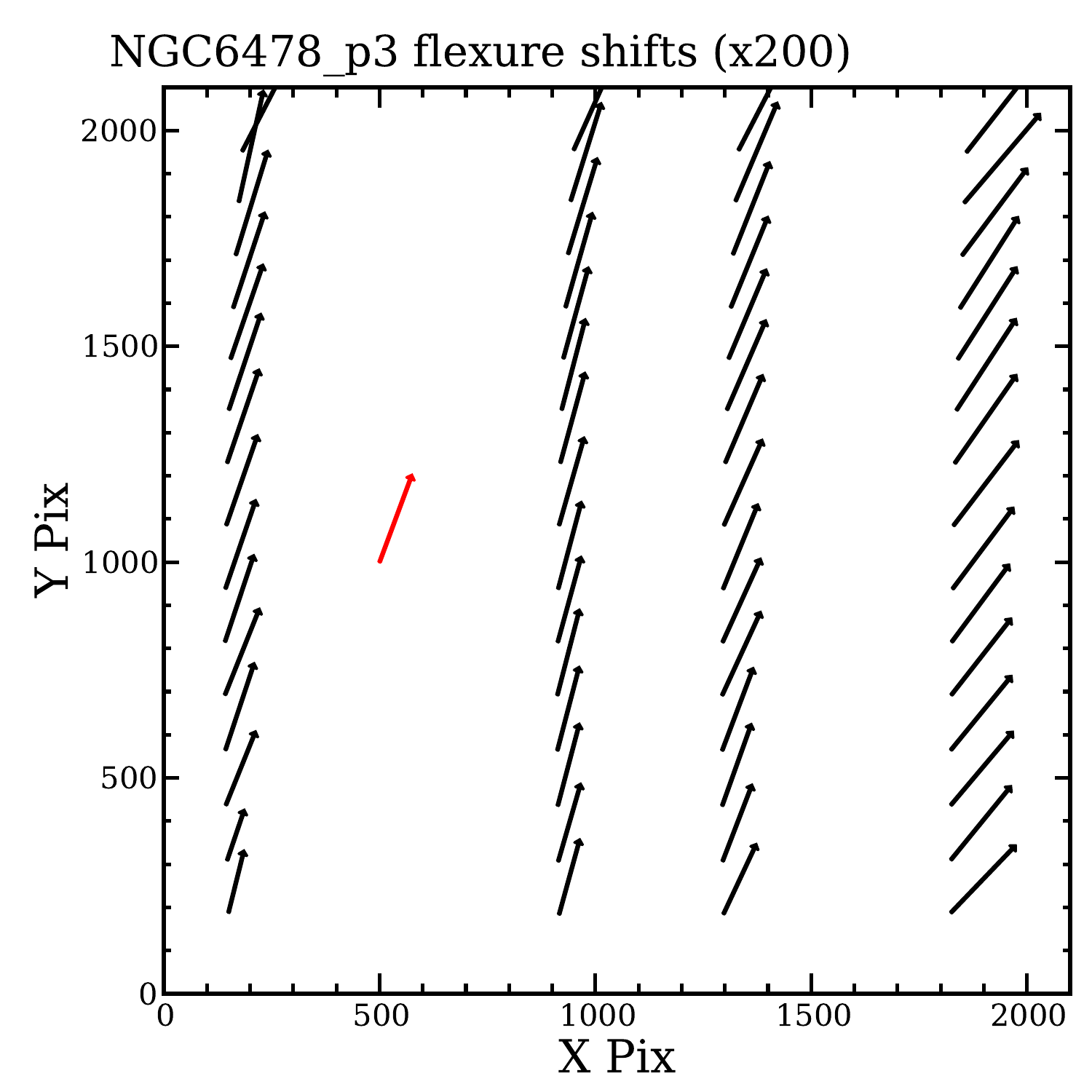}}
  \caption{Example of the measured flexure shifts for the third V500 pointing of NGC~6478. The black arrows indicate the shift in $x$ and $y$ direction in CCD detector coordinates for positions across the CCD in the science exposures with respect to the continuum and arc lamp calibration frames. The length of the arrows represent the shift in pixels but multiplied by a factor of 200 to be visible. The median shift of all measurements is shown by the red arrow.}
  \label{fig:flexure}
\end{figure}

In pipeline V1.3c, we only estimate \emph{relative flexure offsets} with respect to the continuum and arc-lamp calibration frame of each object in a two-stage process. Relative offsets in the cross-dispersion direction are measured by comparing the peak positions of fibers in each science frame close to the brightest night-sky emission lines with respect to their position in continuum lamp frame. After extraction of spectra, the wavelength of prominent sky lines is measured assuming the wavelength solution imposed by the arc-lamp calibration, which is compared with the expected wavelength of those lines. Any systematic deviation is attributed to flexure effects and the wavelength solution is corrected specifically for each individual science frame. These relative flexure offsets are often small ($\leq0.5$\,pixel) because of the relatively small changes in telescope position between the science and calibration exposures for CALIFA. An example of the flexure pattern is shown in Fig.~\ref{fig:flexure}, which indicates that the flexure shifts are nearly constant across the field. A global shift in dispersion and cross-dispersion is adopted for each frame. No resampling of the data is applied to correct for flexures in pipeline V1.3c, because only the calibration information is adjusted.

\subsection{Subtraction of straylight}\label{sect:straylight} 
An undesired feature of PMAS is an almost homogeneous background signal across the CCD image. Scattered light within the spectrograph, highly variable dark currents, or issues with the CCD controller could be potential origins    and we simply refer to this feature as ``straylight''. We find that the straylight contribution to some fibers can exceed 15\% in counts depending on their throughput and input signal.  The  strength of the straylight affects the S/N of an object because it increases the Poisson noise. Straylight was implicitly corrected in pipeline V1.2 by the sky subtraction, because the sky spectra inherit almost the same straylight signal as all other spectra. This approach, however, takes neither the subtle inhomogeneity of the straylight into account, nor assigns the correct error budget to the individual constituents, i.e. straylight, sky and object signal. 

\begin{figure}
 \resizebox{\hsize}{!}{\includegraphics{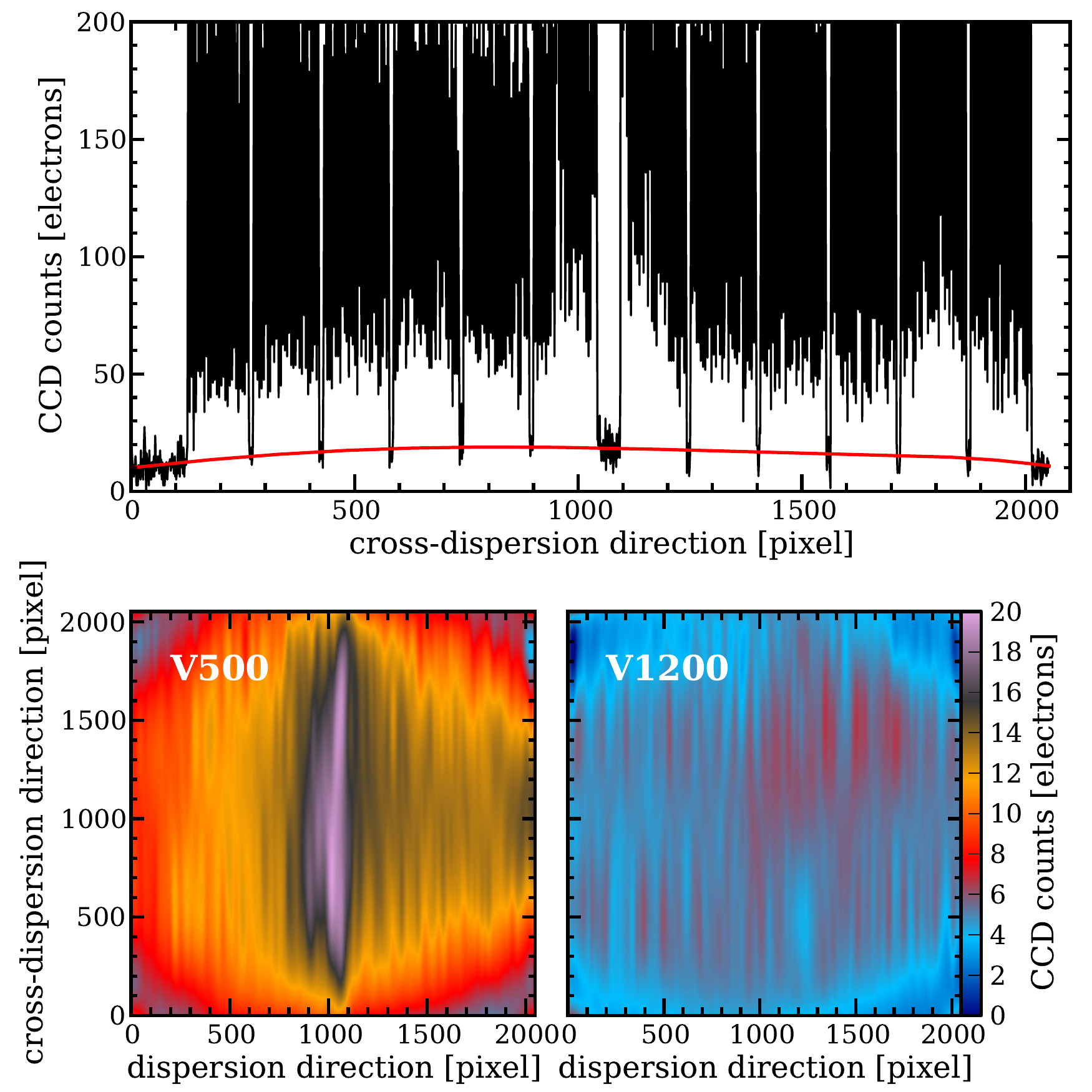}}
  \caption{Example of the straylight subtraction process. \emph{Top panel:} Pixel counts along a CCD cross-dispersion cut of the science observation. A sixth order polynomial fit only through the gaps between the fiber bundles is shown as the red line. \emph{Bottom panel:} Straylight maps for the V500 and V1200 data, respectively, reconstructed from the polynomial fit for each cross-dispersion slice of the CCD and smoothed by a 2D Gaussian filter with a 20\,pixel width.  }
  \label{fig:straylight}
\end{figure}

In pipeline V1.3c, we reconstruct the straylight signal using the few gaps between the fiber traces. First, the data is smoothed in the dispersion direction using a 5\,pixel wide median filter to suppress the noise of individual CCD pixels. In a second step, we fit a sixth order polynomial to each cross-dispersion slice of the CCD, considering only pixels more than 6\,pixels away from any fiber peak. An example is shown in Fig.~\ref{fig:straylight} (upper panel). Finally, the resulting straylight map is smoothed by a 2D Gaussian kernel with a dispersion of 20\,pixels to suppress any artificial high-frequency structure. Examples of the smoothed straylight distribution are shown in Fig.~\ref{fig:straylight} (lower panels) for the V1200 and V500 setup. This distribution is subtracted from the calibration and science exposures before the fiber spectra are extracted. The straylight signal is of the order of 5--30 counts for the majority of the data, but in very few cases it reached several hundred counts or showed a complex pattern potentially due to problems with the cooling of the CCD. The computed Poisson noise of the images is unchanged so that the effect of straylight is properly included in the error budget of the data.

\subsection{Extraction of the fiber spectra}
The amount of cross-talk between the PPak fibers was estimated in pipeline V1.2 during spectral extraction by means of a modified Gaussian suppression technique \citep{Sanchez:2006a}. This scheme achieved an accuracy of less than 1\% for the cross-talk if the width of each fiber profile is a-priori known along the dispersion axis. A constant FWHM of 2.5\,pixels in cross-dispersion direction was assumed in pipeline V1.2, but it changes significantly  across the CCD.  

\begin{figure}
 \resizebox{\hsize}{!}{\includegraphics{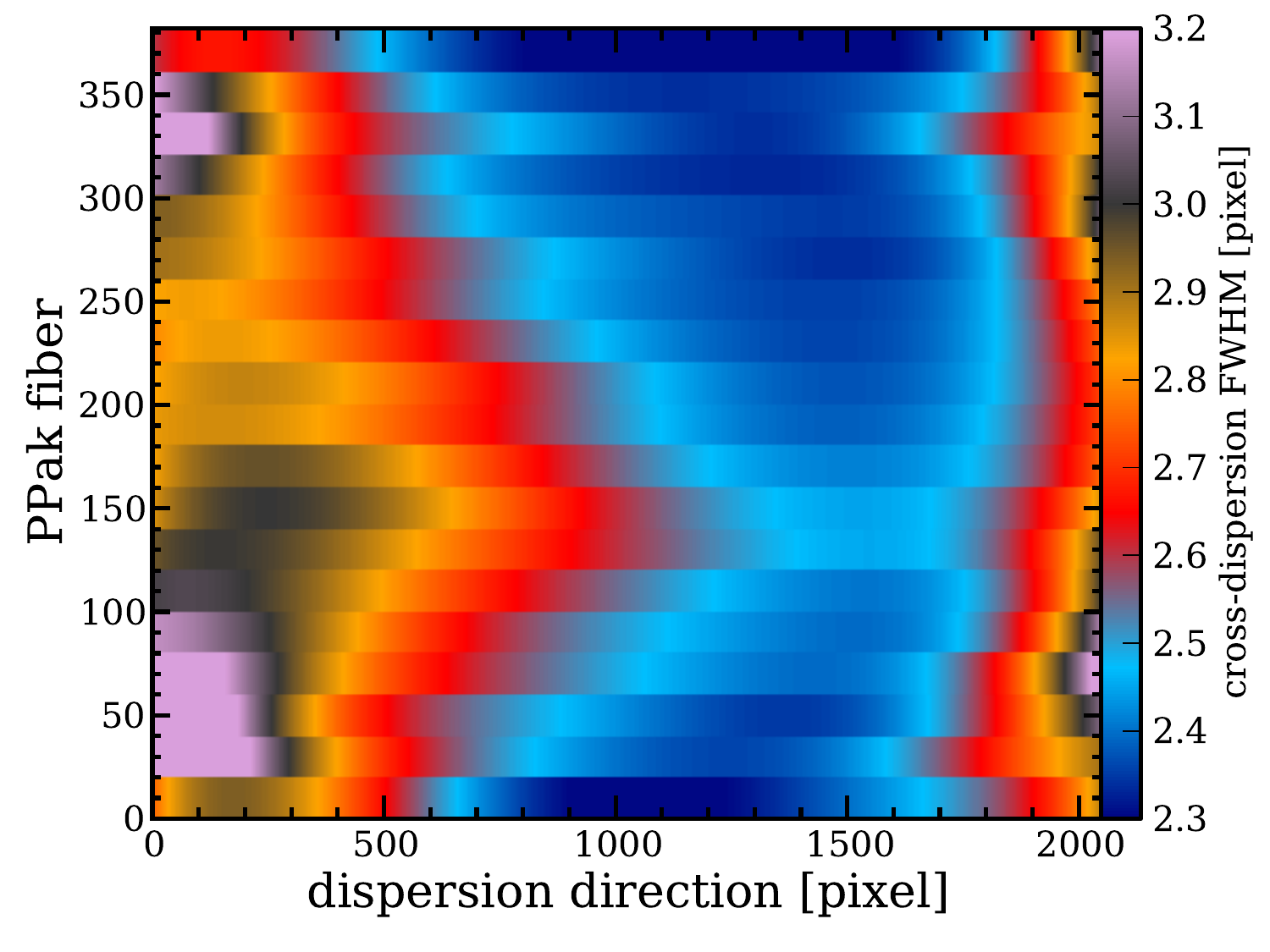}}
  \caption{Example of the measured fiber profile width in cross-dispersion along the dispersion direction for each fiber. The measurements are performed in blocks of 20 fibers and interpolated along the dispersion direction with a fifth order polynomial function.}
  \label{fig:crossFWHM}
\end{figure}

In pipeline V1.3c, we now measure the widths of the fiber profiles averaged over blocks of 20 fibers at each 50th spectral element along the dispersion axis using a $\chi^2$ minimization with the integrated fiber fluxes and a common FWHM per block as free parameters. The position of fibers is fixed based on the information from the continuum lamp (including any flexure offset). Subsequently, the measured fiber widths  are  interpolated by a 5th-order polynomial along the entire dispersion axis. An example of the estimated FWHM along the dispersion axis is shown in Fig.~\ref{fig:crossFWHM}. 

For the spectrum extraction in pipeline V1.3c, we use the optimal extraction algorithm \citep{Horne:1986} using a parallelized code that takes advantage of multi-core processor systems. The positions and widths of each Gaussian fiber profile are fixed to the values of the previous measurements. The advantage of optimal extraction is that errors can be analytically propagated based on the individual pixel errors. Bad pixels are also withdrawn during the extraction, despite a higher associated error. However, the extracted flux would be unreliable if all three of the most central pixels of a fiber are \emph{simultaneously} flagged as bad.  Thus, the corresponding spectral resolution element is flagged as bad in that case to avoid potential artifacts in the data. 

\subsection{Wavelength calibration and homogenization of the spectral resolution} 
The wavelength solution, as well as changes of the spectral resolution with wavelength are measured for each fiber based on the HeHgCd calibration lamp exposure taken for each CALIFA dataset. This information is used to resample the spectra to a linear grid in wavelength and to homogenize the spectral resolution to a common value along the dispersion axis using an adaptive Gaussian convolution. This scheme is unchanged in the current version of the pipeline, but now the flexure offsets in the dispersion direction are explicitly included in the wavelength solution to avoid a further resampling step.

The main goal of the V1.3c pipeline is to properly process the errors and bad pixel masks during the reduction step. The errors are analytically propagated during the Gaussian convolution and a Monte Carlo approach is used to estimate the noise vector after the spline resampling of the spectra. Both processing steps introduce some correlation between neighboring pixels in the data and the noise. This is not a serious problem unless the data are binned in the dispersion direction. The assumption of uncorrelated errors would lead to a higher S/N compared to what will be measured from the real spectra. Bad pixels are completely masked out during both processes. Given the loss of information during the spline resampling at these pixels, we expand the bad pixel mask by 2 neighboring pixels on both sides along the dispersion axis. Finally, the error value of bad pixels is set to a high value ($\sim10^{10}$ counts) and the actual data are replaced with a linear interpolation along the dispersion axis of the nearest unmasked pixels. 

\subsection{Fiber transmission correction}
Applying a correction for the underlying straylight signal (Sect.~\ref{sect:straylight}) also improves the quality of the fiber flat-fielding. This is particularly important for the vignetted areas at the corners of the CCD (see S12 for the details of the origin and effect of the vignetting), where the contribution of the straylight starts to dominate over the actual fiber signal. Consequently, we were able to reduce the threshold in the relative fiber transmission at which the data suffer from very low S/N from 75\% to 30\% in pipeline V1.3c.  Pixels below this threshold are flagged as bad pixels and set to zero value. This decreases the masked area caused by vignetting at the corners of the CCD. It also increases the reconstructed image quality because three low transmission fibers (55--70\% of transmission) within the PPak FoV now remain above this new transmission threshold and do not need to be masked out anymore.

\subsection{Flux calibration}
\begin{figure}
 \resizebox{\hsize}{!}{\includegraphics{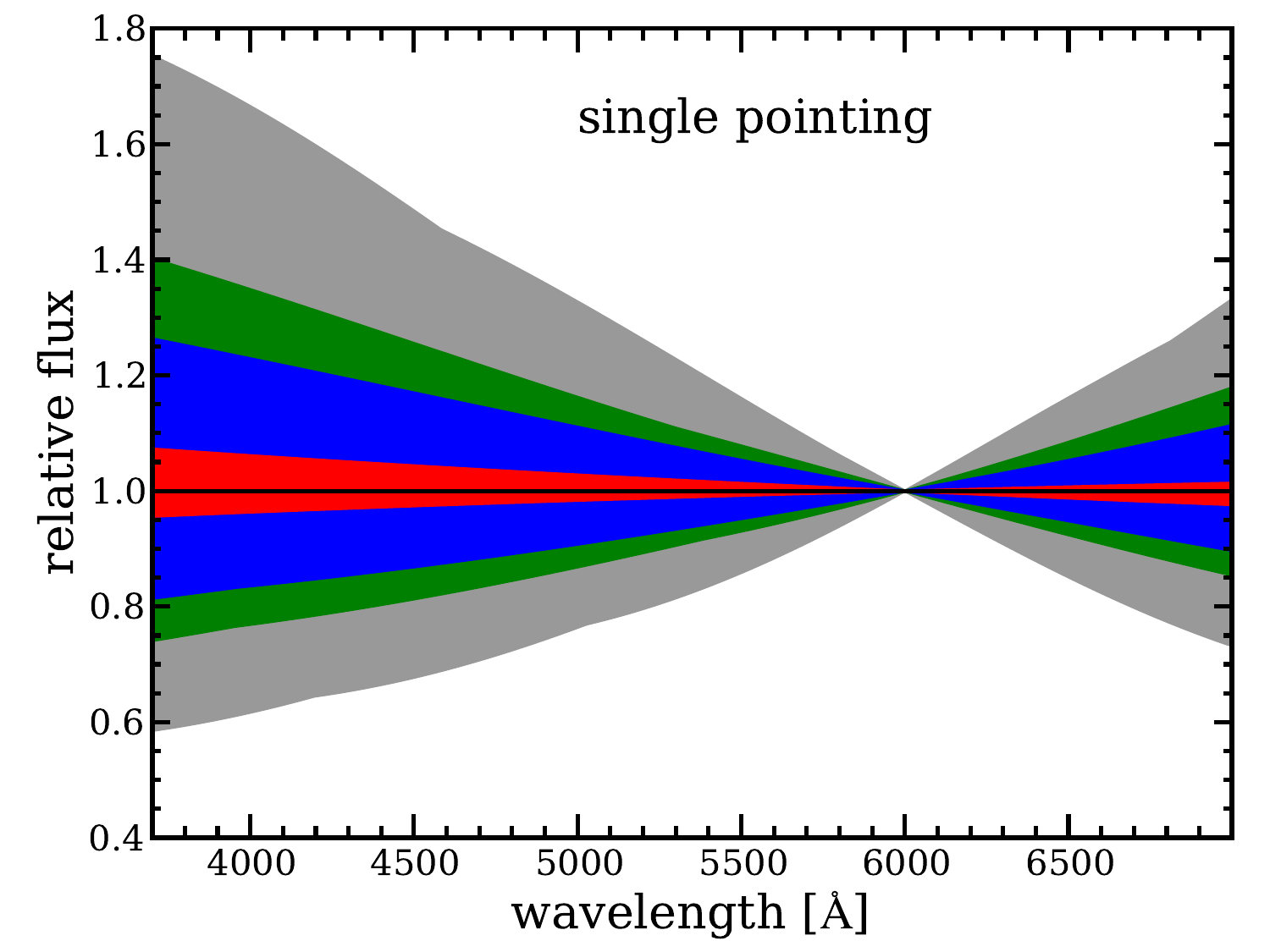}}
  \caption{Relative sensitivity curves normalized at 6000\AA\ from simulated (noise free) standard star observations with PPak at a seeing of 1.2\arcsec. Single pointing observations were simulated for a grid of pointing positions in $\alpha$ and $\delta$ with respect to the standard star. Simulations were performed for airmasses of 1.5 (gray), 1.3 (green),  1.1 (blue), and 1.0 (red) to take into account the change in the differential atmospheric refraction. The blue-to-red photometric calibration accuracy achievable is not sufficient even at low airmasses simply because of wavelength dependent aperture losses.}
  \label{fig:simulation_sens}
\end{figure}
The large fiber size and low filling factor of PPak can complicate obtaining an accurate sensitivity function for the whole wavelength range. The relative blue-to-red sensitivity function depends strongly on the actual pointing of the PPak fiber bundle with respect to the position of the standard star. The strength of this effect can be severe as inferred from a large suite of simulated standard star observations (see Fig.~\ref{fig:simulation_sens}). The reason for this is that the light of a standard star covered by a PPak fiber inevitably changes with wavelength because of atmospheric dispersion, thoroughly discussed in \citet{Garcia-Benito:2009}, \citet{Garcia-Benito:2010} and \citet{Rosales-Ortega:2009}. These classical aperture losses occur even if the standard star has been centered on a single fiber as accurately as possible for panchromatic observation as performed with the V500 setup. 

We therefore empirically calibrated these aperture losses by comparing the CALIFA spectra with available SDSS spectra. We selected two photometric nights (shortly before and after mirror re-coating in September 2011) which had the best conditions: (i) the standard star was almost centered on a single fiber, (ii) the seeing conditions were good ($<$1.5\arcsec), and (iii) the airmass was sufficiently low ($<$1.3). After extracting the calibration star spectra we corrected them for the atmospheric extinction along wavelength \citep{Sanchez:2007b} considering the airmass and the monitored $V$ band extinction at the time of observation. The corresponding sensitivity function was then used for flux calibration of the science data. In contrast to pipeline V1.2, we do not smooth the sensitivity function with a Gaussian kernel, but model it with a high-order polynomial to better preserve its shape while clipping outliers and residuals caused by stellar absorption line mismatches.
\begin{figure}
 \resizebox{\hsize}{!}{\includegraphics{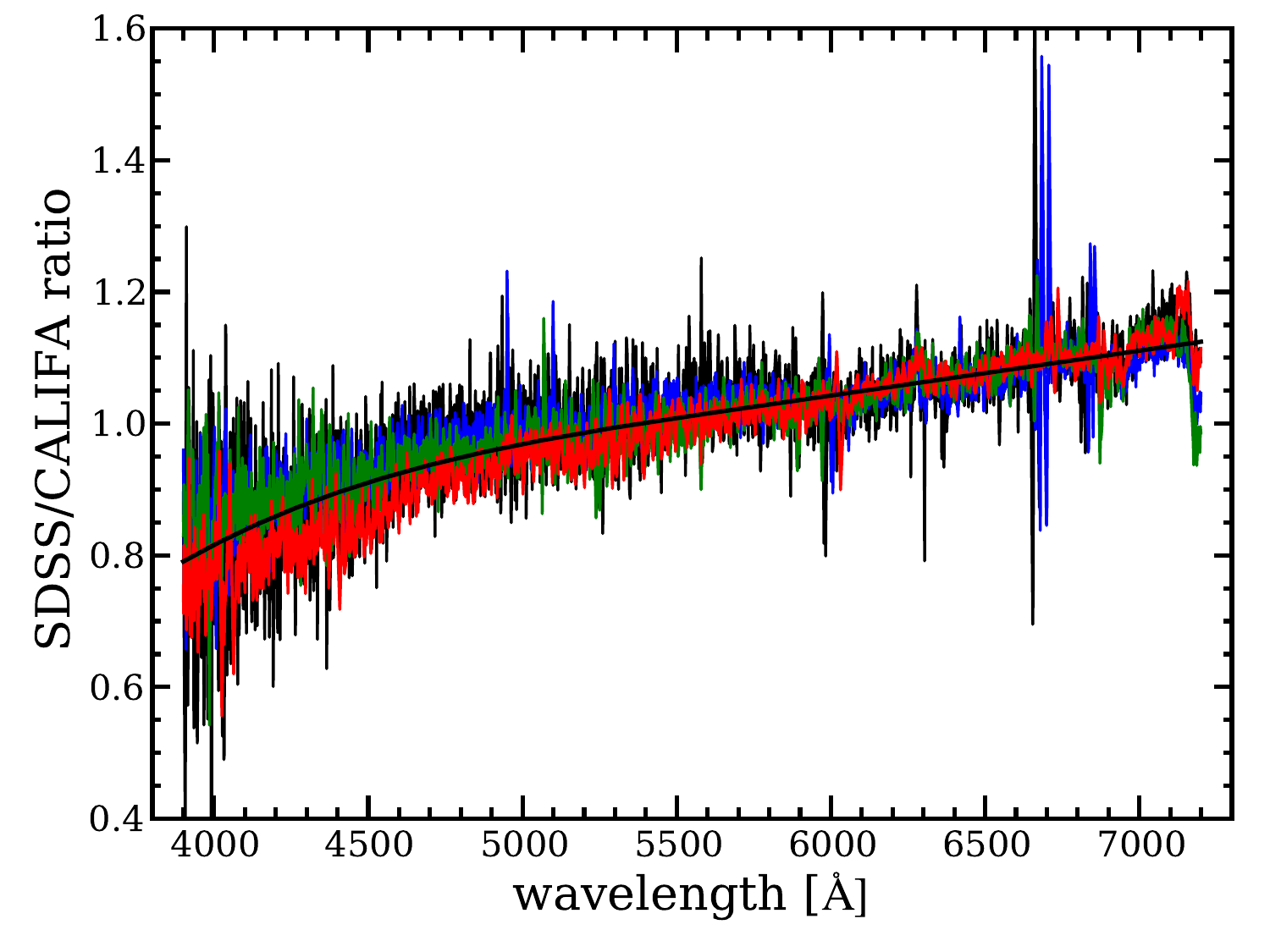}}
  \caption{The relative blue-to-red ratio between SDSS spectra and observed CALIFA V500 galaxy spectra (matched to the SDSS fiber aperture) along wavelength. All CALIFA spectra shown were obtained during the photometric night (2012 January 20th) and flux calibrated using the single pointing standard star observation of that night. A fourth order polynomial functions is sufficient to describe the trend presumably originating from the aperture losses of the standard star observation.}
  \label{fig:correct_sdss}
\end{figure}

SDSS DR7 spectra are available for $\sim$40\% of the CALIFA galaxies observed during those two nights. We extracted the spectrum within a radius of 1.5\arcsec\ around the galaxy center taking fractional pixel coverage into account to properly compare the spectrophotometry of the SDSS spectra and CALIFA data. The ratio of the 3\arcsec-diameter aperture SDSS and CALIFA spectra are shown in Fig.~\ref{fig:correct_sdss} for one of the nights. The significant red-to-blue deviation along the wavelength is attributed purely to aperture losses of standard star observations as discussed before. We corrected the initial sensitivity functions of the two nights correspondingly and used the mean of both as a master sensitivity function for the calibration of all the CALIFA data distributed in DR1. The quality of this approach is evaluated in Sect.~\ref{sect:specphot_cal_depth}.

\subsection{Sky subtraction}
A high S/N sky spectrum is obtained by combining the 36 sky fibers of PPak. In pipeline V1.2, we rejected outlier sky spectra using a 2$\sigma$ clipping. This approach turned out to be unstable when a bright field star or neighboring galaxy filled an entire sky fiber bundle consisting of 6 fibers. Thus, we decided to simply take the mean of the 30 faintest sky fibers for pipeline V1.3c, which effectively avoids this problem except if more than one sky fiber bundle would be affected. In such a rare case it would be difficult anyway to obtain a reliable sky spectrum from the few remaining sky fibers. Another advantage of this procedure is that the error vector for the mean sky spectrum is well-defined and easily propagated through the sky subtraction step. One caveat is that uncertainties in the wavelength solution and the spectral resolution matching are not included in the uncertainty of the sky subtraction.

\subsection{Image reconstruction and DAR correction} 
Since pipeline V1.2 a flux-conserving inverse-distance weighting scheme (see S12 for details) is used to reconstruct a spatial image from the three dithered PPak pointings with a sampling of $1\arcsec$. The effect of differential atmospheric refraction (DAR) on the data can be empirically measured from the spatially resampled datacube by tracing the centroid of the galaxy on the image plane along the wavelength axis. Common practice is often to resample each monochromatic image slice of the datacube again to align the galaxy centroid to a common reference position. 

It is crucial in pipeline V1.3 to avoid this additional resampling step  for a clean error propagation. Therefore, we adopt a two-stage iteration, (i) reconstruct the datacube and estimate the DAR effect, (ii) reconstruct the datacube again where the position of the fiber against the regular grid is changed according to the DAR offset empirically measured in the previous step. Although the error vector is well-defined for each spectrum in the reconstructed cubes, the inverse-distance weighting scheme introduces a strong correlation of the signal in the spatial dimension which is discussed in Sect.~\ref{sect:correlation} in more detail. 

Bad pixels are completely masked out during the weighting scheme so that in principle a full image can be reconstructed. However, a pixel value cannot be appropriately reconstructed if the contrast between the expected flux and the measured flux in the neighboring fibers is too high. We therefore flag each pixel in a reconstructed image as bad if the relative distance-based weight exceeds 5\% for a fiber masked as bad at a given spectral resolution element. This value was chosen by experience to properly flag apparent artifacts along the wavelength axis caused by this effect.

\end{document}